\documentclass[preprint, aps, prd]{revtex4-1}

\def\mc{\mathcal}

\usepackage[dvips]{graphicx}
\usepackage{amssymb}
\usepackage{amssymb,amsmath}
\usepackage{graphicx}
\usepackage{dsfont}
\usepackage{caption}
\usepackage{subcaption}
\makeatother
\begin{document}

\title{Supersymmetric solutions of $7D$ maximal gauged supergravity}

\author{Parinya Karndumri} \author{Patharadanai Nuchino} \email[REVTeX Support:
]{parinya.ka@hotmail.com and danai.nuchino@hotmail.com} 
\affiliation{String Theory and
Supergravity Group, Department of Physics, Faculty of Science,
Chulalongkorn University, 254 Phayathai Road, Pathumwan, Bangkok
10330, Thailand}

\date{\today}
\begin{abstract}
We study a number of supersymmetric solutions in the form of $Mkw_3\times S^3$- and $AdS_3\times S^3$-sliced domain walls in the maximal gauged supergravity in seven dimensions. These solutions require non-vanishing three-form fluxes to support the $AdS_3$ and $S^3$ subspaces. We consider solutions with $SO(4)$, $SO(3)$, $SO(2)\times SO(2)$ and $SO(2)$ symmetries in $CSO(p,q,5-p-q)$, $CSO(p,q,4-p-q)$ and $SO(2,1)\ltimes \mathbf{R}^4$ gauge groups. All of these solutions can be analytically obtained. For $SO(5)$ and $CSO(4,0,1)$ gauge groups, the complete truncation ansatze in terms of eleven-dimensional supergravity on $S^4$ and type IIA theory on $S^3$ are known. We give the full uplifted solutions to eleven and ten dimensions in this case. The solutions with an $AdS_3\times S^3$ slice are interpreted as two-dimensional surface defects in six-dimensional $N=(2,0)$ superconformal field theory in the case of $SO(5)$ gauge group or $N=(2,0)$ nonconformal field theories for other gauge groups. For $SO(4)$ symmetric solutions, it is possible to find solutions with both the three-form fluxes and $SO(3)$ gauge fields turned on. However, in this case, the solutions can be found only numerically. For $SO(3)$ symmetric solutions, the three-form fluxes and $SO(3)$ gauge fields cannot be non-vanishing simultaneously.
\end{abstract}
\maketitle

\section{Introduction}
Gauged supergravities in various space-time dimensions have become a useful tool for studying different aspects of the AdS/CFT correspondence \cite{maldacena,Gubser_AdS_CFT,Witten_AdS_CFT} and the DW/QFT correspondence \cite{DW_QFT1,DW_QFT2,DW_QFT3}. Solutions to gauged supergravities provide some insight to the dynamics of stongly-coupled conformal and non-conformal field theories via holographic descriptions, see for example \cite{holo_weyl,Skenderis_recon,go_with_RG,holo_renor,precise_non-conformal}. The study along this line is particularly fruitful in the presence of supersymmetry. In this case, many aspects of both the gravity and field theory sides are more controllable even at strong coupling. This makes finding various types of supersymmetric solutions in gauged supergravities worth considering.
\\
\indent In this paper, we are interested in supersymmetric solutions in the maximal gauged supergravity in seven dimensions. The solutions under consideration here take the form of $Mkw_3\times S^3$ and $AdS_3\times S^3$-sliced domain walls. This type of solutions has originally been considered in the minimal $N=2$ gauged supergravity in \cite{7D_sol_Dibitetto}, see also \cite{7D_N2_DW_3_form} for similar solutions in the matter-coupled $N=2$ gauged supergravity. Some of these solutions have been interpreted as surface defects within $N=(1,0)$ superconformal field theory (SCFT) in six dimensions in \cite{6D_surface_Dibitetto}, see \cite{6D_charged_DW,6D_charged_DW2} for similar solutions in six dimensions and \cite{defect1,defect2,defect3,defect4,defect5,defect6} for examples of another holographic description of conformal defects in terms of Janus solutions.
\\
\indent We will find these $Mkw_3\times S^3$ and $AdS_3\times S^3$-sliced domain walls in the maximal $N=4$ gauged supergravity with various types of gauge groups. The most general gaugings of the $N=4$ supergravity can be constructed by using the embedding tensor formalism \cite{N4_7D_Henning}, for an earlier construction see \cite{7DN4_gauged1} and \cite{7DN4_gauged2}. The embedding tensor describes the embedding of an admissible gauge group $G_0$ in the global symmetry group $SL(5)$ and encodes all information about the resulting gauged supergravity. Supersymmetry allows for two components of the embedding tensor transforming in $\mathbf{15}$ and $\overline{\mathbf{40}}$ representations of $SL(5)$. We will consider $CSO(p,q,5-p-q)$ and $CSO(p,q,4-p-q)$ gauge groups obtained from the embedding tensor in $\mathbf{15}$ and $\overline{\mathbf{40}}$ representations, respectively. We will also study similar solutions in $SO(2,1)\ltimes \mathbf{R}^4$ gauge group from the embedding tensor in both $\mathbf{15}$ and $\overline{\mathbf{40}}$ representations. Vacuum solutions in terms of half-supersymmetric domain walls for all these gauge groups have already been studied in \cite{our_7D_DW}. In this paper, we will extend these solutions, which involve only the metric and scalars, by including non-vanishing two- and three-form fields. In some cases, in addition to two- and three-form fields, it is also possible to couple $SO(3)$ gauge fields to the solutions.
\\
\indent As shown in \cite{Henning_Hohm1} using the framework of exceptional field theory, seven-dimensional gauged supergravity in $\mathbf{15}$ representation with $CSO(p,q,5-p-q)$ gauge group can be obtained from a consistent truncation of eleven-dimensional supergravity on $H^{p,q}\circ T^{5-p-q}$. On the other hand, a consistent truncation of type IIB theory on $H^{p,q}\circ T^{4-p-q}$ gives rise to $CSO(p,q,4-p-q)$ gauging from $\overline{\mathbf{40}}$ representation. This has been shown in \cite{Henning_Emanuel} along with a partial result on the corresponding truncation ansatze. In particular, internal components of all the ten-dimensional fields have been given.
\\
\indent For $SO(5)$ and $CSO(4,0,1)$ gauge groups, the complete truncation ansatze have already been constructed long ago in \cite{11D_to_7D_Nastase1,11D_to_7D_Nastase2} and \cite{S3_S4_typeIIA}. In this work, we will mainly consider uplifted solutions from these two gauge groups using the truncation ansatze given in \cite{11D_to_7D_Nastase1,11D_to_7D_Nastase2,S3_S4_typeIIA} which are more useful for solutions involving two- and three-form fields in seven dimensions. We leave uplifting solutions from other gauge groups for future work.
\\
\indent The paper is organized as follows. In section \ref{N4_7D_SUGRA}, we give a brief review of the
maximal gauged supergravity in seven dimensions. Supersymmetric $Mkw_3\times S^3$- and $AdS_3\times S^3$-sliced domain walls in $CSO(p,q,5-p-q)$ gauge group together with the uplifted solutions to eleven and ten dimensions in the case of $SO(5)$ and $CSO(4,0,1)$ gauge groups are presented in section \ref{Y_gauging}. Similar solutions for $CSO(p,q,4-p-q)$ and $SO(2,1)\ltimes \mathbf{R}^4$ gauge groups obtained from gaugings in $\overline{\mathbf{40}}$ and $(\mathbf{15},\overline{\mathbf{40}})$ representations are given in sections \ref{Z_gauging} and \ref{YZ_gauging}, respectively. Conclusions and comments are given in section \ref{conclusion}. In the two appendices, all bosonic field equations of the maximal gauged supergravity and consistent truncation ansatze for eleven-dimensional supergravity on $S^4$  and type IIA theory on $S^3$ are given.

\section{Maximal gauged supergravity in seven dimensions}\label{N4_7D_SUGRA}
In this section, we briefly review $N = 4$ gauged supergravity in seven dimensions in the embedding tensor formalism. We mainly focus on the bosonic Lagrangian and fermionic supersymmetry transformations which are relevant for finding supersymmetric solutions. The reader is referred to \cite{N4_7D_Henning} for the detailed construction of the maximal gauged supergravity.
\\
\indent As in other dimensions, the maximal $N=4$ supersymmetry in seven dimensions allows only the supergravity multiplet with the field content
\begin{equation}\label{fieldcon}
(e^{\hat{\mu}}_\mu, \psi^a_\mu, A^{MN}_\mu, B_{\mu\nu M}, \chi^{abc}, {\mathcal{V}_M}^A).
\end{equation}
This multiplet consists of the graviton $e^{\hat{\mu}}_\mu$, four gravitini $\psi^a_\mu$, ten vectors $A^{MN}_\mu=A^{[MN]}_\mu$, five two-form fields $B_{\mu\nu M}$, sixteen spin-$\frac{1}{2}$ fermions $\chi^{abc}=\chi^{[ab]c}$, and fourteen scalar fields described by the $SL(5)/SO(5)$ coset representative ${\mathcal{V}_M}^A$.
\\
\indent Throughout the paper, we will use the following convention on various types of indices. Curved and flat space-time indices are denoted by $\mu$, $\nu,\ldots$ and $\hat{\mu}$, $\hat{\nu},\ldots$, respectively. Lower (upper) $M,N=1,...,5$ indices refer to the (anti-) fundamental representation \textbf{5} ($\overline{\mathbf{5}}$) of the global $SL(5)$ symmetry. Accordingly, the vector $A^{MN}_\mu$ and two-form $B_{\mu\nu M}$ fields transform in the representations $\overline{\mathbf{10}}$ and $\mathbf{5}$, respectively.
\\
\indent On the other hand, fermionic fields transform in representations of the local $SO(5)\sim USp(4)$ R-symmetry with $USp(4)$ fundamental or $SO(5)$ spinor indices $a,b,\ldots =1,...,4$. The gravitini then transform as $\mathbf{4}$ while the spin-$\frac{1}{2}$ fields $\chi^{abc}$ transform as $\mathbf{16}$ of $USp(4)$. The latter satisfy the following conditions
\begin{equation}
\chi^{[abc]}=0\qquad \textrm{and}\qquad \Omega_{ab}\chi^{abc}=0
\end{equation}
with $\Omega_{ab}=\Omega_{[ab]}$ being the $USp(4)$ symplectic form satisfying the properties
\begin{equation}
(\Omega_{ab})^*=\Omega^{ab}\qquad \textrm{and}\qquad \Omega_{ac}\Omega^{bc}=\delta^b_a\, .
\end{equation}
\indent It should also be noted that raising and lowering of $USp(4)$ indices by $\Omega^{ab}$ and $\Omega_{ab}$ correspond to complex conjugation. Furthermore, all fermions are symplectic Majorana spinors subject to the conditions
\begin{equation}
\bar{\psi}^T_{\mu a}=\Omega_{ab}C\psi_\mu^b\qquad \textrm{and}\qquad \bar{\chi}^T_{abc}=\Omega_{ad}\Omega_{be}\Omega_{cf}C\chi^{def}
\end{equation}
where $C$ denotes the charge conjugation matrix obeying
\begin{equation}
C=C^T=-C^{-1}=-C^\dagger\, .
\end{equation}
With the space-time gamma matrices denoted by $\gamma^\mu$, the Dirac conjugate on a spinor $\Psi$ is defined by $\overline{\Psi}=\Psi^\dagger \gamma^0$.
\\
\indent The fourteen scalars parametrizing $SL(5)/SO(5)$ coset are described by the coset representative ${\mc{V}_M}^A$, transforming under the global $SL(5)$ and local $SO(5)$ symmetries by left and right multiplications. Indices $M=1,2,\ldots, 5$ and $A=1,2,\ldots ,5$ are accordingly $SL(5)$ and $SO(5)$ fundamental indices, respectively. In order to couple fermions which transform under $USp(4)$, we write the $SO(5)$ vector indices of ${\mc{V}_M}^A$ as a pair of antisymmetric $USp(4)$ fundamental indices in the form of ${\mc{V}_M}^{ab}={\mc{V}_M}^{[ab]}$. In addition, the coset representative ${\mathcal{V}_M}^{ab}$ satisfies the relation
\begin{equation}
{\mathcal{V}_M}^{ab}\Omega_{ab}=0\, .
\end{equation}
Similarly, the inverse of ${\mathcal{V}_M}^A$ denoted by ${\mc{V}_A}^M$ will be written as ${\mathcal{V}_{ab}}^M$. We then have the following relations
\begin{equation}
{\mathcal{V}_M}^{ab}{\mathcal{V}_{ab}}^N=\delta^N_M \qquad \textrm{and} \qquad {\mathcal{V}_{ab}}^M{\mathcal{V}_M}^{cd}=\delta^{[c}_{a}\delta^{d]}_{b}-\frac{1}{4}\Omega_{ab}\Omega^{cd}\, .
\end{equation}
\indent Gaugings are deformations of the $N=4$ supergravity by promoting a subgroup $G_0\subset SL(5)$ to be a local symmetry. The most general gaugings of a supergravity theory can be efficiently described by using the embedding tensor formalism. The embedding of $G_0$ within $SL(5)$ is achieved by using a constant $SL(5)$ tensor ${\Theta_{MN,P}}^Q={\Theta_{[MN],P}}^Q$ living in the product representation \cite{N4_7D_Henning}
\begin{equation}\label{Dectensor}
\mathbf{10}\otimes\mathbf{24}=\mathbf{10}+\mathbf{15}+\overline{\mathbf{40}}+\mathbf{175}\, .
\end{equation}
It turns out that supersymmetry allows only the embedding tensor in the $\mathbf{15}$ and $\overline{\mathbf{40}}$  representations. These two representations can be described by the tensors $Y_{MN}$ and $Z^{MN,P}$ with $Y_{MN}=Y_{(MN)}$, $Z^{MN,P}=Z^{[MN],P}$ and $Z^{[MN,P]}=0$ in terms of which the embedding tensor can be written as
\begin{equation}\label{embedding tensor}
{\Theta_{MN,P}}^Q=\delta^Q_{[M}Y_{N]P}-2\epsilon_{MNPRS}Z^{RS,Q}\, .
\end{equation}
\indent In term of the embedding tensor, gauge generators are given by
\begin{equation}\label{gaugeGen}
X_{MN}={\Theta_{MN,P}}^Q{t^P}_Q
\end{equation}
in which ${t^M}_N$, satisfying ${t^M}_M=0$, are $SL(5)$ generators. In particular, the gauge generators in the fundamental $\mathbf{5}$ and $\mathbf{10}$ representations are given by
\begin{eqnarray}
{X_{MN,P}}^Q&=& {\Theta_{MN,P}}^Q=\delta^Q_{[M}Y_{N]P}-2\epsilon_{MNPRS}Z^{RS,Q},\\
\textrm{and}\qquad {(X_{MN})_{PQ}}^{RS}&=&2{X_{MN,[P}}^{[R}\delta^{S]}_{Q]}
\end{eqnarray}
with $\epsilon_{MNPQR}$ being the invariant tensor of $SL(5)$. To ensure that the gauge generators form a closed subalgebra of $SL(5)$
\begin{equation}
\left[X_{MN},X_{PQ}\right]=-{(X_{MN})_{PQ}}^{RS}X_{RS},
\end{equation}
the embedding tensor needs to satisfy the quadratic constraint
\begin{equation}\label{QuadCon}
Y_{MQ}Z^{QN,P}+2\epsilon_{MRSTU}Z^{RS,N}Z^{TU,P}=0\, .
\end{equation}
\indent Gaugings introduce minimal coupling between the gauge fields and other fields via the covariant derivative
\begin{equation}\label{GenCovDiv}
D_\mu=\nabla_\mu-gA^{MN}_\mu{\Theta_{MN,P}}^Q{t^P}_Q
\end{equation}
where $\nabla_\mu$ is the spacetime covariant derivative including (possibly) composite $SO(5)$ connections. To restore supersymmetry of the original $N=4$ supergravity, fermionic mass-like terms and the scalar potential at first and second orders in the gauge coupling constant are needed. In addition, to ensure gauge covariance, the field strength tensors of vector and two-form fields need to be modified as
\begin{eqnarray}
\mathcal{H}_{\mu\nu}^{(2)MN}&=&F^{MN}_{\mu\nu}+gZ^{MN,P}B_{\mu\nu P},\label{ModTensor1}\\
\mathcal{H}_{\mu\nu\rho M}^{(3)}&=&gY_{MN}S^N_{\mu\nu\rho}+3D_{[\mu}B_{\nu\rho]M}\nonumber\\
&&+6\epsilon_{MNPQR}A^{NP}_{[\mu}(\partial_\nu A^{QR}_{\rho]}+\frac{2}{3}g {X_{ST,U}}^QA^{RU}_\nu A^{ST}_{\rho]})\label{ModTensor2}
\end{eqnarray}
where the non-abelian gauge field strength tensor is defined by
\begin{equation}\label{Ful2Form}
F^{MN}_{\mu\nu}=2\partial_{[\mu}A^{MN}_{\nu]}+g{(X_{PQ})_{RS}}^{MN}A^{PQ}_{[\mu}A^{RS}_{\nu]}\, .
\end{equation}
Note that the three-form fields $S^M_{\mu\nu\rho}$ in $\mc{H}^{(3)}_{\mu\nu\rho}$ only appear under the projection of $Y_{MN}$. In ungauged supergravity, all of the three-form fields can be dualized to two-form fields. However, this is not the case in the gauged supergravity. Therefore, different gaugings lead to different field contents in the resulting gauged supergravity.
\\
\indent Following \cite{N4_7D_Henning}, we first define $s\equiv\text{rank}\ Z$ and $t\equiv\text{rank}\ Y$. In a given gauging, $t$ two-forms can be set to zero by tensor gauge transformations of the three-form fields. This results in $t$ self-dual massive three-forms. Similarly, $s$ gauge fields can be set to zero by tensor gauge transformations of the two-forms giving rise to $s$ massive two-form fields. It should also be pointed out that there can be massive vector fields arising from broken gauge symmetry via the usual Higgs mechanism. We can see that the numbers of two- and three-form tensor fields depend on the gauging under consideration. However, the quadratic constraint ensures that $t+s\leq 5$, so the degrees of freedom from the ten vector and five two-form fields in the ungauged supergravity are redistributed into two- and three-form fields in the gauged theory. This fact will affect our ansatz for finding supersymmetric solutions in subsequent sections. To summarize, we repeat the distribution of degrees of freedom after gauge fixing from \cite{N4_7D_Henning} in table \ref{DisTensor}.
\begin{table}[h]
\centering
\begin{tabular}{|r | c | c|}
\hline
fields & \# & \# d.o.f\\
\hline
massless vectors & $10-s$ & 5\\
massless 2-forms & $5-s-t$ & 10\\
massive 2-forms & $s$ & 15\\
massive sd. 3-forms & $t$ & 10 \\
\hline
\end{tabular}
\caption{Distribution of the tensor fields' degrees of freedom after gauge fixing.}\label{DisTensor}
\end{table}
\\
\indent The covariance two- and three-form field strengths satisfy the following modified Bianchi identities
\begin{eqnarray}
D_{[\mu}\mathcal{H}^{(2)MN}_{\nu\rho]}&=&\frac{1}{3}gZ^{MN,P}\mathcal{H}^{(3)}_{\mu\nu\rho P}, \label{DefBianchi1}\\
D_{[\mu}\mathcal{H}^{(3)}_{\nu\rho\lambda] M}&=&\frac{3}{2}\epsilon_{MNPQR}\mathcal{H}^{(2)NP}_{[\mu\nu}\mathcal{H}^{(2)QR}_{\rho\lambda]}+\frac{1}{4}gY_{MN}\mathcal{H}^{(4)N}_{\mu\nu\rho\lambda}\label{DefBianchi2}
\end{eqnarray}
where the covariant field strengths of the three-form fields are given by
\begin{eqnarray}\label{4-form}
Y_{MN}\mathcal{H}^{(4)N}_{\mu\nu\rho\lambda}&=&Y_{MN}\left[4D_{[\mu}S^N_{\nu\rho\lambda]}+6F^{NP}_{[\mu\nu}B_{\rho\lambda] P}+3gZ^{NP,Q}B_{[\mu\nu P}B_{\rho\lambda] Q}\right.\nonumber \\
& &\left.+4g\epsilon_{PQRVW}{X_{ST,U}}^VA^{NP}_{[\mu}A^{QR}_\nu A^{ST}_\rho A^{UW}_{\lambda]}+8\epsilon_{PQRST}A^{NP}_{[\mu}A^{QR}_\nu\partial_\rho A^{ST}_{\lambda]}\right].\nonumber \\
& &
\end{eqnarray}
It should be emphasized that the three-forms $S^M_{\mu\nu\rho}$ and its field strength tensors always appear under the projection by $Y_{MN}$.
\\
\indent With all these ingredients, the bosonic Lagrangian of the seven-dimensional maximal gauged supergravity can be written as
\begin{eqnarray}
e^{-1}\mathcal{L}&=&\frac{1}{2}R-\mathcal{M}_{MP}\mathcal{M}_{NQ}\mathcal{H}_{\mu\nu}^{(2)MN}\mathcal{H}^{(2)PQ\mu\nu }-\frac{1}{6}\mathcal{M}^{MN}\mathcal{H}_{\mu\nu\rho M}^{(3)}{\mathcal{H}^{(3)\mu\nu\rho}}_ N\nonumber \\
& &+\frac{1}{8}(D_\mu\mathcal{M}_{MN})(D^\mu\mathcal{M}^{MN})-e^{-1}\mathcal{L}_{VT}-\mathbf{V}\, .\label{BosLag}
\end{eqnarray}
In this equation, the scalar fields are described by a unimodular symmetric matrix
\begin{equation}\label{fullM}
\mathcal{M}_{MN}={\mathcal{V}_M}^{ab}{\mathcal{V}_N}^{cd}\Omega_{ac}\Omega_{bd}\, .
\end{equation}
Its inverse is given by
\begin{equation}
\mathcal{M}^{MN}={\mathcal{V}_{ab}}^M{\mathcal{V}_{cd}}^N\Omega^{ac}\Omega^{bd}\, .
\end{equation}
We will not give the explicit form of the vector-tensor topological term $\mathcal{L}_{VT}$ here due to its complexity but refer the reader to \cite{N4_7D_Henning}. Finally, the scalar potential is given by
\begin{eqnarray}\label{scalarPot}
\mathbf{V}&=& \frac{g^2}{64}\left[2\mathcal{M}^{MN}Y_{NP}\mathcal{M}^{PQ}Y_{QM}-(\mathcal{M}^{MN}Y_{MN})^2\right]\nonumber \\
& &+g^2Z^{MN,P}Z^{QR,S}\left(\mathcal{M}_{MQ}\mathcal{M}_{NR}\mathcal{M}_{PS}-\mathcal{M}_{MQ}\mathcal{M}_{NP}\mathcal{M}_{RS}\right).
\end{eqnarray}
\indent The supersymmetry transformations of fermionic fields which are essential for finding supersymmetric solutions read
\begin{eqnarray}
\delta\psi^a_\mu&=& D_\mu\epsilon^a-g\gamma_\mu A^{ab}_1\Omega_{bc}\epsilon^c+\frac{1}{15}\mathcal{H}_{\nu\rho\lambda M}^{(3)}({\gamma_\mu}^{\nu\rho\lambda}-\frac{9}{2}\delta^\nu_\mu\gamma^{\rho\lambda})\Omega^{ab}{\mathcal{V}_{bc}}^M\epsilon^c\nonumber \\ & &+\frac{1}{5}\mathcal{H}_{\nu\rho}^{(2)MN}({\gamma_\mu}^{\nu\rho}-8\delta^\nu_\mu\gamma^{\rho}){\mathcal{V}_M}^{ad}\Omega_{de}{\mathcal{V}_N}^{eb}\Omega_{bc}\epsilon^c,\\
\delta\chi^{abc}&=& 2\Omega^{cd}{P_{\mu de}}^{ab}\gamma^\mu\epsilon^e+gA^{d,abc}_2\Omega_{de}\epsilon^e\nonumber \\ & &
+2\mathcal{H}_{\mu\nu}^{(2)MN}\gamma^{\mu\nu}\Omega_{de}\left[{\mathcal{V}_M}^{cd}{\mathcal{V}_N}^{e[a}\epsilon^{b]}-\frac{1}{5}(\Omega^{ab}\delta^c_g-\Omega^{c[a}\delta^{b]}_g){\mathcal{V}_M}^{gf}\Omega_{fh}{\mathcal{V}_N}^{hd}\epsilon^{e}\right] \nonumber \\ & &
-\frac{1}{6}\mathcal{H}_{\mu\nu\rho M}^{(3)}\gamma^{\mu\nu\rho}{\mathcal{V}_{fe}}^M\left[\Omega^{af}\Omega^{be}\epsilon^c-\frac{1}{5}(\Omega^{ab}\Omega^{cf}+4\Omega^{c[a}\Omega^{b]f})\epsilon^e\right].
\end{eqnarray}
The covariant derivative of the supersymmetry parameters is defined by
\begin{equation}
D_\mu\epsilon^a=\nabla_\mu\epsilon^a-{Q_{\mu b}}^a\epsilon^b\, .
\end{equation}
The composite connection ${Q_{\mu a}}^b$ and the vielbein on the $SL(5)/SO(5)$ coset ${P_{\mu ab}}^{cd}$ are obtained from the following relation
\begin{equation}
{P_{\mu ab}}^{cd}+2{Q_{\mu [a}}^{[c}\delta^{d]}_{b]}= {\mathcal{V}_{ab}}^M(\partial_\mu{\mathcal{V}_M}^{cd}-gA^{PQ}_\mu{X_{PQ,M}}^N{\mathcal{V}_N}^{cd}).
\end{equation}
\indent The fermion shift matrices $A_1$ and $A_2$ are given by
\begin{eqnarray}
A^{ab}_1&=& -\frac{1}{4\sqrt{2}}\left(\frac{1}{4}B\Omega^{ab}+\frac{1}{5}C^{ab}\right),\label{A1}\\
A^{d,abc}_2&=&\frac{1}{2\sqrt{2}}\left[\Omega^{ec}\Omega^{fd}({C^{ab}}_{ef}-{B^{ab}}_{ef})\right. \nonumber \\
& &\left.+\frac{1}{4}(C^{ab}\Omega^{cd}+\frac{1}{5}\Omega^{ab}C^{cd}+\frac{4}{5}\Omega^{c[a}C^{b]d})\right]\label{A2}
\end{eqnarray}
with various components of $B$ and $C$ tensors defined by
\begin{eqnarray}\label{BCfunctions}
B&=&\frac{\sqrt{2}}{5}\Omega^{ac}\Omega^{bd}Y_{ab,cd},\\
{{B^{ab}}_{cd}}&=&\sqrt{2}\left[\Omega^{ae}\Omega^{bf}\delta^{[g}_c \delta^{h]}_d-\frac{1}{5}(\delta^{[a}_c \delta^{b]}_d-\frac{1}{4}\Omega^{ab}\Omega_{cd})\Omega^{eg}\Omega^{fh}\right]Y_{ef,gh},\\
C^{ab}&=&8\Omega_{cd}Z^{(ac)[bd]},\\
{C^{ab}}_{cd}&=&8\left(-\Omega_{ce}\Omega_{df}\delta^{[a}_g \delta^{b]}_h+\Omega_{g(c}\delta^{[a}_{d)} \delta^{b]}_e\Omega_{fh}\right)Z^{(ef)[gh]}\, .
\end{eqnarray}
In the above equations, we have introduced ``dressed'' components of the embedding tensor defined by
\begin{eqnarray}
Y_{ab,cd}&=&{\mathcal{V}_{ab}}^M{\mathcal{V}_{cd}}^NY_{MN},\\
\textrm{and}\qquad Z^{(ac)[ef]}&=&\sqrt{2}{\mathcal{V}_M}^{ab}{\mathcal{V}_N}^{cd}{\mathcal{V}_P}^{ef}\Omega_{bd}Z^{MN,P}\, .
\end{eqnarray}
\indent Finally, we note that the scalar potential can also be written in terms of the fermion-shift matrices $A_1$ and $A_2$ as
\begin{equation}
V=-15A_1^{ab}A_{1 ab}+ \frac{1}{8}A_2^{a,bcd}A_{2 a,bcd}=-15|A_1|^2+\frac{1}{8}|A_2|^2\, .
\end{equation}
In the following sections, we will find supersymmetric solutions in a number of possible gauge groups.

\section{Supersymmetric solutions from gaugings in $\mathbf{15}$ representation}\label{Y_gauging}
We begin with gaugings in $\mathbf{15}$ representation with $Z^{MN,P}=0$. The $SL(5)$ symmetry can be used to bring $Y_{MN}$ to the form
\begin{equation}
Y_{MN}=\text{diag}(\underbrace{1,..,1}_p,\underbrace{-1,..,-1}_q,\underbrace{0,..,0}_r),\qquad p+q+r=5\, .
\end{equation}
This corresponds to the gauge group
\begin{equation}\label{CSOgaugegroup}
CSO(p,q,r)\sim SO(p,q)\ltimes\mathbf{R}^{(p+q)r}\, .
\end{equation}
\indent To give an explicit parametrization of the $SL(5)/SO(5)$ coset, we first introduce $GL(5)$ matrices
\begin{equation}\label{GL(5,R)}
{(e_{MN})_K}^L=\delta_{MK}\delta_{N}^L\, .
\end{equation}
 We will use the following choice of $SO(5)$ gamma matrices to convert an $SO(5)$ vector index to a pair of antisymmetric spinor indices
\begin{eqnarray}
\Gamma_1&=&-\sigma_2\otimes\sigma_2, \qquad  \Gamma_2=\mathbf{I}_2\otimes\sigma_1,\qquad \Gamma_3=\mathbf{I}_2\otimes\sigma_3,\nonumber  \\
\Gamma_4&=&\sigma_1\otimes\sigma_2, \qquad \Gamma_5=\sigma_3\otimes\sigma_2
\end{eqnarray}
where $\sigma_i$ are the usual Pauli matrices. $\Gamma_A$ satisfy the following relations
\begin{eqnarray}
\{\Gamma_A,\Gamma_B\}&=&2\delta_{AB}\mathbf{I}_4,\qquad  (\Gamma_A)^{ab}=-(\Gamma_A)^{ba}, \nonumber \\
\Omega_{ab}(\Gamma_A)^{ab}&=&0, \qquad ((\Gamma_A)^{ab})^*=\Omega_{ac}\Omega_{bd}(\Gamma_A)^{cd}\, .
\end{eqnarray}
The symplectic form of $USp(4)$ is chosen to be
\begin{equation}
\Omega_{ab}=\Omega^{ab}=\mathbf{I}_2\otimes i\sigma_2\, .
\end{equation}
The coset representative of the form ${\mc{V}_M}^{ab}$ and the inverse ${\mc{V}_{ab}}^M$ are then obtained from the following relations
\begin{equation}
{\mathcal{V}_M}^{ab}=\frac{1}{2} {\mathcal{V}_M}^A(\Gamma_A)^{ab}\qquad \textrm{and}\qquad
{\mathcal{V}_{ab}}^M=\frac{1}{2} {\mathcal{V}_A}^M(\Gamma^A)_{ab}\, .
\end{equation}
\indent We will use the metric ansatz in the form of an $AdS_3\times S^3$-sliced domain wall
\begin{equation}
ds_7^2=e^{2U(r)}ds_{AdS_3}^2+e^{2V(r)}dr^2+e^{2W(r)}ds_{S^3}^2\, .
\end{equation}
The seven-dimensional coordinates are taken to be $x^\mu=(x^m,r,x^i)$ with $m=0,1,2$ and $i=4,5,6$. Note that $V(r)$ is an arbitrary non-dynamical function that can be set to zero with a suitable gauge choice. The explicit forms for the metrics on $AdS_3$ and $S^3$ are given in Hopf coordinates by
\begin{eqnarray}
ds_{AdS_3}^2&=&\frac{1}{\tau^2}\left[-dt^2+(dx^1)^2+(dx^2)^2+2\sinh{x^1}dtdx^2\right],\label{AdS3_metric}\\
ds_{S^3}^2&=&\frac{1}{\kappa^2}\left[(dx^4)^2+(dx^5)^2+(dx^6)^2+2\sin{x^5}dx^4dx^6\right]\label{S3_metric}
\end{eqnarray}
in which $\tau$ and $\kappa$ are constants. In the limit $\tau\rightarrow0$ and $\kappa\rightarrow0$, the $AdS_3$ and $S^3$ parts become flat Minkowski space $Mkw_3$ and flat space $\mathbb{R}^3$, respectively.
\\
\indent With the following choice of vielbeins
\begin{eqnarray}
e^{\hat{0}}&=&\frac{1}{\tau}e^{U(r)}(dt-\sinh{x^1}dx^2),\qquad e^{\hat{1}}=\frac{1}{\tau}e^{U(r)}(\cos{t}dx^1-\sin{t}\cosh{x^1}dx^2),\nonumber \\
e^{\hat{2}}&=&\frac{1}{\tau}e^{U(r)}(\sin{t}dx^1+\cos{t}\cosh{x^1}dx^2),\qquad
e^{\hat{3}}= e^{V(r)}dr,\nonumber \\
e^{\hat{4}}&=&\frac{1}{\kappa}e^{W(r)}(dx^4+\sin{x^5}dx^6),\qquad e^{\hat{5}}=\frac{1}{\kappa}e^{W(r)}(\cos{x^4}dx^5-\sin{x^4}\cos{x^5}dx^6),\nonumber \\
e^{\hat{6}}&=&\frac{1}{\kappa}e^{W(r)}(\sin{x^4}dx^5+\cos{x^4}\cos{x^5}dx^6),
\end{eqnarray}
we find the following non-vanishing components of the spin connection
\begin{eqnarray}
{{\omega_{\hat{n}}}^{\hat{m}}}_{\hat{3}}&=& e^{-V(r)}U'(r)\delta^{\hat{m}}_{\hat{n}}, \qquad \omega_{\hat{m}\hat{n}\hat{p}}=\frac{\tau}{2}e^{-U(r)}\varepsilon_{\hat{m}\hat{n}\hat{p}},\nonumber \\
{{\omega_{\hat{j}}}^{\hat{i}}}_{\hat{3}}&=& e^{-V(r)}W'(r)\delta^{\hat{i}}_{\hat{j}}, \qquad \omega_{\hat{i}\hat{j}\hat{k}}=\frac{\kappa}{2}e^{-W(r)}\varepsilon_{\hat{i}\hat{j}\hat{k}}
\end{eqnarray}
with the convention that $\varepsilon_{\hat{0}\hat{1}\hat{2}}=-\varepsilon^{\hat{0}\hat{1}\hat{2}}=\varepsilon_{\hat{4}\hat{5}\hat{6}}=\varepsilon^{\hat{4}\hat{5}\hat{6}}=1$. Throughout this paper, we will use a prime to denote the $r$-derivative.
\\
\indent Following \cite{7D_sol_Dibitetto}, we take the ansatz for the Killing spinors to be
\begin{equation}\label{DefDWKilling}
\epsilon^a=e^{U(r)/2}\left[\cos{\theta(r)}\mathbf{I}_8+\sin{\theta(r)}\gamma^{\hat{0}\hat{1}\hat{2}}\right]\epsilon^a_0
\end{equation}
with $\epsilon^a_0$ being constant spinors. In addition, we will use the following ansatz for the three-form field strength tensors
\begin{equation}\label{fulldyonic3form}
\mathcal{H}^{(3)}_{\hat{m}\hat{n}\hat{p} M}=k_M(r)e^{-3U(r)}\varepsilon_{\hat{m}\hat{n}\hat{p}} \qquad \textrm{and} \qquad
\mathcal{H}^{(3)}_{\hat{i}\hat{j}\hat{k} M}=l_M(r)e^{-3W(r)}\varepsilon_{\hat{i}\hat{j}\hat{k}}
\end{equation}
or, equivalently,
\begin{equation}
\mc{H}^{(3)}_M=k_M\textrm{vol}_{AdS_3}+l_M\textrm{vol}_{S^3}\, .
\end{equation}
In subsequent analysis, we will call the solutions with non-vanishing $\mc{H}^{(3)}_M$ ``charged'' domain walls.

\subsection{$SO(4)$ symmetric charged domain walls}
We first consider charged domain wall solutions with $SO(4)$ symmetry. As in \cite{our_7D_DW}, we will find supersymmetric solutions with a given unbroken symmetry from many gauge groups within a single framework. Gauge groups that can give rise to $SO(4)$ symmetric solutions are $SO(5)$, $SO(4,1)$ and $CSO(4,0,1)$. We will accordingly write $Y_{MN}$ in the following form
\begin{equation}\label{SO(4)Ytensor}
Y_{MN}=\text{diag}(+1,+1,+1,+1,\rho)
\end{equation}
where $\rho=+1,-1,0$ corresponding to $SO(5)$, $SO(4,1)$, and $CSO(4,0,1)$ gauge groups, respectively. With this embedding tensor, the $SO(4)$ residual symmetry is generated by $X_{MN}$ with $M,N=1,2,3,4$.
\\
\indent Among the fourteen scalars in $SL(5)/SO(5)$ coset, there is one $SO(4)$ invariant scalar corresponding to the noncompact generator
\begin{equation}\label{YSO(4)Ys}
\hat{Y}=e_{1,1}+e_{2,2}+e_{3,3}+e_{4,4}-4e_{5,5}\, .
\end{equation}
With the coset representative
\begin{equation}\label{YSO(4)coset}
\mathcal{V}=e^{\phi\hat{Y}},
\end{equation}
the scalar potential is given by
\begin{equation}\label{YSO(4)Pot}
\mathbf{V}=-\frac{g^2}{64}e^{-4\phi}(8+8\rho e^{10\phi}-\rho^2e^{20\phi}).
\end{equation}
For $\rho=1$, this potential admits two $AdS_7$ critical points with $SO(5)$ and $SO(4)$ unbroken symmetries. The former preserves all supersymmetry while the latter is non-supersymmetric. These vacua are given respectively by
\begin{equation}\label{SO(5)Cripoint}
\phi=0\qquad \textrm{ and }\qquad  \mathbf{V}_0=-\frac{15}{64}g^2
\end{equation}
and
\begin{equation}\label{SO(4)Cripoint}
\phi=\frac{1}{10}\ln{2}\qquad \text{ and }\qquad \mathbf{V}_0=-\frac{5g^2}{16\times2^{2/5}}\, .
\end{equation}
The cosmological constant is denoted by $\mathbf{V}_0$, the value of the scalar potential at the vacuum.
\\
\indent To preserve $SO(4)$ symmetry, we will keep only the following components of $\mc{H}_M^{(3)}$ nonvanishing
\begin{equation}\label{sdSO(3)dyonic3form}
\mathcal{H}^{(3)}_{\hat{m}\hat{n}\hat{p} 5}=k(r)e^{-3U(r)}\varepsilon_{\hat{m}\hat{n}\hat{p}} \qquad \textrm{and} \qquad
\mathcal{H}^{(3)}_{\hat{i}\hat{j}\hat{k} 5}=l(r)e^{-3W(r)}\varepsilon_{\hat{i}\hat{j}\hat{k}}\, .
\end{equation}
At this point, it is useful to consider the $\mc{H}^{(3)}_M$ contribution in more detail. For $SO(5)$ and $SO(4,1)$ gauge groups corresponding to a non-degenerate $Y_{MN}$, the field content of the gauged supergravity contains $t=5$ massive three-form fields $S^M_{\mu\nu\rho }$. For vanishing gauge and two form fields, the field strength tensor $\mc{H}^{(3)}_M$ is then given by
\begin{equation}
\mc{H}^{(3)}_{\mu \nu\rho M}=gY_{MN}S^N_{\mu\nu\rho}\, .
\end{equation}
Since the four-form field strengths do not enter the supersymmetry transformations of fermionic fields, the functions $k_M(r)$ and $l_M(r)$ will appear, in this case, algebraically in the resulting BPS equations. This is in contrast to the pure $N=2$ gauged supergravity considered in \cite{7D_sol_Dibitetto} in which the four-form field strength of the massive three-form field appears in the supersymmetry transformations. Therefore, in that case, the BPS conditions result in differential equations for $k(r)$ and $l(r)$.
\\
\indent For $CSO(4,0,1)$ gauge group with $Y_{55}=0$, $S^5_{\mu\nu\rho}$ does not contribute to $\mc{H}^{(3)}_M$, but, in this case with $s=0$ and $t=4$, there is $5-t=1$ massless two-form field $B_{\mu\nu 5}$ with the field strength
\begin{equation}
\mc{H}^{(3)}_{\mu\nu\rho 5}=3D_{[\mu}B_{\nu\rho]5}\, .
\end{equation}
To satisfy the Bianchi's identity $D\mc{H}^{(3)}=0$, we need $k'=l'=0$ or constant three-form fluxes. We will see that this is indeed the case for our BPS solutions. Taking this condition into account, we can write the ansatz for the two-form field as
\begin{equation}
B_M=k_M(r)\omega_2+l_M(r)\tilde{\omega}_2
\end{equation}
with $\textrm{vol}_{AdS_3}=d\omega_2$ and $\textrm{vol}_{S^3}=d\tilde{\omega}_2$. With the metrics given in \eqref{AdS3_metric} and \eqref{S3_metric}, the explicit form of $\omega_2$ and $\tilde{\omega}_2$ is given by
\begin{equation}
\omega_2=-\frac{1}{\tau^3}\sinh x^1 dt\wedge dx^2\qquad \textrm{and}\qquad \tilde{\omega}_2=-\frac{1}{\kappa^3}\sin x^5 dx^4\wedge dx^6\, .
\end{equation}
\indent After imposing two projection conditions
\begin{equation}\label{DefDWProj}
\gamma_{\hat{3}}\epsilon^a_0={(\Gamma_5)^a}_b\epsilon^b_0=\epsilon^a_0,
\end{equation}
we find the following BPS equations from the conditions $\delta\psi^a_\mu=0$ and $\delta\chi^{abc}=0$
\begin{eqnarray}
U'&=&\frac{e^{V-2\phi}}{80\cos{2\theta}}\left[g(8-\rho e^{10\phi})+3g\rho e^{10\phi}\cos{4\theta}-16\tau e^{2\phi-U}\sin{2\theta}\right],\label{YDefDWUflow}\\
W'&=&\frac{e^{V-2\phi}}{40\cos{2\theta}}\left[g(4+2\rho e^{10\phi})-g\rho e^{10\phi}\cos{4\theta}-8\tau e^{2\phi-U}\sin{2\theta}\right],\\
\phi'&=&\frac{e^{V-2\phi}}{80\cos{2\theta}}\left[g(4-3\rho e^{10\phi})-g\rho e^{10\phi}\cos{4\theta}-8\tau e^{2\phi-U}\sin{2\theta}\right],\\
\theta'&=&-\frac{1}{16}g\rho e^{V+8\phi}\sin{2\theta},\label{theta_eq1}\\
k&=&\frac{1}{8}e^{2U-4\phi}(4\tau-g\rho e^{U+8\phi}\sin{2\theta})\label{Yksol},\\
l&=&\frac{1}{8}e^{3W-6\phi}\left[g(\rho e^{10\phi}-2)\tan{2\theta}+4\tau e^{2\phi-U}\sec{2\theta}\right]\label{Ylsol}
\end{eqnarray}
together with an algebraic constraint
\begin{equation}\label{DefDWconstraint}
0=e^{-W}\kappa-e^{-U}\tau\sec{2\theta}+\frac{1}{2}ge^{-2\phi}\tan{2\theta}\, .
\end{equation}
We note here that the appearance of the $SO(5)$ gamma matrix $\Gamma_5$ in the projection conditions is due to the non-vanishing $\mc{H}^{(3)}_{\mu\nu\rho 5}$. Note also that the solutions are $\frac{1}{4}$-BPS since the Killing spinors $\epsilon^a_0$ are subject to two projectors. We now consider various possible solutions to these BPS equations.

\subsubsection{$Mkw_3\times\mathbb{R}^3$-sliced domain walls}
We begin with a simple case of $Mkw_3\times\mathbb{R}^3$-sliced domain walls with vanishing $\tau$ and $\kappa$. Imposing $\tau=\kappa=0$ into the constraint \eqref{DefDWconstraint} gives
\begin{equation}
0=\frac{1}{2}ge^{-2\phi}\tan{2\theta}\, .\label{con1}
\end{equation}
Setting $g=0$ corresponds to ungauged $N=4$ supergravity and gives rise to a supersymmetric $Mkw_3\times\mathbb{R}\times\mathbb{R}^3\sim Mkw_7$ background as expected.
\\
\indent Another possibility to satisfy the condition \eqref{con1} is to set $\tan 2\theta=0$ which implies $\theta=\frac{n\pi}{2}$, $n=0,1,2,3,\ldots$. For even $n$, we have $\sin\theta=0$ and, from \eqref{DefDWKilling}, the Killing spinors take the form
\begin{equation}
\epsilon^a=e^{U(r)/2}\epsilon^a_0
\end{equation}
with $\epsilon^a_0$ satisfying the projection conditions given in \eqref{DefDWProj}. For odd $n$ with $\cos\theta=0$, the Killing spinors become
\begin{equation}
\epsilon^a=e^{U(r)/2}\gamma^{\hat{0}\hat{1}\hat{2}}\epsilon^a_0\, .
\end{equation}
We can redefine $\epsilon^a_0$ to $\tilde{\epsilon}^a_0=\gamma^{\hat{0}\hat{1}\hat{2}}\epsilon^a_0$ satisfying the projection conditions
\begin{equation}
-\gamma_{\hat{3}}\epsilon^a_0={(\Gamma_5)^a}_b\epsilon^b_0=\epsilon^a_0\, .
\end{equation}
This differs from the projectors in \eqref{DefDWProj} only by a minus sign in the $\gamma_{\hat{3}}$ projector. Therefore, the two possibilities obtained from the condition $\tan2\theta=0$ are equivalent by flipping the sign of $\gamma_{\hat{3}}$ projector. We can accordingly choose $\theta=0$ without losing any generality.
\\
\indent With $\theta=0$, the BPS equations \eqref{YDefDWUflow} to \eqref{Ylsol} become
\begin{eqnarray}
U'&=&W'=\frac{1}{40}ge^{V-2\phi}(4+\rho e^{10\phi}),\\
\phi'&=&\frac{1}{20}ge^{V-2\phi}(1-\rho e^{10\phi}),\\
k&=&l=0\, .
\end{eqnarray}
By choosing $V=-3\phi$, we find the following solution
\begin{eqnarray}
U&=&W=2\phi-\frac{1}{4}\ln\left[1-\rho e^{10\phi}\right],\label{WarpDW}\\
e^{5\phi}&=&\frac{1}{\sqrt{\rho}}\tanh\left[\frac{\sqrt{\rho}}{4}(gr+C)\right]\label{scalarDW}
\end{eqnarray}
with an integration constant $C$. Since $k=l=\theta=0$, the $\Gamma_5$ projection in \eqref{DefDWProj} is not needed. This is then a half-supersymmetric solution with vanishing three-form fluxes and is exactly the $SO(4)$ symmetric domain wall studied in \cite{our_7D_DW}. Therefore, the $Mkw_3\times\mathbb{R}^3$-sliced solution is just the standard flat domain wall.

\subsubsection{$Mkw_3\times S^3$-sliced domain walls}
In this case, we look for domain wall solutions with $Mkw_3\times S^3$ slice. Following \cite{7D_sol_Dibitetto}, we choose the follwing gauge choice
\begin{equation}\label{Dibiettetogaugechoice}
e^{-V}=\frac{1}{16}e^{8\phi}\, .
\end{equation}
By setting $\tau=0$, we can solve the BPS equations \eqref{YDefDWUflow} - \eqref{Ylsol} and obtain the following solution, for  $\rho=\pm 1$,
\begin{eqnarray}
U&=&2\phi-\ln\left(\sin{2\theta}\right),\\
W&=&2\phi-\ln\left(\tan{2\theta}\right),\\
e^{10\phi}&=&2C(\cos{4\theta}-3)+(4C+\rho)\sec^2{2\theta},\\
k&=&-\frac{g}{8}\left(4\rho C+\csc^4{2\theta}\right)\tan^2{2\theta},\\
l&=&\frac{g}{16}\left[\rho C(\cos{8\theta}+3)-2(2\rho C+1)\cos{4\theta}\right]\csc^2{2\theta},\\
\theta&=&\arctan{\left(e^{-2g\rho r}\right)}
\end{eqnarray}
with $\kappa=-g/2$. $C$ is an integration constant in the solution for $\phi$.
\\
\indent For $SO(5)$ gauge group with $\rho=1$, the solution is locally asymptotic to the $N=4$ supersymmetric $AdS_7$ in the limit $r\rightarrow\infty$ with
\begin{equation}\label{LocAdS7DW}
U\sim W\sim 2gr,\qquad  \phi\sim \theta\sim0\, .
\end{equation}
It should be noted that in this limit, the main contribution to the solution is obtained from the scalar. The contribution from the three-form field strength is highly suppressed as can be seen from its components in flat basis given in \eqref{sdSO(3)dyonic3form}. In the limit $r\rightarrow0$, the solution is singular similar to the solution studied in \cite{7D_sol_Dibitetto}.
\\
\indent For $SO(4,1)$ gauge group with $\rho=-1$, there is no $AdS_7$ asymptotic since this gauge group does not admit a supersymmetric $AdS_7$ vacuum. In this case, the solution is the $SO(4)$ symmetric domain wall studied in \cite{our_7D_DW} with a dyonic profile of the three-form flux.
\\
\indent For $CSO(4,0,1)$ gauge group with $\rho=0$, the BPS equations \eqref{YDefDWUflow} - \eqref{Ylsol}, with $\tau=0$, become
\begin{eqnarray}
U'&=&W'=\frac{1}{10}ge^{V-2\phi}\sec{2\theta},\\
\phi'&=&\frac{1}{20}ge^{V-2\phi}\sec{2\theta},\\
\theta'&=&k=0,\label{theta_eq_1} \\
l&=&-\frac{1}{4}ge^{3W-6\phi}\tan{2\theta}\label{l_eq_CSO4_1}
\end{eqnarray}
together with the constraint
\begin{equation}
\kappa=-\frac{1}{2}ge^{W-2\phi}\tan{2\theta}\, .\label{con_1}
\end{equation}
Equation \eqref{theta_eq_1} implies that $\theta$ is constant. Note that for $\theta=0$, these equations reduce to those of the $Mkw_3\times\mathbb{R}^3$-sliced domain wall.
\\
\indent In the present case, the constraint \eqref{con_1} implies that $\theta$ cannot be zero since $\kappa\neq0$. Furthermore, a non-vanishing $\theta$ gives a non-trivial three-form flux according to \eqref{l_eq_CSO4_1} to support the $S^3$ part. For constant $\theta\neq 0$, we can find the following solution, after choosing $V=0$ gauge choice,
\begin{eqnarray}
U&=&W=2\phi, \qquad
k=0,\\
l&=&-\frac{1}{4}g\tan{2\theta},\\
e^{2\phi}&=&\frac{1}{10}gr\sec{2\theta}+2C
\end{eqnarray}
with an integration constant $C$. The constant $\theta$ is given by
\begin{equation}
\theta=-\frac{1}{2}\tan^{-1}\frac{2\kappa}{g}\, .
\end{equation}
As in the $SO(4,1)$ gauge group, it can be verified that for a given constant $\theta$, this solution is the $SO(4)$ symmetric domain wall of $CSO(4,0,1)$ gauge group given in \cite{our_7D_DW} with a magnetic profile of a constant three-form flux.

\subsubsection{$AdS_3\times S^3$-sliced domain walls}
We now consider more complicated solutions with an $AdS_3\times S^3$ slice. As in \cite{7D_sol_Dibitetto}, we begin with a simpler solution with a single warp factor $U=W$. From the BPS equations \eqref{YDefDWUflow} - \eqref{Ylsol}, imposing $U'=W'$ gives
\begin{equation}
\theta=0, \qquad k=l, \qquad \tau=\kappa\, .
\end{equation}
Setting $\theta=0$, we find that the BPS equations become
\begin{eqnarray}
U'&=&\frac{g}{40}e^{V-2\phi}(4+\rho e^{10\phi}),\\
\phi'&=&\frac{g}{20}e^{V-2\phi}(1-\rho e^{10\phi}),\\
k&=&\frac{1}{2}e^{2U-4\phi}\tau\, .
\end{eqnarray}
By choosing $V=-3\phi$, we obtain the following solution
\begin{eqnarray}
U&=&2\phi-\frac{1}{4}\ln\left(1-\rho e^{10\phi}\right),\label{fSO(4)DW}\\
e^{5\phi}&=&\frac{1}{\sqrt{\rho}}\tanh\left[\frac{\sqrt{\rho}}{4}(gr+C)\right],\\
k&=&\frac{1}{2}\tau\cosh\left[\frac{\sqrt{\rho}}{4}(gr+C)\right]\label{lSO(4)DW}
\end{eqnarray}
with an integration constant $C$. This solution is the $SO(4)$ symmetric domain wall coupled to a dyonic profile of the three-form flux.
\\
\indent For $SO(5)$ gauge group, the solution is locally asymptotic to the supersymmetric $AdS_7$ dual to $N=(2,0)$ SCFT in six dimensions. This solution is then expected to describe a surface defect, corresponding to the $AdS_3$ part, within the six-dimensional $N=(2,0)$ SCFT. Similarly, according to the DW/QFT correspondence, the usual $Mkw_6$-sliced domain wall without the three-form flux is dual to an $N=(2,0)$ non-conformal field theory in six dimensions. We then interpret the solutions for $SO(4,1)$ and $CSO(4,0,1)$ gauge groups as describing a surface defect within a non-conformal $N=(2,0)$ field theory in six dimensions.
\\
\indent We now consider more general solutions with the $AdS_3\times S^3$ slice. We will find the solutions for the cases of $\rho=\pm 1$ and $\rho=0$, separately. With the same gauge choice given in \eqref{Dibiettetogaugechoice}, the BPS equations \eqref{YDefDWUflow} - \eqref{Ylsol} for $\rho\neq0$ are solved by
\begin{eqnarray}
U&=&2\phi-\ln\left(\sin{2\theta}\right),\\
W&=&2\phi-\ln\left(\tan{2\theta}\right),\\
e^{10\phi}&=&\frac{3gC+2g\rho-4\tau\rho+4(\tau\rho-gC)\cos{4\theta}+gC\cos{8\theta}}{g(\cos{4\theta}+1)},\\
k&=&\frac{1}{8}\left(4\tau\csc^2{2\theta}-g\csc^4{2\theta}-4g\rho C\right)\tan^2{2\theta},\\
l&=&\frac{1}{8}\left(g\csc^2{2\theta}-2g\cot^2{2\theta}-4\tau+4g\rho C \sin^2{2\theta}\right),\\
\theta&=&\arctan{\left(e^{-2g\rho r}\right)}
\end{eqnarray}
together with the following relation obtained from the constraint \eqref{DefDWconstraint}
\begin{equation}
\kappa=-\frac{g}{2}+\tau\, .
\end{equation}
\indent As in the previous case, for $SO(5)$ gauge group, the solution is locally asymptotically $AdS_7$ given in \eqref{LocAdS7DW} as $r\rightarrow\infty$. For $SO(4,1)$ gauge group, the solution is a charged domain wall with a non-vanishing three-form flux. In general, these solutions describe respectively holographic RG flows from an $N=(2,0)$ SCFT and $N=(2,0)$ non-conformal field theory to a singularity at $r=0$ except for a special case with $\tau=g(\rho C+1)/4$. This is very similar to the solutions of pure $N=2$ gauged supergravity studied in \cite{7D_sol_Dibitetto}
\\
\indent For the particular value of $\tau=g(\rho C+1)/4$, the scalar potential is constant as $r\rightarrow0$, and the solution turns out to be described by a locally $AdS_3\times T^4$ geometry with the following leading profile
\begin{eqnarray}
e^{2U}&\sim &\left(\rho-4C\right)^{\frac{2}{5}},\qquad  e^{2W}\sim0,\qquad \phi\sim\frac{1}{10}\ln\left(\rho-4C\right),\nonumber \\
\theta&\sim&\frac{\pi}{4},\qquad k\sim \frac{g}{8}(4\rho C-1),\qquad l\sim 0\, .
\end{eqnarray}
To obtain real solutions, we choose the integration constant $C<\frac{1}{4}$ and $C<-\frac{1}{4}$ for $SO(5)$ and $SO(4,1)$ gauge groups, respectively.
\\
\indent For $CSO(4,0,1)$ gauge group with $\rho=0$, we find the following solution, after setting $V=0$,
\begin{eqnarray}
U&=&W=2\phi,\\
k&=&\frac{1}{2}\tau,\\
l&=&\frac{1}{4}(2\tau-g\sin{2\theta})\sec{2\theta},\\
e^{2\phi}&=&\frac{1}{10}r\left(g\sec{2\theta}-2\tau\tan{2\theta}\right)+2C
\end{eqnarray}
where the constant $\kappa$ is given by
\begin{equation}
\kappa=\tau\sec{2\theta}-\frac{1}{2}g\tan{2\theta}.
\end{equation}
Note also that, in this case, $\theta$ is constant since the corresponding BPS equation gives $\theta'=0$ as can be seen from equation \eqref{theta_eq1}.

\subsubsection{Coupling to $SO(3)$ gauge fields}
\begin{figure}[h!]
  \centering
  \begin{subfigure}[b]{0.32\linewidth}
    \includegraphics[width=\linewidth]{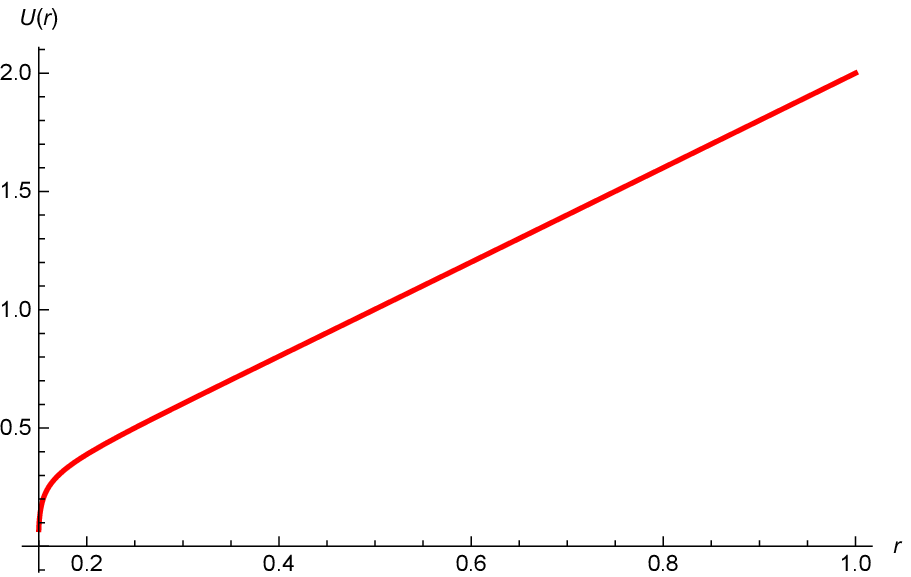}
  \caption{$U$ solution}
  \end{subfigure}
  \begin{subfigure}[b]{0.32\linewidth}
    \includegraphics[width=\linewidth]{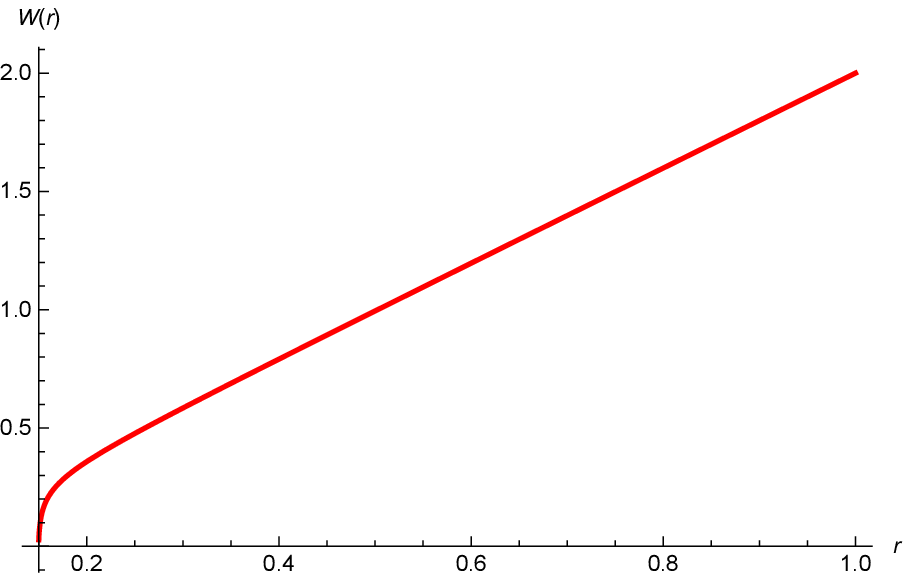}
  \caption{$W$ solution}
  \end{subfigure}
  \begin{subfigure}[b]{0.32\linewidth}
    \includegraphics[width=\linewidth]{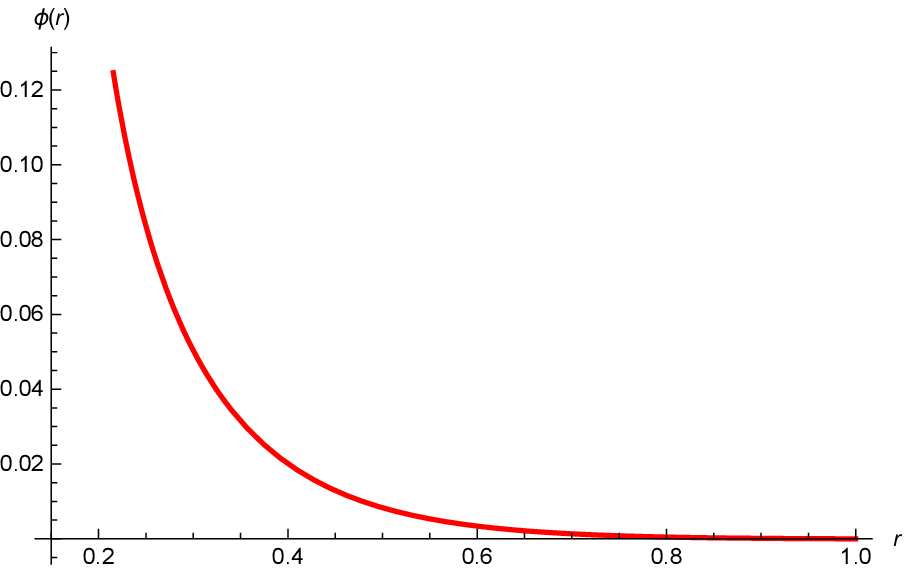}
  \caption{$\phi$ solution}
  \end{subfigure}
  \begin{subfigure}[b]{0.32\linewidth}
    \includegraphics[width=\linewidth]{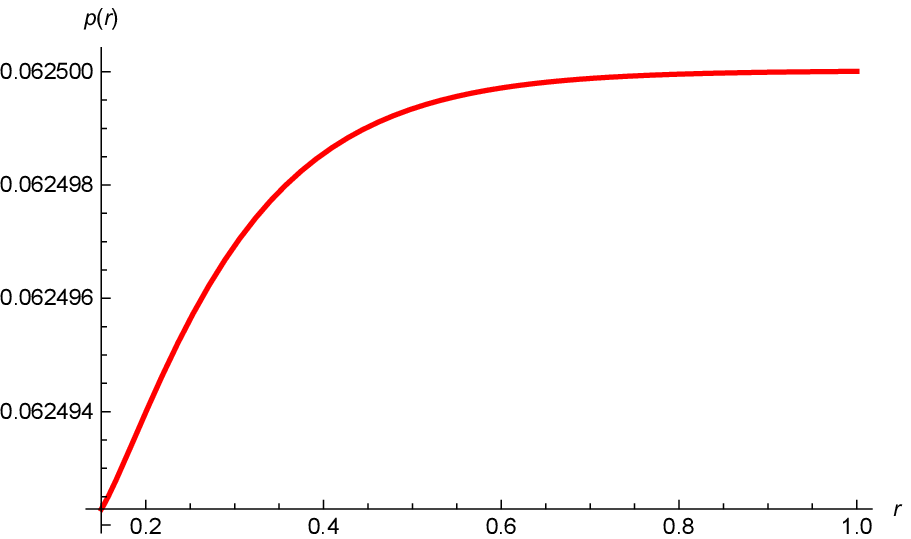}
  \caption{$p$ solution}
  \end{subfigure}
  \begin{subfigure}[b]{0.32\linewidth}
    \includegraphics[width=\linewidth]{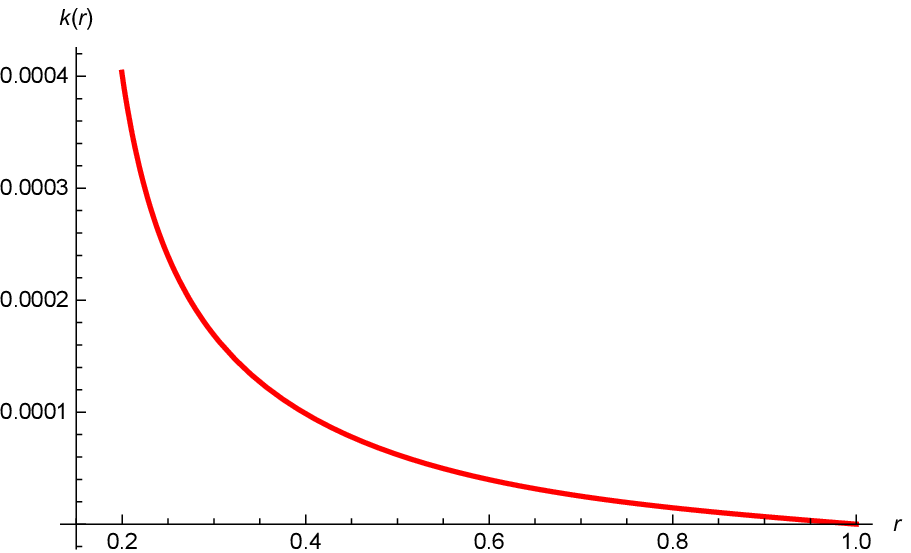}
  \caption{$k$ solution}
  \end{subfigure}
  \begin{subfigure}[b]{0.32\linewidth}
    \includegraphics[width=\linewidth]{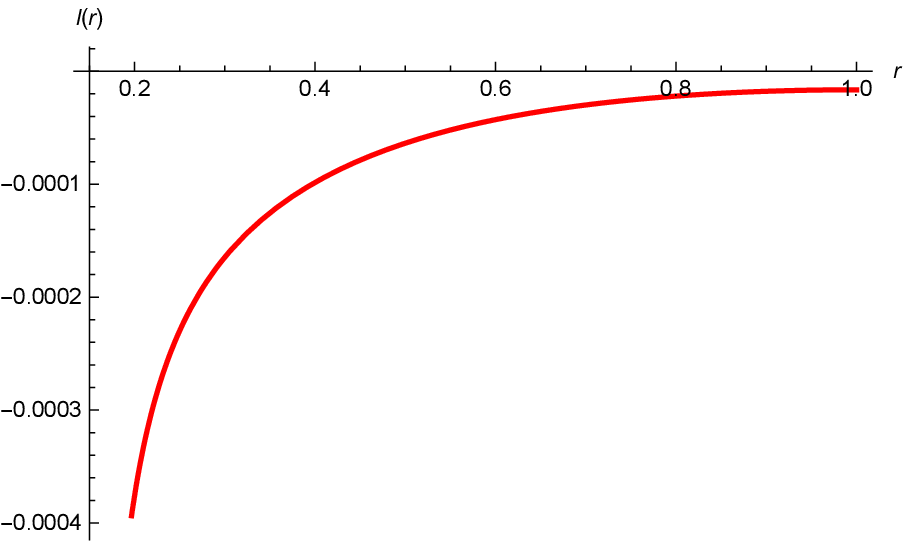}
  \caption{$l$ solution}
  \end{subfigure}
  \caption{A BPS flow from a locally $AdS_7$ geometry at $r\rightarrow\infty$ to the singularity at $r=0$ for the $Mkw_3\times S^3$-sliced domain wall with $\tau=0$.}
  \label{YDefDWflow1}
\end{figure}

\begin{figure}[h!]
  \centering
  \begin{subfigure}[b]{0.32\linewidth}
    \includegraphics[width=\linewidth]{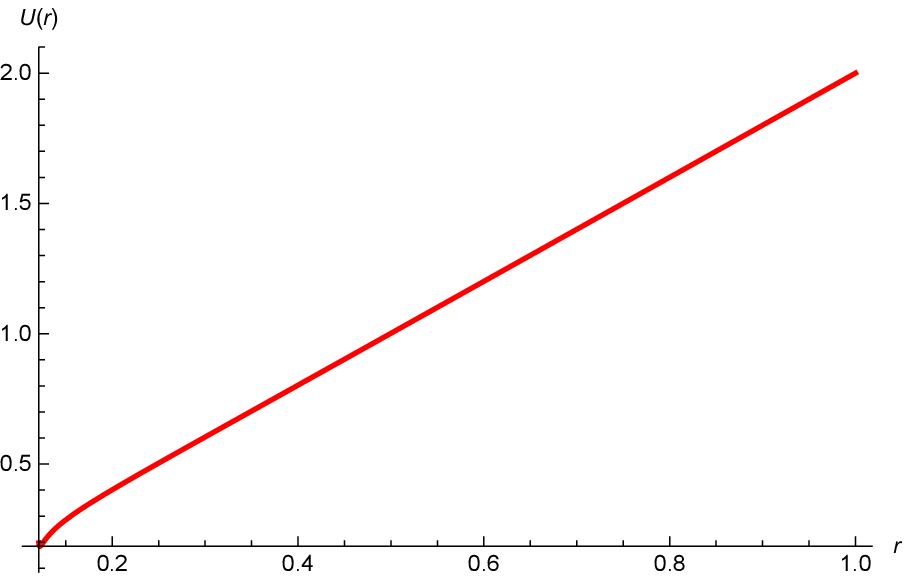}
  \caption{$U$ solution}
  \end{subfigure}
  \begin{subfigure}[b]{0.32\linewidth}
    \includegraphics[width=\linewidth]{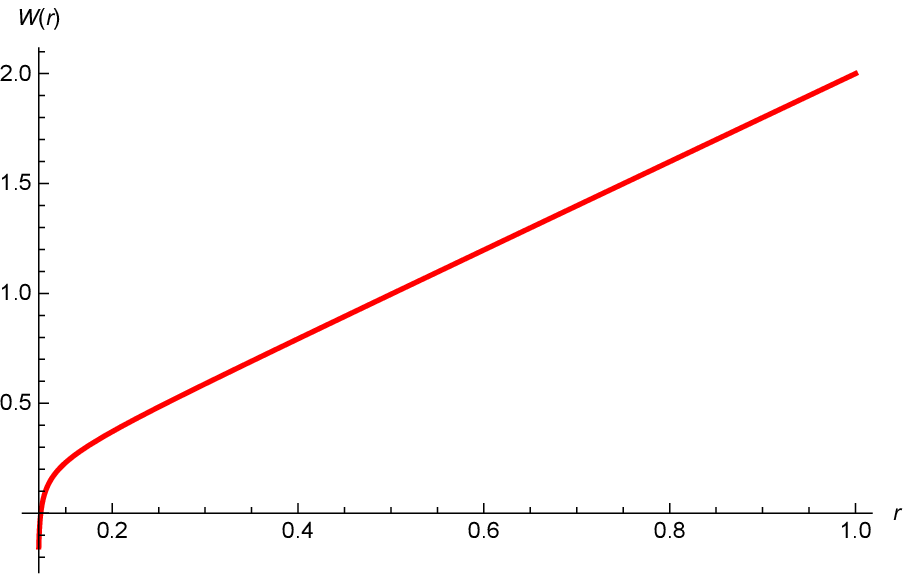}
  \caption{$W$ solution}
  \end{subfigure}
  \begin{subfigure}[b]{0.32\linewidth}
    \includegraphics[width=\linewidth]{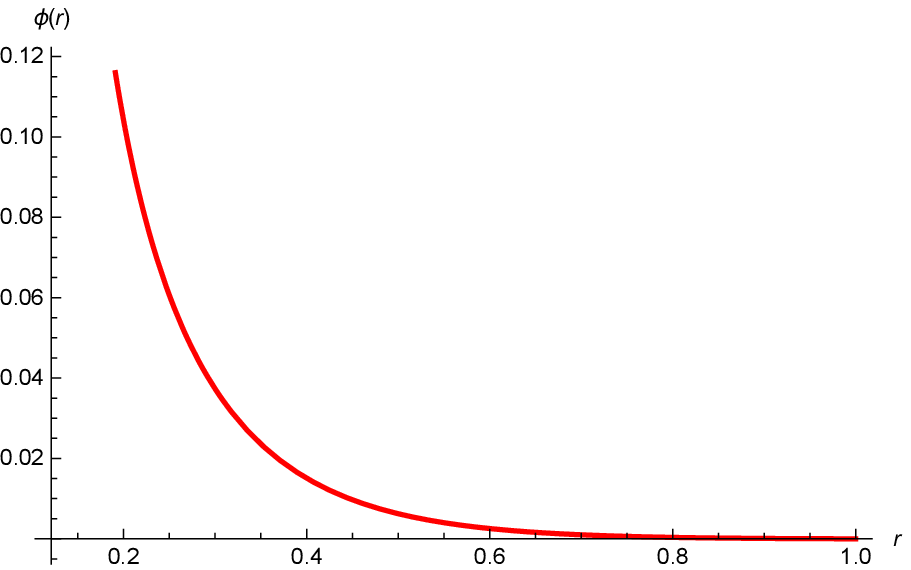}
  \caption{$\phi$ solution}
  \end{subfigure}
  \begin{subfigure}[b]{0.32\linewidth}
    \includegraphics[width=\linewidth]{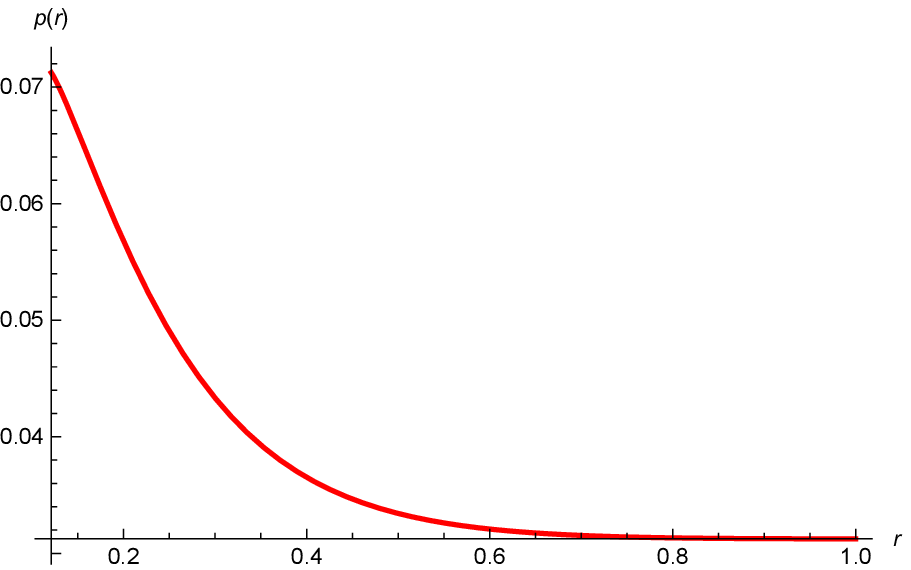}
  \caption{$p$ solution}
  \end{subfigure}
  \begin{subfigure}[b]{0.32\linewidth}
    \includegraphics[width=\linewidth]{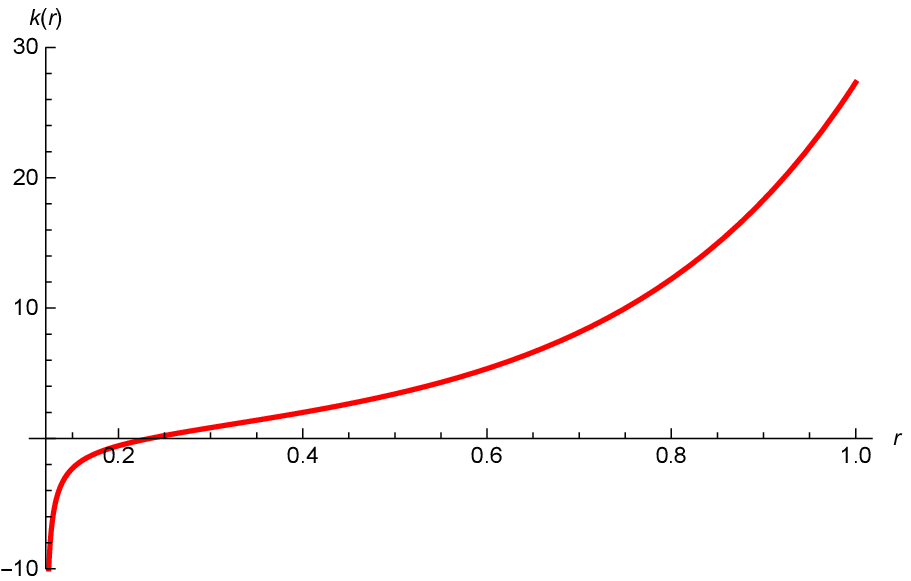}
  \caption{$k$ solution}
  \end{subfigure}
  \begin{subfigure}[b]{0.32\linewidth}
    \includegraphics[width=\linewidth]{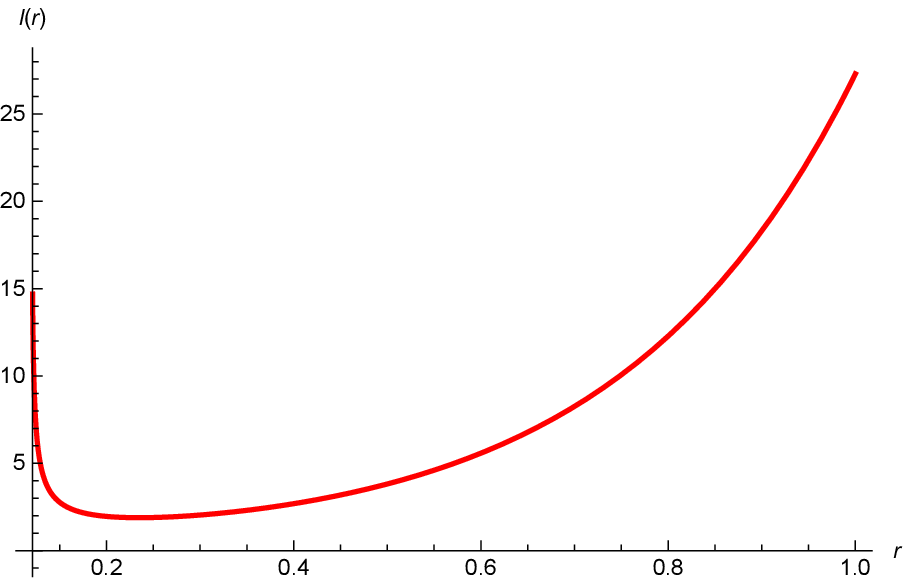}
  \caption{$l$ solution}
  \end{subfigure}
  \caption{A BPS flow from a locally $AdS_7$ geometry at $r\rightarrow\infty$ to the singularity at $r=0$ for the $AdS_3\times S^3$-sliced domain wall with $\tau=1$.}
  \label{YDefDWflow3}
\end{figure}

\begin{figure}[h!]
  \centering
  \begin{subfigure}[b]{0.32\linewidth}
    \includegraphics[width=\linewidth]{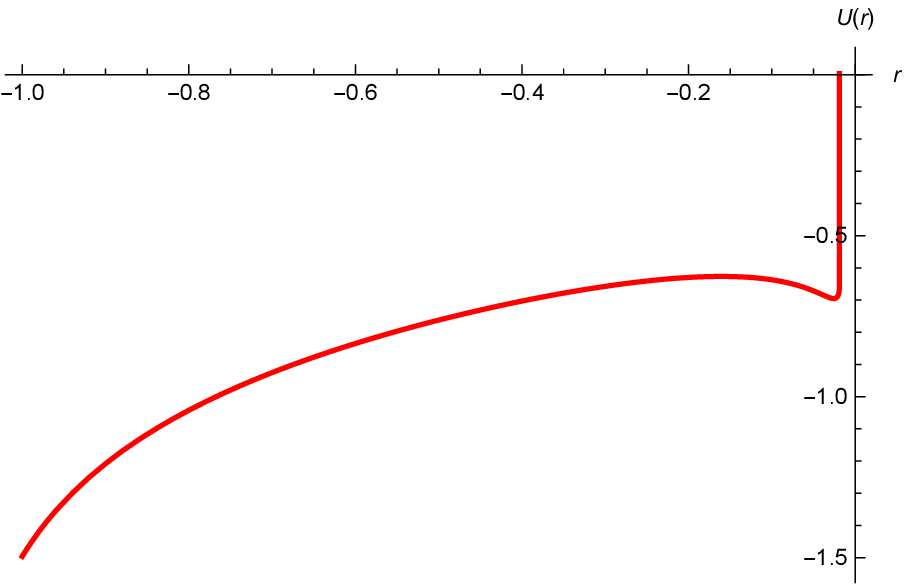}
  \caption{$U$ solution}
  \end{subfigure}
  \begin{subfigure}[b]{0.32\linewidth}
    \includegraphics[width=\linewidth]{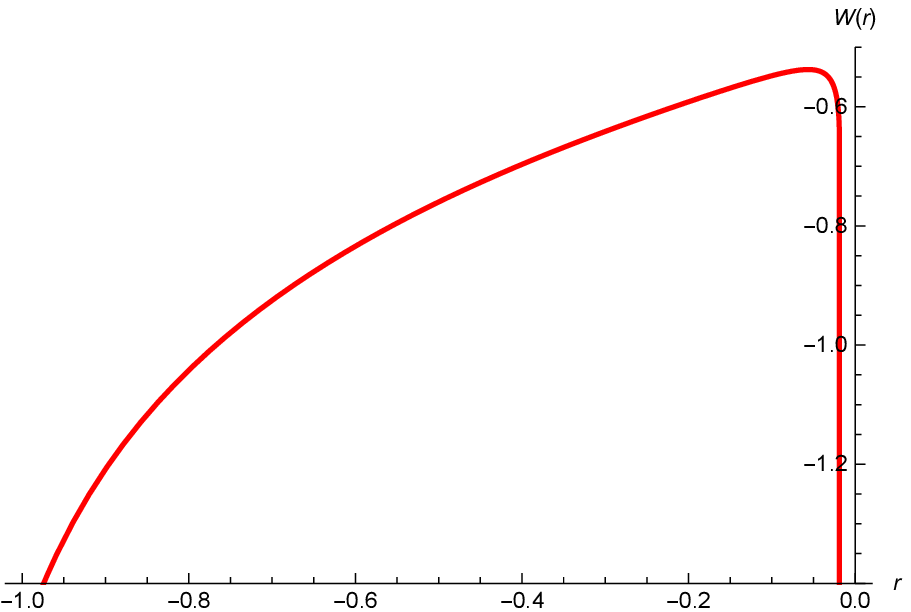}
  \caption{$W$ solution}
  \end{subfigure}
  \begin{subfigure}[b]{0.32\linewidth}
    \includegraphics[width=\linewidth]{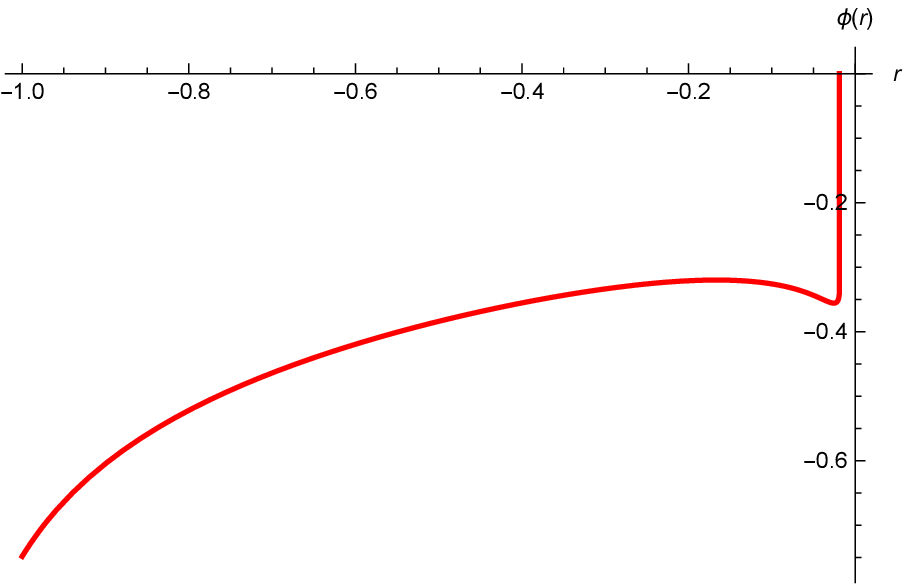}
  \caption{$\phi$ solution}
  \end{subfigure}
  \begin{subfigure}[b]{0.32\linewidth}
    \includegraphics[width=\linewidth]{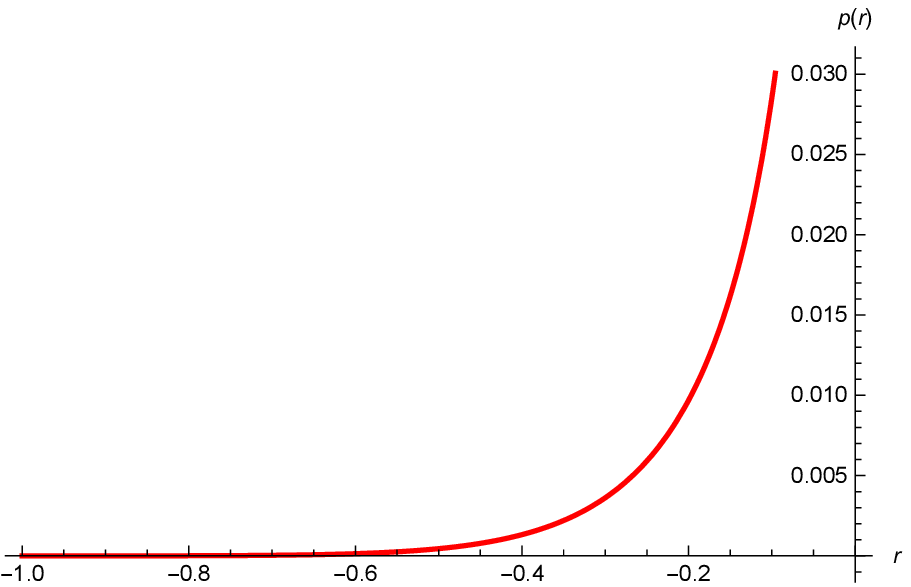}
  \caption{$p$ solution}
  \end{subfigure}
  \begin{subfigure}[b]{0.32\linewidth}
    \includegraphics[width=\linewidth]{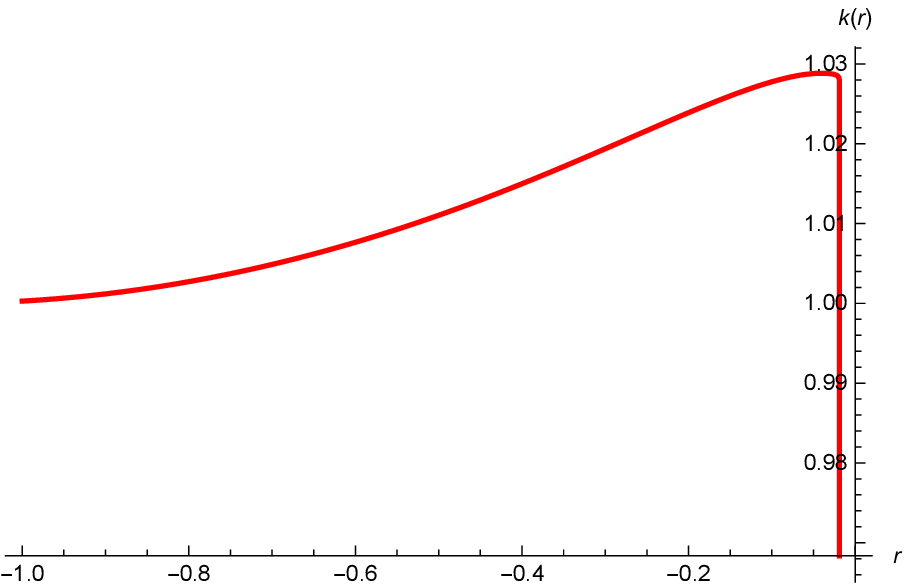}
  \caption{$k$ solution}
  \end{subfigure}
  \begin{subfigure}[b]{0.32\linewidth}
    \includegraphics[width=\linewidth]{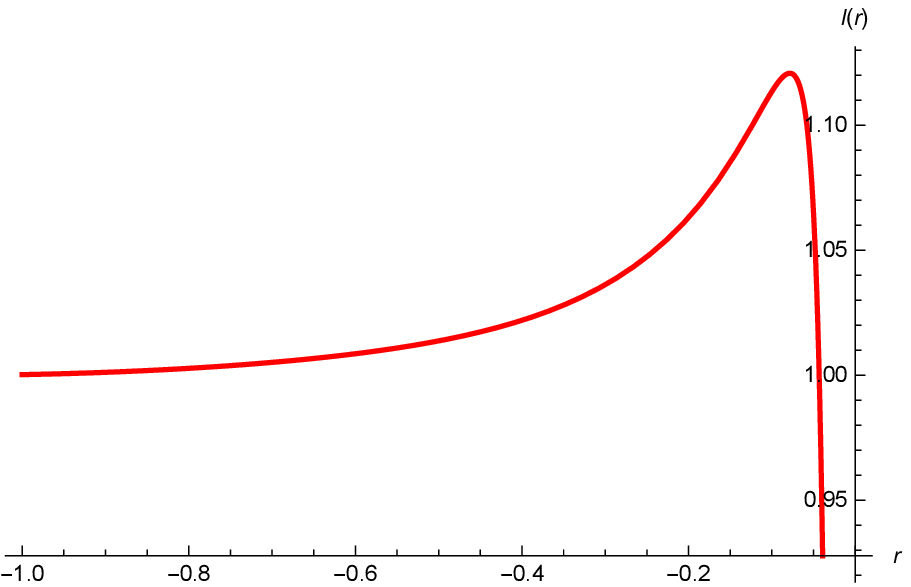}
  \caption{$l$ solution}
  \end{subfigure}
  \caption{A BPS flow from a charged domain wall at $r= -1$ to the singularity at $r=0$ in $SO(5)$ gauge group.}
  \label{YSO5_new_flow}
\end{figure}

\begin{figure}[h!]
  \centering
  \begin{subfigure}[b]{0.32\linewidth}
    \includegraphics[width=\linewidth]{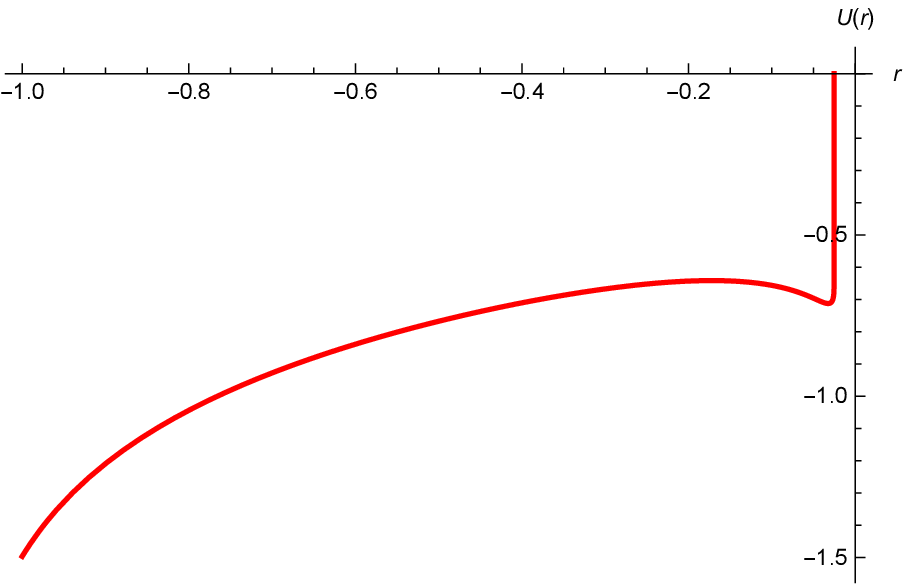}
  \caption{$U$ solution}
  \end{subfigure}
  \begin{subfigure}[b]{0.32\linewidth}
    \includegraphics[width=\linewidth]{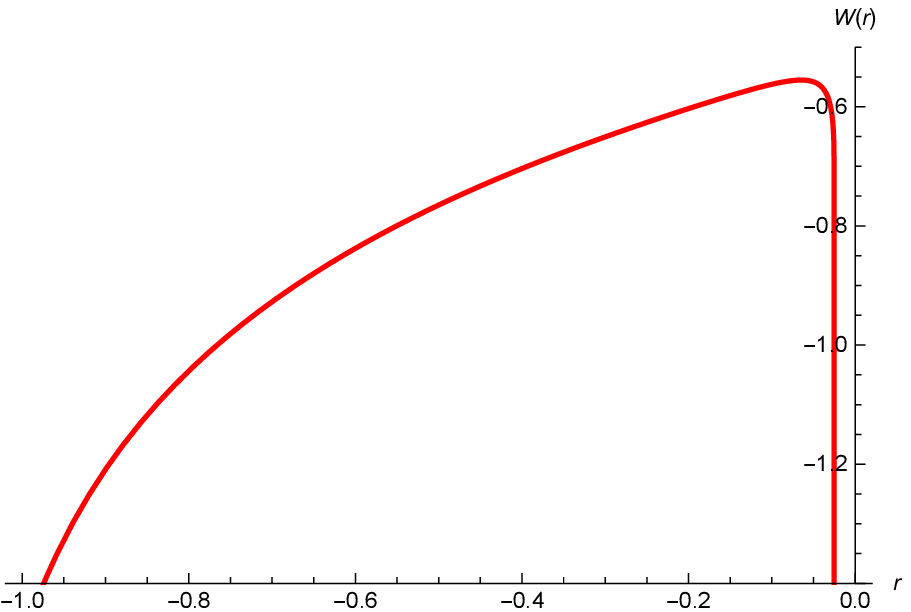}
  \caption{$W$ solution}
  \end{subfigure}
  \begin{subfigure}[b]{0.32\linewidth}
    \includegraphics[width=\linewidth]{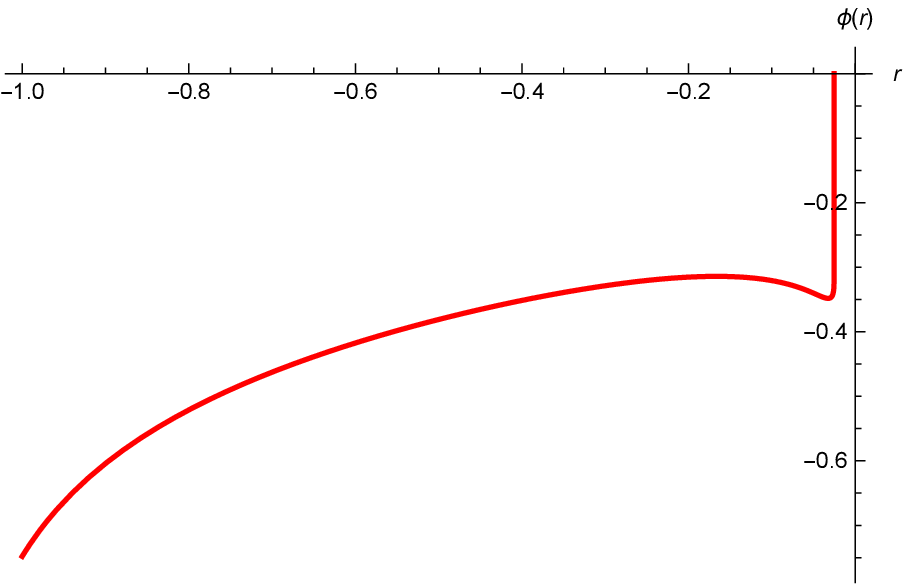}
  \caption{$\phi$ solution}
  \end{subfigure}
  \begin{subfigure}[b]{0.32\linewidth}
    \includegraphics[width=\linewidth]{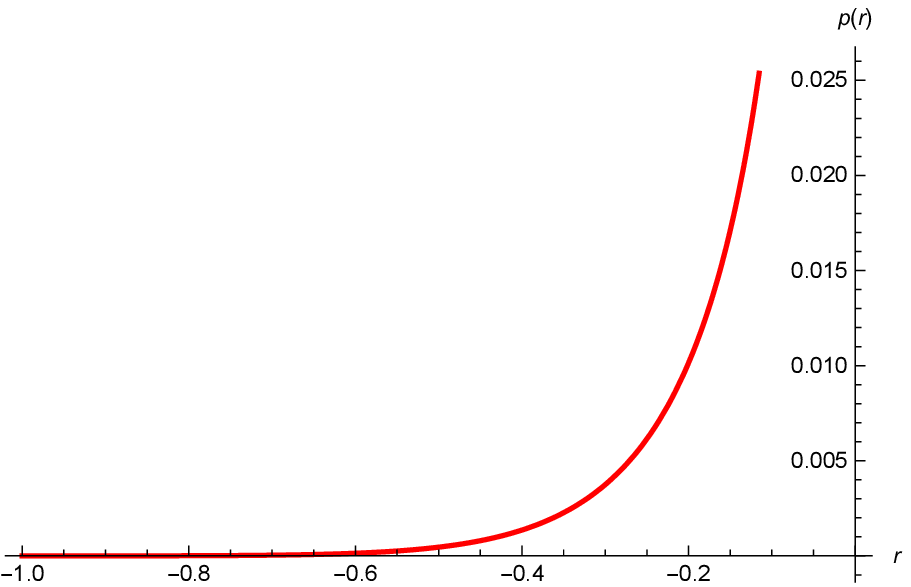}
  \caption{$p$ solution}
  \end{subfigure}
  \begin{subfigure}[b]{0.32\linewidth}
    \includegraphics[width=\linewidth]{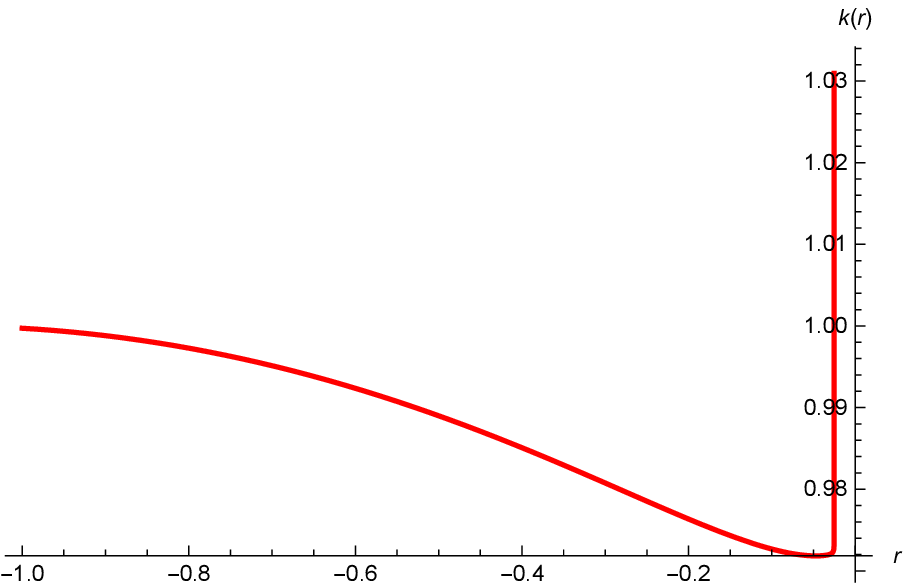}
  \caption{$k$ solution}
  \end{subfigure}
  \begin{subfigure}[b]{0.32\linewidth}
    \includegraphics[width=\linewidth]{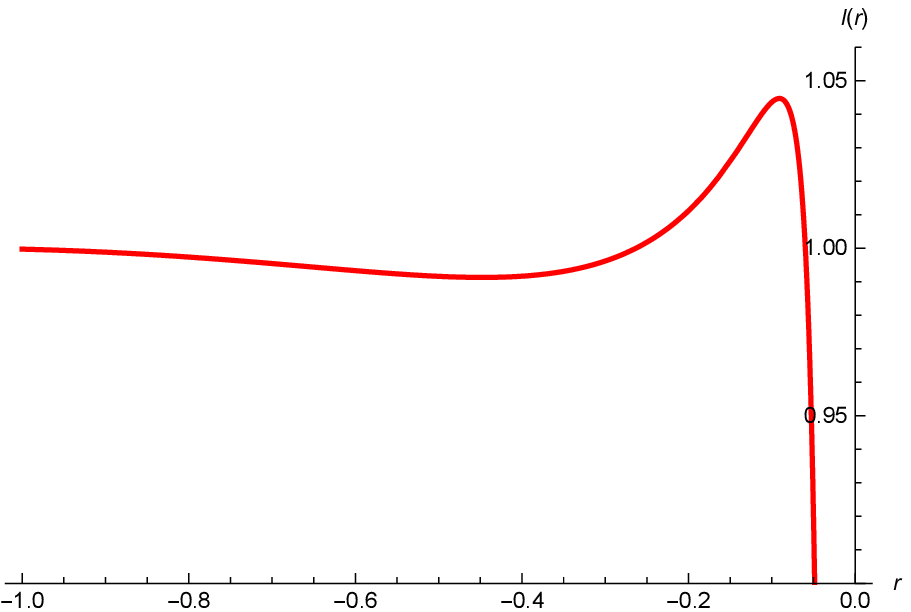}
  \caption{$l$ solution}
  \end{subfigure}
  \caption{A BPS flow from a charged domain wall at $r= -1$ to the singularity at $r=0$ in $SO(4,1)$ gauge group.}
  \label{YSO41_new_flow}
\end{figure}

\begin{figure}[h!]
  \centering
  \begin{subfigure}[b]{0.32\linewidth}
    \includegraphics[width=\linewidth]{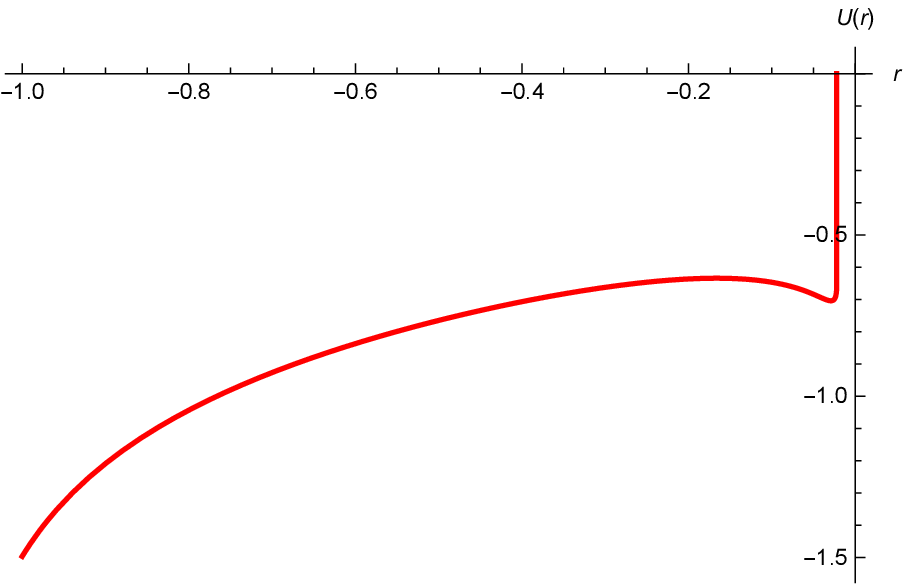}
  \caption{$U$ solution}
  \end{subfigure}
  \begin{subfigure}[b]{0.32\linewidth}
    \includegraphics[width=\linewidth]{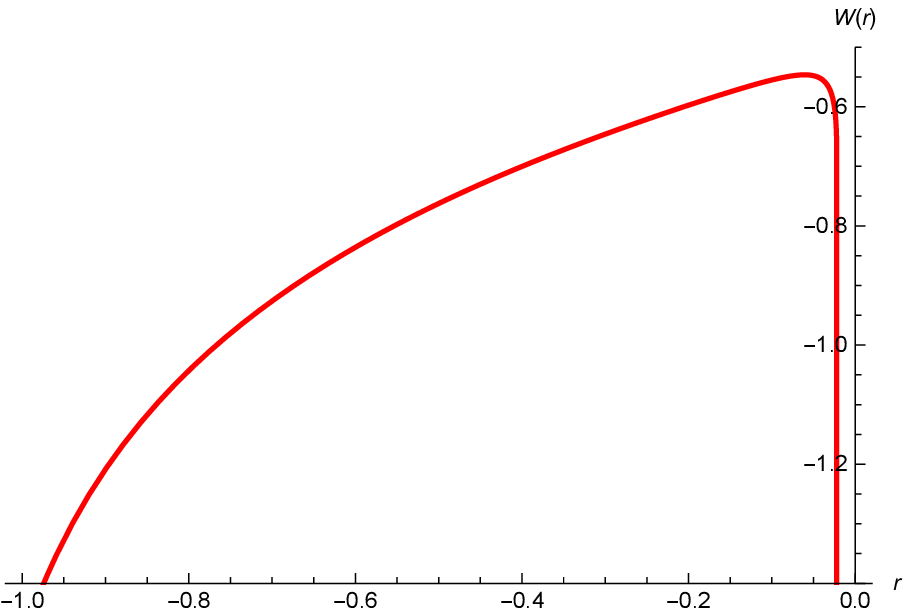}
  \caption{$W$ solution}
  \end{subfigure}
  \begin{subfigure}[b]{0.32\linewidth}
    \includegraphics[width=\linewidth]{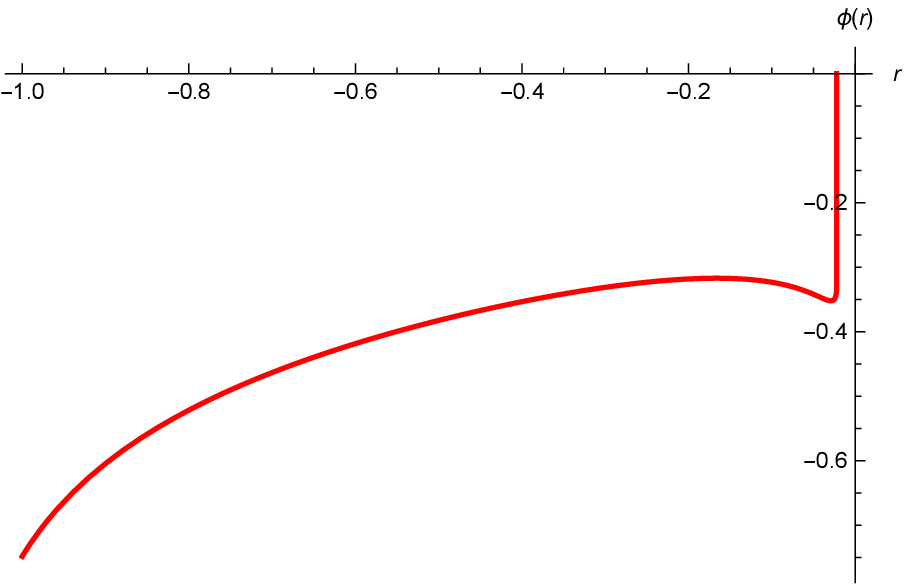}
  \caption{$\phi$ solution}
  \end{subfigure}
  \begin{subfigure}[b]{0.32\linewidth}
    \includegraphics[width=\linewidth]{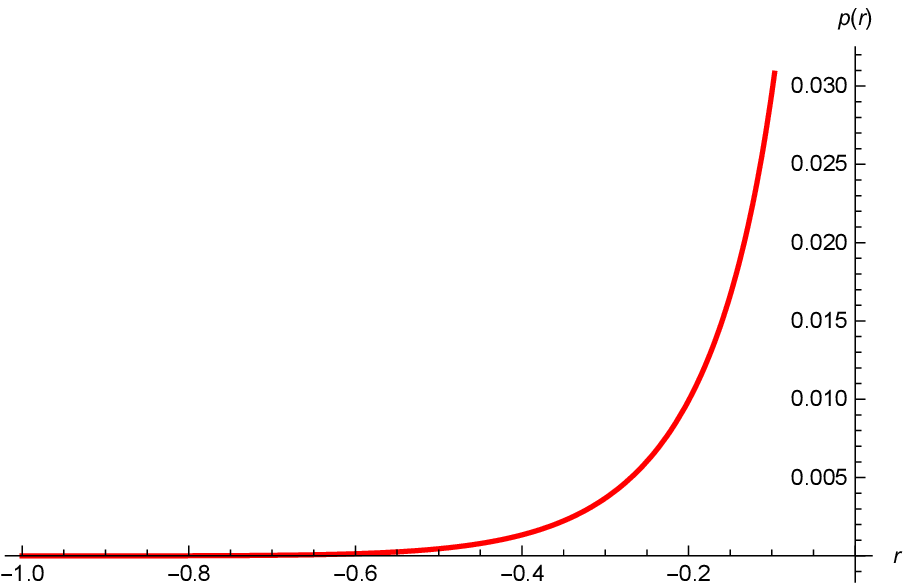}
  \caption{$p$ solution}
  \end{subfigure}
  \begin{subfigure}[b]{0.32\linewidth}
    \includegraphics[width=\linewidth]{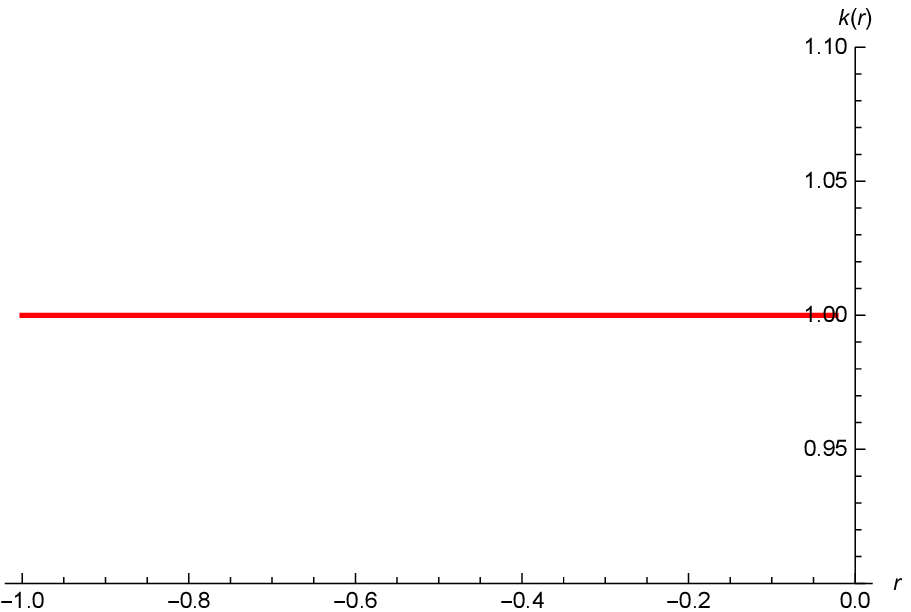}
  \caption{$k$ solution}
  \end{subfigure}
  \begin{subfigure}[b]{0.32\linewidth}
    \includegraphics[width=\linewidth]{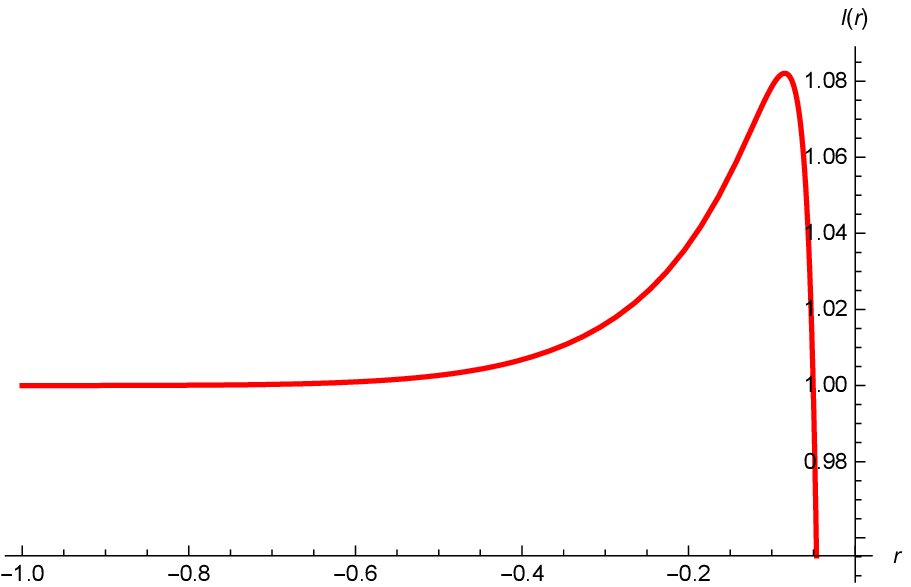}
  \caption{$l$ solution}
  \end{subfigure}
  \caption{A BPS flow from a charged domain wall at $r= -1$ to the singularity at $r=0$ in $CSO(4,0,1)$ gauge group.}
  \label{YCSO401_new_flow}
\end{figure}

\begin{figure}[h!]
  \centering
  \begin{subfigure}[b]{0.32\linewidth}
    \includegraphics[width=\linewidth]{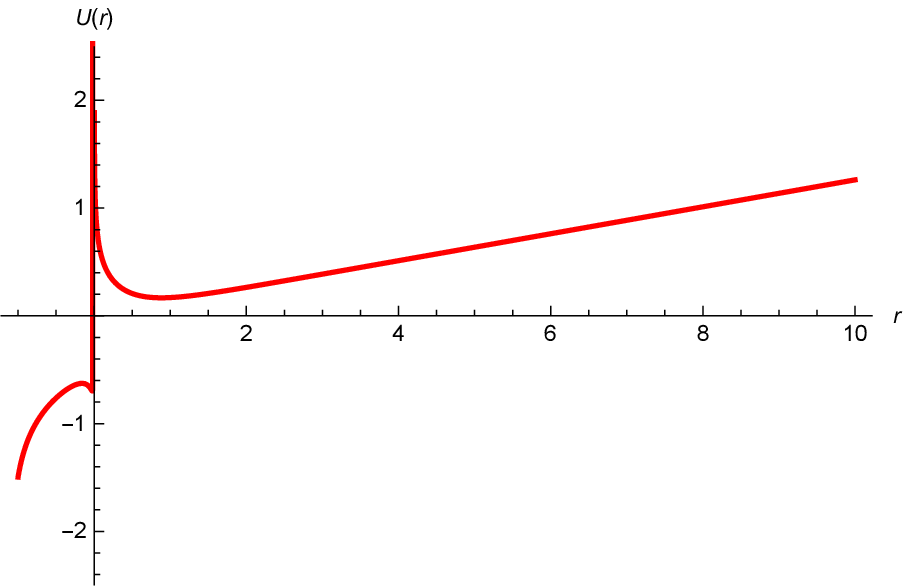}
  \caption{$U$ solution}
  \end{subfigure}
  \begin{subfigure}[b]{0.32\linewidth}
    \includegraphics[width=\linewidth]{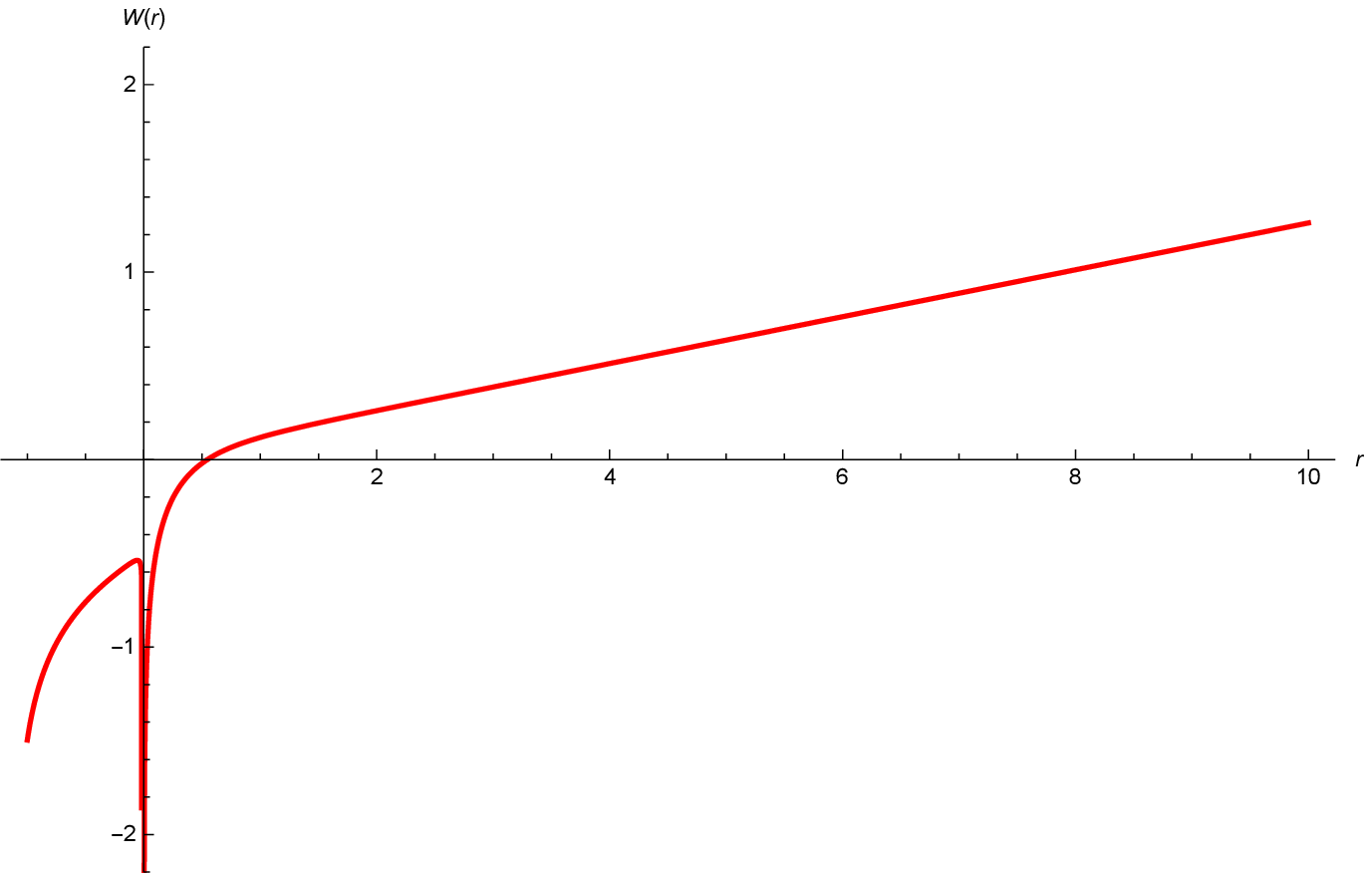}
  \caption{$W$ solution}
  \end{subfigure}
  \begin{subfigure}[b]{0.32\linewidth}
    \includegraphics[width=\linewidth]{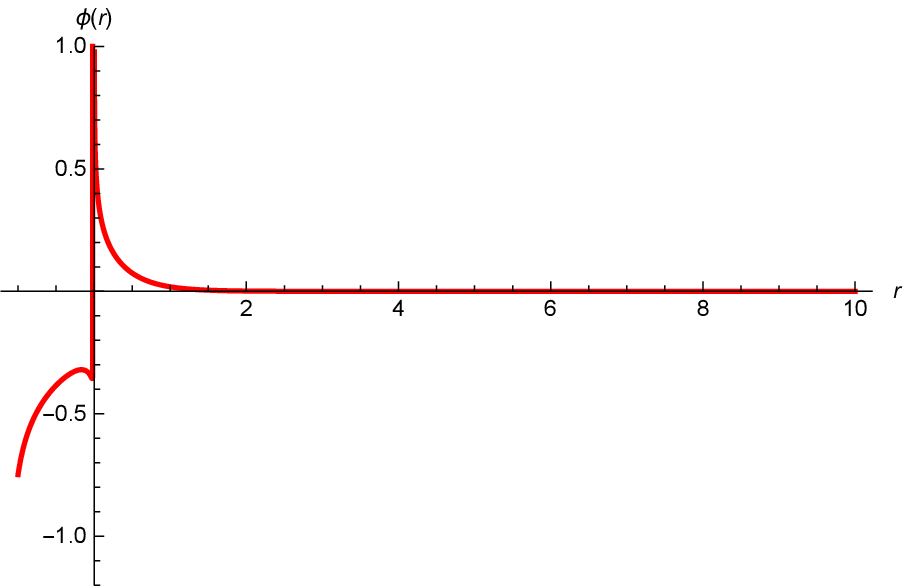}
  \caption{$\phi$ solution}
  \end{subfigure}
  \begin{subfigure}[b]{0.32\linewidth}
    \includegraphics[width=\linewidth]{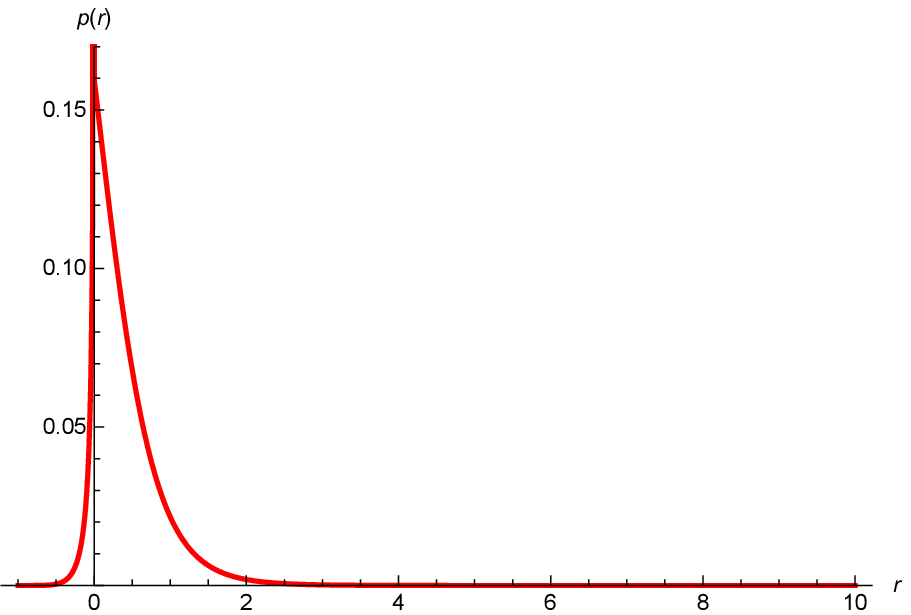}
  \caption{$p$ solution}
  \end{subfigure}
  \begin{subfigure}[b]{0.32\linewidth}
    \includegraphics[width=\linewidth]{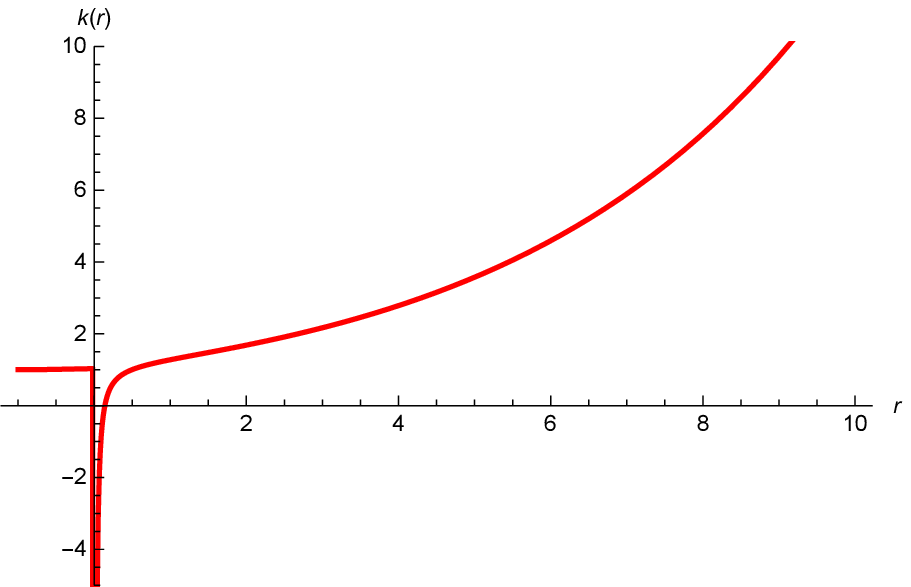}
  \caption{$k$ solution}
  \end{subfigure}
  \begin{subfigure}[b]{0.32\linewidth}
    \includegraphics[width=\linewidth]{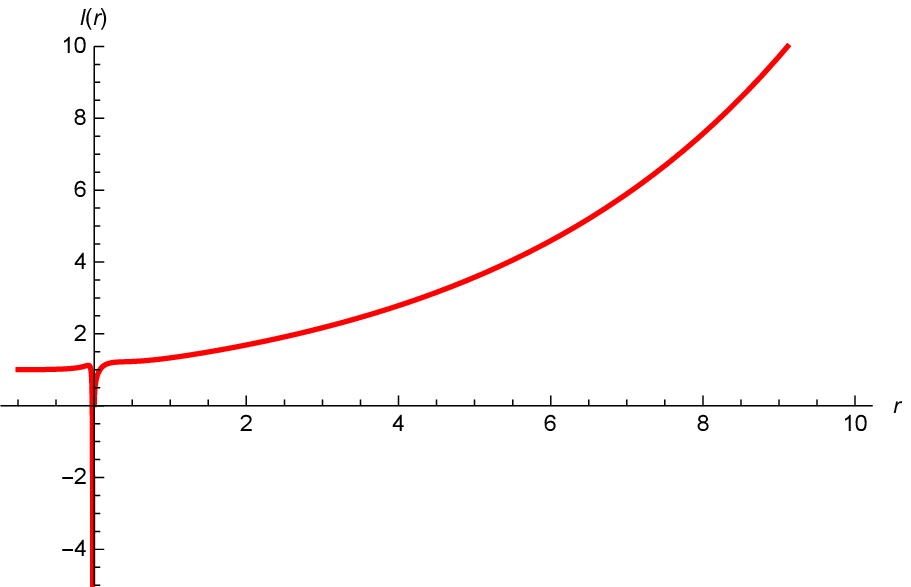}
  \caption{$l$ solution}
  \end{subfigure}
  \caption{A BPS flow between a charged domain wall at $r=-1$ and an asymptotically locally $AdS_7$ geometry as $r\rightarrow\infty$ with an intermediate singularity at $r=0$ in $SO(5)$ gauge group.}
  \label{Connected_flow}
\end{figure}

In this section, we extend the analysis by coupling the previously obtained solutions to $SO(3)$ vectors describing a Hopf fibration of the three-sphere. With the projector ${(\Gamma_5)^a}_b\epsilon^b_0=\epsilon^a_0$ and the identity $\Gamma_1\ldots\Gamma_5=\mathbf{I}_4$, we turn on the gauge fields corresponding to the anti-self-dual $SO(3)\subset SO(4)$. The ansatz for these gauge fields is chosen to be
\begin{eqnarray}
A^{23}_{(1)}=-A^{14}_{(1)}=e^{-W(r)}\frac{\kappa}{4}p(r) e^{\hat{4}},\\
A^{31}_{(1)}=-A^{24}_{(1)}=e^{-W(r)}\frac{\kappa}{4}p(r) e^{\hat{5}},\\
A^{12}_{(1)}=-A^{34}_{(1)}=e^{-W(r)}\frac{\kappa}{4}p(r) e^{\hat{6}}\, .
\end{eqnarray}
The function $p(r)$ is the magnetic charge with the dependence on the radial coordinate. The corresponding two-form field strengths can be computed to be
\begin{eqnarray}
F^{23}_{(2)}=-F^{14}_{(2)}=e^{-V-W}\frac{\kappa}{4}p' e^{\hat{3}}\wedge e^{\hat{4}}+e^{-2W}\frac{\kappa^2}{8}p(2-gp)e^{\hat{5}}\wedge e^{\hat{6}},\\
F^{31}_{(2)}=-F^{24}_{(2)}=e^{-V-W}\frac{\kappa}{4}p' e^{\hat{3}}\wedge e^{\hat{5}}+e^{-2W}\frac{\kappa^2}{8}p(2-gp)e^{\hat{6}}\wedge e^{\hat{4}},\\
F^{12}_{(2)}=-F^{34}_{(2)}=e^{-V-W}\frac{\kappa}{4}p' e^{\hat{3}}\wedge e^{\hat{6}}+e^{-2W}\frac{\kappa^2}{8}p(2-gp)e^{\hat{4}}\wedge e^{\hat{5}}\, .
\end{eqnarray}
For gaugings in the \textbf{15} representation, there are no massive two-form fields due to the vanishing $Z^{MN,P}$. The modified two-form field strengths $\mc{H}^{(2)MN}_{\mu\nu}$ are simply given by the $SO(3)$ gauge field strengths $F^{MN}_{\mu\nu}$.
\\
\indent To preserve some amount of supersymmetry, we need to impose additional projectors on the constant spinors $\epsilon^a_0$ as follow
\begin{equation}\label{DefDWSO(3)Proj}
\gamma_{\hat{4}\hat{5}}\epsilon^a_0=-{(\Gamma_{12})^a}_b\epsilon^b_0, \qquad \gamma_{\hat{5}\hat{6}}\epsilon^a_0=-{(\Gamma_{23})^a}_b\epsilon^b_0, \qquad \gamma_{\hat{6}\hat{4}}\epsilon^a_0=-{(\Gamma_{31})^a}_b\epsilon^b_0\, .
\end{equation}
It should be noted that the last projector is not independent of the first two. Therefore, together with the projectors given in \eqref{DefDWProj}, there are four independent projectors on $\epsilon^a_0$, and the residual supersymmetry consists of two supercharges.
\\
\indent With all these, the resulting BPS equations for the $AdS_3\times S^3$-sliced domain wall are given by
\begin{eqnarray}
U'&\hspace{-0.2cm}=&\hspace{-0.2cm}\frac{e^{V-2(W+\phi)}}{80\cos{2\theta}}\left[e^{2W}\left(g(4+\rho e^{10\phi})(3\cos{4\theta}-1)+32e^{2\phi-U}\tau\sin{2\theta}\right)\right.\nonumber\\&&\hspace{-0.2cm}\left.+12e^{4\phi}\left(\kappa^2p(gp-2)(\cos{4\theta}-3)+2e^{W-2\phi}\kappa(gp-1)\sin{4\theta}\right)\right],\label{U_eq_vec1}\\
W'&\hspace{-0.2cm}=&\hspace{-0.2cm}\frac{e^{V-2(W+\phi)}}{40\cos{2\theta}}\left[e^{2W}\left(g(4+\rho e^{10\phi})(2-\cos{4\theta})+24e^{2\phi-U}\tau\sin{2\theta}\right)\right.\nonumber\\&&\hspace{-0.2cm}\left.+4e^{4\phi}\left(\kappa^2p(gp-2)(\cos{4\theta}-8)-2e^{W-2\phi}\kappa(gp-1)\sin{4\theta}\right)\right],\\
\phi'&\hspace{-0.2cm}=&\hspace{-0.2cm}\frac{e^{V-2(W+\phi)}}{80\cos{2\theta}}\left[e^{2W}\left(g(6\cos{4\theta}-2-\rho e^{10\phi}(\cos{4\theta}+3))+16e^{2\phi-U}\tau\sin{2\theta}\right)\right.\nonumber\\&&\hspace{-0.2cm}\left.+6e^{4\phi}\left(\kappa^2p(gp-2)(3-\cos{4\theta})+2e^{W-2\phi}\kappa(gp-1)\sin{4\theta}\right)\right],\label{phi_eq_vec1}\\
\theta'&\hspace{-0.2cm}=&\hspace{-0.2cm}\frac{e^{V-2(W+\phi)}}{16}\left[24e^{W+2\phi}\left(e^{W-U}\tau+\kappa(gp-1)\cos{2\theta}\right)\right.\nonumber\\&&\hspace{-0.2cm}\left.-\left(ge^{2W}(12+\rho e^{10\phi})-12e^{4\phi}\kappa^2p(gp-2)\right)\sin{2\theta}\right],\\
k&\hspace{-0.2cm}=&\hspace{-0.2cm}\frac{1}{8}e^{3U-4\phi}(4e^{-U}\tau-g\rho e^{8\phi}\sin{2\theta}),\\
l&\hspace{-0.2cm}=&\hspace{-0.2cm}\frac{1}{8}e^{3W-6\phi}\left[g(4+\rho e^{10\phi})\tan{2\theta}-8e^{2\phi-U}\tau\sec{2\theta}\right.\nonumber\\&&\hspace{-0.2cm}\left.-12e^{4\phi-2W}\left(\kappa^2p(gp-2)\tan{2\theta}+e^{W-2\phi}\kappa(gp-1)\right)\right],\\
p'&\hspace{-0.2cm}=&\hspace{-0.2cm}\frac{e^{V-W-4\phi}}{2\kappa}\left[2e^{W+2\phi}\left(e^{W-U}\tau+\kappa(gp-1)\cos{2\theta}\right)\right.\nonumber\\&&\hspace{-0.2cm}\left.-\left(ge^{2W}-e^{4\phi}\kappa^2p(gp-2)\right)\sin{2\theta}\right].\label{p_eq_vec1}
\end{eqnarray}
In contrast to the previous case, it can also be verified that these equations satisfy the second-order field equations without imposing any constraint. By setting $\tau=0$, we can obtain the BPS equations for a $Mkw_3\times S^3$-sliced domain wall. For $p(r)=0$, we obtain the BPS equations \eqref{YDefDWUflow} - \eqref{Ylsol} for charged domain walls without gauge fields. In this case, equation \eqref{p_eq_vec1} becomes the algebraic constraint \eqref{DefDWconstraint}.
\\
\indent The BPS equations in this case are much more complicated, and we are not able to find analytic flow solutions. We then look for numerical solutions with some appropriate boundary conditions. We first consider the solutions in $SO(5)$ gauge group with an $AdS_7$ asymptotic at large $r$. With $\rho=1$, we find that the following locally $AdS_7$ configuration solves the BPS equations at the leading order as $r\rightarrow\infty$
\begin{equation}\label{LocAdS7}
U\sim W\sim\frac{r}{L},\qquad \phi\sim\theta\sim 0,\qquad p\sim\frac{1}{g}\left(1-\frac{\tau}{\kappa}\right)
\end{equation}
with $L=\frac{8}{g}$. With this boundary condition and $V=0$ gauge choice, we find some examples of the BPS flows from this locally $AdS_7$ geometry as $r\rightarrow\infty$ to the singularity at $r=0$ as shown in figures \ref{YDefDWflow1} and \ref{YDefDWflow3} for $g=16$ and $\kappa=2$. It should be noted that we have not imposed the boundary conditions on $k$ and $l$ since the corresponding BPS equations are algebraic. This is rather different from the solutions in \cite{7D_sol_Dibitetto} in which the BPS equations for $k$ and $l$ are differential. 
\\
\indent From the numerical solution in figure \ref{YDefDWflow3}, the solutions for $k$ and $l$ appear to be diverging as $k\sim e^{2U}$ and $l\sim e^{2W}$ for $r\rightarrow \infty$. However, the contribution from the three-form flux is sufficiently suppressed for $r\rightarrow \infty$ since the terms involving $\mc{H}^{(3)}_5$ in the BPS equations behave as $ke^{-3U}+le^{-3W}$.  
\\
\indent For $SO(4,1)$ and $CSO(4,0,1)$ gauge groups, there is no locally asymptotic $AdS_7$ configuration. However, we can look for solutions of the BPS equations \eqref{U_eq_vec1} -\eqref{p_eq_vec1} in the form of a flow from the charged domain wall without vector fields given previously to the singularity at $r=0$. We first choose the gauge choice $V=-3\phi$ and consider the following behavior at the leading order when $gr\rightarrow C$, for a constant $C$,
\begin{eqnarray}\label{YDefDWprofile}
& &U\sim W\sim\frac{2}{5}\ln(gr-C),\qquad \phi\sim\frac{1}{5}\ln(gr-C),\nonumber \\
& &\theta\sim p\sim 0\qquad \textrm{and}\qquad k\sim l\sim\frac{\tau}{2}
\end{eqnarray}
with $\tau=\kappa$. It can be verified that this configuration solves the BPS equations \eqref{YDefDWUflow}-\eqref{Ylsol} and \eqref{DefDWconstraint} in the limit $gr\rightarrow C$. Since this configuration also appears in $SO(5)$ gauge group, we will consider the solutions for $SO(5)$ gauge group as well.
\\
\indent Examples of the BPS flows from the charged domain wall in \eqref{YDefDWprofile} as $gr\rightarrow C$ to the singularity at $r=0$ in $SO(5)$, $SO(4,1)$, and $CSO(4,0,1)$ gauge groups are shown in figures \ref{YSO5_new_flow}, \ref{YSO41_new_flow}, and \ref{YCSO401_new_flow}, respectively. In these solutions, we have chosen the following numerical values $g=1$, $\kappa=\tau=2$ and $C=-1$. These solutions should describe surface defects within $N=(2,0)$ nonconformal field theories in six dimensions. For the solution in figure \ref{YCSO401_new_flow}, $k$ is constant since, for $\rho=0$, the BPS equations \eqref{U_eq_vec1} and \eqref{phi_eq_vec1} give constant $U-2\phi$. 
\\
\indent For $SO(5)$ gauge group, it is also possible to find flow solutions between the asymptotically locally $AdS_7$ geometry and the charged domain wall configuration with an intermediate singularity in the presence of non-vanishing vector fields at $r=0$. With the gauge choice $V=-3\phi$ and $g=1$, $\kappa=\tau=2$ and $C=-1$, an example of these solutions is shown in figure \ref{Connected_flow}. In this solution, it is clearly seen that the vector fields vanish at both ends of the flow with a singularity at $r=0$.

\subsection{$SO(3)$ symmetric charged domain walls}\label{SO3_Y_DW}
In this section, we consider charged domain walls preserving $SO(3)$ residual symmetry. There are three singlet scalars corresponding to the following noncompact generators
\begin{eqnarray}
\hat{Y}_1&=&2e_{1,1}+2e_{2,2}+2e_{3,3}-3e_{4,4}-3e_{5,5},\nonumber  \\
\hat{Y}_2&=&e_{4,5}+e_{5,4},\nonumber  \\
\hat{Y}_3&=&e_{4,4}-e_{5,5}\, .
\end{eqnarray}
There are many possible gauge groups with an $SO(3)$ subgroup. To accommodate all of these gauge groups in a single framework, we use the embedding tensor of the form
\begin{equation}
Y_{MN}=\text{diag}(+1,+1,+1,\sigma,\rho).
\end{equation}
For different values of $\rho,\sigma=0,\pm 1$, this embedding tensor gives rise to the following gauge groups, $SO(5)$ ($\rho=\sigma=1$), $SO(4,1)$ ($-\rho=\sigma=1$), $SO(3,2)$ ($\rho=\sigma=-1$), $CSO(4,0,1)$ ($\rho=0$, $\sigma=1$), $CSO(3,1,1)$ ($\rho=0$, $\sigma=-1$) and $CSO(3,0,2)$ ($\rho=\sigma=0$). The unbroken $SO(3)$ symmetry is generated by $X_{MN}$, $M,N=1,2,3$, generators.
\\
\indent With the $SL(5)/SO(5)$ coset representative of the form
\begin{equation}
\mathcal{V}=e^{\phi_1\hat{Y}_1+\phi_2\hat{Y}_2+\phi_3\hat{Y}_3},\label{SO3_Y_coset}
\end{equation}
the scalar potential reads
\begin{eqnarray}
\mathbf{V}&=&-\frac{g^2}{64}\left[3e^{-8\phi_1}+6e^{2\phi_1}\left[(\rho+\sigma)\cosh{2\phi_2}\cosh{2\phi_3}+(\rho-\sigma)\sinh{2\phi_3}\right]\right.\nonumber \\
& &+\frac{1}{4}e^{12\phi_1}\left[\rho^2+10\rho\sigma+\sigma^2-(3\rho^2-2\rho\sigma+3\sigma^2)\cosh{4\phi_3}\right.\nonumber \\
& &
\left.\left.-(\rho+\sigma)^2\cosh{4\phi_2}(1+\cosh{4\phi_3})-4(\rho^2-\sigma^2)\cosh{2\phi_2}\sinh{4\phi_3}\right]\right].\qquad\label{YSO(3)Pot}
\end{eqnarray}
For $SO(5)$ gauge group, this potential admits a supersymmetric $AdS_7$ vacuum given in \eqref{SO(5)Cripoint} at $\phi_1=\phi_2=\phi_3=0$ and a non-supersymmetric $AdS_7$ given in \eqref{SO(4)Cripoint} at $\phi_1=\frac{1}{20}\ln{2}$, $\phi_2=\pm\frac{1}{4}\ln{2}$ and $\phi_3=0$.
\\
\indent We now repeat the same procedure as in the previous section to set up the BPS equations. The $SO(3)$ residual symmetry allows for two three-form field strengths, $\mc{H}^{(3)}_{\mu\nu\rho M}$ with $M=4,5$. We will choose the following ansatz
\begin{eqnarray}
\mathcal{H}^{(3)}_{\hat{m}\hat{n}\hat{p} 4}&=& k_4(r)e^{-3U(r)}\varepsilon_{\hat{m}\hat{n}\hat{p}}, \qquad
\mathcal{H}^{(3)}_{\hat{i}\hat{j}\hat{k} 4}=l_4(r)e^{-3W(r)}\varepsilon_{\hat{i}\hat{j}\hat{k}},\\
\mathcal{H}^{(3)}_{\hat{m}\hat{n}\hat{p} 5}&=& k_5(r)e^{-3U(r)}\varepsilon_{\hat{m}\hat{n}\hat{p}}, \qquad
\mathcal{H}^{(3)}_{\hat{i}\hat{j}\hat{k} 5}=l_5(r)e^{-3W(r)}\varepsilon_{\hat{i}\hat{j}\hat{k}}\, .
\end{eqnarray}
With $\mc{H}^{(3)}_{\mu\nu\rho 4}$ non-vanishing, the $SO(5)$ gamma matrix $\Gamma_4$ will appear in the BPS conditions. To avoid an additional projector, which will break more supersymmetry, we impose the following condition
\begin{equation}
k_4(r)=\tanh{\phi_2}k_5(r)\qquad  \textrm{and}\qquad l_4(r)=\tanh{\phi_2}l_5(r)\, .\label{k4_k5_con}
\end{equation}
This simply makes the coefficient of $\Gamma_4$ vanish. It would also be interesting to consider a more general projector.
\\
\indent With the projection conditions in \eqref{DefDWProj}, we can find a consistent set of BPS equations for
\begin{equation}
\theta=0\qquad \textrm{and} \qquad \tau=e^{U-W}\kappa\, .\label{SO3_SO2_con}
\end{equation}
The latter forbids the possibility of setting either $\tau=0$ or $\kappa=0$ without ending up with $\kappa=\tau=0$. Therefore, the solutions in this case can only be $AdS_3\times S^3$-sliced domain walls.
\\
\indent The resulting BPS equations take the form
\begin{eqnarray}
U'&\hspace{-0.2cm}=&\hspace{-0.2cm}\frac{g}{40}e^{V+6\phi_1}\left(3e^{-10\phi_1}+(\rho+\sigma)\cosh{2\phi_2}\cosh{2\phi_3}+(\rho-\sigma)\sinh{2\phi_3}\right),\quad\\
W'&\hspace{-0.2cm}=&\hspace{-0.2cm}\frac{g}{40}e^{V+6\phi_1}\left(3e^{-10\phi_1}+(\rho+\sigma)\cosh{2\phi_2}\cosh{2\phi_3}+(\rho-\sigma)\sinh{2\phi_3}\right),\quad\\
\phi'_1&\hspace{-0.2cm}=&\hspace{-0.2cm}\frac{g}{40}e^{V+6\phi_1}\left(2e^{-10\phi_1}-(\rho+\sigma)\cosh{2\phi_2}\cosh{2\phi_3}-(\rho-\sigma)\sinh{2\phi_3}\right),\quad\\
\phi'_2&\hspace{-0.2cm}=&\hspace{-0.2cm}-\frac{g}{8}e^{V+6\phi_1}(\rho+\sigma)\sinh{2\phi_2}\text{ sech }{2\phi_3},\\
\phi'_3&\hspace{-0.2cm}=&\hspace{-0.2cm}-\frac{g}{8}e^{V+6\phi_1}\left((\rho+\sigma)\cosh{2\phi_2}\sinh{2\phi_3}+(\rho-\sigma)\cosh{2\phi_3}\right),\label{phi3_eq}\\
k_5&\hspace{-0.2cm}=&\hspace{-0.2cm}\frac{1}{2}e^{3U-W-3\phi_1-\phi_3}\cosh{\phi_2}\kappa,\\
l_5&\hspace{-0.2cm}=&\hspace{-0.2cm}\frac{1}{2}e^{2W-3\phi_1-\phi_3}\cosh{\phi_2}\kappa.
\end{eqnarray}
However, the compatibility between these BPS equations and the corresponding field equations requires either $\phi_2=0$ or $\phi_3=0$. It should be noted that setting $\phi_3=0$ is consistent with equation \eqref{phi3_eq}, namely $\phi'_3=0$, only for $\sigma=\rho$, so solutions with vanishing $\phi_3$ can only be obtained in $SO(5)$, $SO(3,2)$ and $CSO(3,0,2)$ gauge groups. To find explicit solutions, we separately consider various possible values of $\rho$ and $\sigma$.

\subsubsection{Charged domain walls in $CSO(3,0,2)$ gauge group}
For the simplest $CSO(3,0,2)$ gauge group corresponding to $\rho=\sigma=0$, we find $\phi'_2=\phi'_3=0$, so we can consistently set $\phi_3=0$ and $\phi_2=0$. With $\phi_2=0$, equation \eqref{k4_k5_con} gives $k_4=l_4=0$. Choosing $V=0$ gauge choice, we find the following charged domain wall solution
\begin{eqnarray}
U&=&W=\frac{3}{8}\ln\left[\frac{gr}{5}+C\right], \\
\phi_1&=&\frac{1}{4}\ln\left[\frac{gr}{5}+C\right], \\
 \textrm{and}\qquad k_5&=&l_5=\frac{1}{2}\tau
\end{eqnarray}
with an integration constant $C$.

\subsubsection{Charged domain walls in $CSO(4,0,1)$ and $CSO(3,1,1)$ gauge groups}
In this case, we have $\rho=0$ and $\sigma=\pm 1$ corresponding to $CSO(4,0,1)$ ($\sigma=+1$) and $CSO(3,1,1)$ ($\sigma=-1$) gauge groups. Choosing $V=-6\phi_1$ gauge choice, we can find a charged domain wall solution, with $\phi_2=0$,
\begin{eqnarray}
\phi_3&\hspace{-0.2cm}=&\hspace{-0.2cm}\frac{1}{2}\ln\left[\frac{g\sigma r}{4}+C_1\right],\\
\phi_1&\hspace{-0.2cm}=&\hspace{-0.2cm}-\frac{1}{5}\phi_3+\frac{1}{10}\ln\left[C_2+e^{4\phi_3}\right],\\
U&\hspace{-0.2cm}=&\hspace{-0.2cm}W=\frac{1}{5}\phi_3+\frac{3}{20}\ln\left[C_2+e^{4\phi_3}\right],\\
k_4&\hspace{-0.2cm}=&\hspace{-0.2cm}l_4=0\qquad \textrm{and}  \qquad  k_5=l_5=\frac{1}{2}\tau
\end{eqnarray}
where $C_1$ and $C_2$ are integration constants. For these gauge groups, it is not possible to find solutions with $\phi_3=0$.

\subsubsection{Charged domain walls in $SO(4,1)$ gauge group}
In this case, the gauge group is a non-compact $SO(4,1)$ with $\sigma=-\rho=1$. As in the previous case, it is not possible to set $\phi_3=0$, so we only consider solutions with $\phi_2=0$. Using the same gauge choice $V=-6\phi_1$, we find the following solution
\begin{eqnarray}
e^{2\phi_3}&\hspace{-0.2cm}=&\hspace{-0.2cm}\tan\left[\frac{gr}{4}+C_1\right],\\
\phi_1&\hspace{-0.2cm}=&\hspace{-0.2cm}-\frac{1}{5}\phi_3+\frac{1}{10}\ln\left[C_2(e^{4\phi_3}+1)-1\right],\\
U&\hspace{-0.2cm}=&\hspace{-0.2cm}W=\frac{1}{5}\phi_3-\frac{1}{4}\ln\left[e^{4\phi_3}+1\right]+\frac{3}{20}\ln\left[C_2(e^{4\phi_3}+1)-1\right],\\
k_4&\hspace{-0.2cm}=&\hspace{-0.2cm}l_4=0,\\
k_5&\hspace{-0.2cm}=&\hspace{-0.2cm}l_5=\frac{1}{2}\tau\cos\left[\frac{gr}{4}+C_1\right].
\end{eqnarray}

\subsubsection{Charged domain walls in $SO(5)$ and $SO(3,2)$ gauge groups}
We now look at the last possibility with $\rho=\sigma=\pm1$ corresponding to $SO(5)$ and $SO(3,2)$ gauge groups. In this case, it is possible to set $\phi_2=0$ or $\phi_3=0$. With $\phi_2=0$ and $V=-6\phi_1$, we find the following solution
\begin{eqnarray}
\phi_3&\hspace{-0.2cm}=&\hspace{-0.2cm}\frac{1}{2}\ln\left[\frac{e^{\frac{g\rho r}{2}}-C_1}{e^{\frac{g\rho r}{2}}+C_1}\right],\label{phi2=0sol1}\\
\phi_1&\hspace{-0.2cm}=&\hspace{-0.2cm}-\frac{1}{5}\phi_3+\frac{1}{10}\ln\left[C_2(e^{4\phi_3}-1)+1\right],\\
U&\hspace{-0.2cm}=&\hspace{-0.2cm}W=\frac{1}{5}\phi_3-\frac{1}{4}\ln\left[e^{4\phi_3}-1\right]+\frac{3}{20}\ln\left[C_2(e^{4\phi_3}-1)+1\right]\label{phi2=0sol3}
\end{eqnarray}
together with
\begin{equation}
k_4=l_4=0\qquad \textrm{ and } \qquad k_5=l_5=\frac{\tau}{2\sqrt{e^{4\phi_3}-1}}\, .
\end{equation}
\indent For $\phi_3=0$, we find the same solution as in \eqref{phi2=0sol1} - \eqref{phi2=0sol3} with $\phi_3$ replaced by $\phi_2$, but the solution for $k_{4,5}$ and $l_{4,5}$ are now given by
\begin{equation}
k_4=l_4=\frac{(e^{2\phi_2}-1)\tau}{4\sqrt{e^{4\phi_2}-1}}\qquad \text{ and } \qquad k_5=l_5=\frac{(e^{2\phi_2}+1)\tau}{4\sqrt{e^{4\phi_2}-1}}\, .
\end{equation}
Unlike the previous cases, this solution has two non-vanishing three-form fluxes.
\\
\indent We end this section by giving a comment on solutions with non-vanishing $SO(3)$ gauge fields. Repeating the same procedure as in the $SO(4)$ symmetric solutions leads to a set of BPS equations together with the following constraints 
\begin{equation}
p'=0 \qquad \textrm{and}\qquad p=\frac{\kappa-\tau e^{W-U}}{g\kappa}\, .\label{SO3_sym_con}
\end{equation}
It turns out that, in this case, the compatibility between the resulting BPS equations and the corresponding field equations requires that
\begin{equation}
\tau(e^W\tau-e^U\kappa)=0\, .
\end{equation}
For $\tau=0$, we can have a constant magnetic charge $p$ as required by the conditions in \eqref{SO3_sym_con}, but in this case, the three-form flux vanishes unless $e^{W}\tau=e^U\kappa$ as required by \eqref{SO3_SO2_con}. This case corresponds to performing a topological twist along the $S^3$ part. Since this type of solutions is not the main aim of this paper, we will not consider them here. On the other hand, setting $e^W\tau=e^U\kappa$ does lead to non-vanishing three-form fluxes, but equation \eqref{SO3_sym_con} gives vanishing gauge fields. This corresponds to the charged domain walls given above. Therefore, there does not seem to be solutions with both $SO(3)$ gauge fields and three-form fluxes non-vanishing at least for the ansatz considered here. This is very similar to the result of \cite{7D_N2_DW_3_form} in the matter-coupled $N=2$ gauged supergravity.

\subsection{$SO(2)\times SO(2)$ symmetric charged domain walls}
We finally consider charged domain walls with $SO(2)\times SO(2)$ symmetry generated by $X_{12}$ and $X_{34}$. There are two $SO(2)\times SO(2)$ invariant scalars corresponding to the noncompact generators
\begin{equation}\label{YSO(2)xSO(2)Ys}
\tilde{Y}_1=e_{1,1}+e_{2,2}-2e_{5,5}\qquad \textrm{ and }\qquad
\tilde{Y}_2=e_{3,3}+e_{4,4}-2e_{5,5}\, .
\end{equation}
The $SL(5)/SO(5)$ coset representative can be written as
\begin{equation}\label{YSO(2)xSO(2)Ys}
\mathcal{V}=e^{\phi_1\tilde{Y}_1+\phi_2\tilde{Y}_2}\, .
\end{equation}
The embedding tensor giving rise to gauge groups with an $SO(2)\times SO(2)$ subgroup is given by
\begin{equation}\label{SO(2)xSO(2)Ytensor}
Y_{MN}=\text{diag}(+1,+1,\sigma,\sigma,\rho)
\end{equation}
with $\rho,\sigma=0,\pm 1$. These gauge groups are $SO(5)$ ($\rho=\sigma=1$), $SO(4,1)$ ($-\rho=\sigma=1$), $SO(3,2)$ ($\rho=-\sigma=1$), $CSO(4,0,1)$ ($\rho=0$, $\sigma=1$) and $CSO(2,2,1)$ ($\rho=0$, $\sigma=-1$).
\\
\indent Using the coset representative \eqref{YSO(2)xSO(2)Ys}, we obtain the scalar potential
\begin{equation}\label{YSO(2)xSO(2)Pot}
\mathbf{V}=-\frac{1}{64}g^2e^{-2(\phi_1+\phi_2)}\left[8\sigma-\rho^2e^{10(\phi_1+\phi_2)}+4\rho(e^{4\phi_1+6\phi_2}+\sigma e^{6\phi_1+4\phi_2})\right].
\end{equation}
As in the previous case, a consistent set of BPS equations can be found only for $\theta=0$ and $\tau e^{W}=\kappa e^{U}$. With the three-form flux \eqref{sdSO(3)dyonic3form}, which is manifestly invariant under $SO(2)\times SO(2)$, and the projectors given in \eqref{DefDWProj}, the resulting BPS equations read
\begin{eqnarray}
U'&\hspace{-0.2cm}=&\hspace{-0.2cm}W'=\frac{g}{40}e^V(2e^{-2\phi_1}+\rho e^{4(\phi_1+\phi_2)}+2\sigma e^{-2\phi_2}),\\
\phi'_1&\hspace{-0.2cm}=&\hspace{-0.2cm}\frac{g}{20}e^V(3e^{-2\phi_1}-\rho e^{4(\phi_1+\phi_2)}-2\sigma e^{-2\phi_2}),\\
\phi'_2&\hspace{-0.2cm}=&\hspace{-0.2cm}\frac{g}{20}e^V(3\sigma e^{-2\phi_2}-\rho e^{4(\phi_1+\phi_2)}-2 e^{-2\phi_1}),\\
k&\hspace{-0.2cm}=&\hspace{-0.2cm}\frac{1}{2}e^{2U-2(\phi_1+\phi_2)}\tau,\\
l&\hspace{-0.2cm}=&\hspace{-0.2cm}\frac{1}{2}e^{3W-U-2(\phi_1+\phi_2)}\tau\, .
\end{eqnarray}
\indent By choosing $V=2\phi_1$, we obtain the solution
\begin{eqnarray}
\phi_1&\hspace{-0.2cm}=&\hspace{-0.2cm}-\frac{1}{10}\ln\left[e^{C_1-\frac{gr}{2}}+\rho\right]-\frac{1}{5}\ln\left[e^{C_2-\frac{gr}{2}}+\sigma\right],\\
\phi_2&\hspace{-0.2cm}=&\hspace{-0.2cm}-\frac{3}{2}\phi_1-\frac{1}{4}\ln\left[e^{C_1-\frac{gr}{2}}+\rho\right],\\
U=W&\hspace{-0.2cm}=&\hspace{-0.2cm}\frac{1}{8}gr+\frac{1}{20}\ln\left[e^{C_1-\frac{gr}{2}}+\rho\right]+\frac{1}{10}\ln\left[e^{C_2-\frac{gr}{2}}+\sigma\right],\\
k=l&\hspace{-0.2cm}=&\hspace{-0.2cm}\frac{1}{2}\tau e^{\frac{gr}{4}}\sqrt{e^{C_1-\frac{gr}{2}}+\rho}
\end{eqnarray}
with the integration constants $C_1$ and $C_2$. This solution is just the $SO(2)\times SO(2)$ symmetric domain wall found in \cite{our_7D_DW} with a dyonic profile for the three-form flux. In this case, coupling to $SO(3)$ gauge fields is not possible due to the absence of any unbroken $SO(3)$ gauge symmetry.

\subsection{Uplifted solutions in ten and eleven dimensions}
We now give the uplifted solutions in the case of $SO(5)$ and $CSO(4,0,1)$ which can be obtained from consistent truncations of eleven-dimensional supergravity on $S^4$ and type IIA theory on $S^3$, respectively. As shown in \cite{Henning_Hohm1}, other gauge groups of the form $CSO(p,q,5-p-q)$ with the embedding tensor in $\mathbf{15}$ representation can also be obtained from truncations of eleven-dimensional supergravity on $H^{p,q}\circ T^{5-p-q}$. However, in this paper, we will not consider uplifted solutions for these gauge groups since the complete truncation ansatze have not been constructed so far. Furthermore, we will not consider uplifting solutions with non-vanishing vector fields since, in this case, the uplifted solutions are not very useful due to the lack of analytic solutions.

\subsubsection{Uplift to eleven dimensions}
We first consider uplifting the seven-dimensional solutions in $SO(5)$ gauge group  to eleven-dimensional supergravity. We begin with the $SO(4)$ symmetric solution with the $SL(5)/SO(5)$ scalar matrix
\begin{equation}
\mc{M}_{MN}=\textrm{diag}(e^{2\phi},e^{2\phi},e^{2\phi},e^{2\phi},e^{-8\phi})
\end{equation}
and the coordinates on $S^4$ given by
\begin{equation}
\mu^M=(\mu^i,\mu^5)=(\sin\xi \hat{\mu}^i,\cos\xi),\qquad i=1,2,3,4
\end{equation}
with $\hat{\mu}^i$ being coordinates on $S^3$ satisfying $\hat{\mu}^i\hat{\mu}^i=1$. With the formulae given in appendix \ref{truncation_ansatz}, the eleven-dimensional metric and the four-form field strength are given by
\begin{eqnarray}
d\hat{s}^2_{11}&=&\Delta^{\frac{1}{3}}\left(e^{2U(r)}ds_{M_3}^2+e^{2V(r)}dr^2+e^{2W(r)}ds_{S^3}^2 \right)
\nonumber \\
& &+\frac{16}{g^2}\Delta^{-\frac{2}{3}}\left[e^{-8\phi}\sin^2\xi d\xi^2+e^{2\phi}(\cos^2\xi d\xi^2+\sin^2\xi d\Omega^2_{(3)})\right],\\
\hat{F}_{(4)}&=&\frac{64}{g^3}\Delta^{-2}\sin^4\xi\left(U\sin\xi d\xi -10e^{6\phi}\phi'\cos\xi dr\right)\wedge \epsilon_{(3)}\nonumber \\
& &-2\cos\xi e^{8\phi}\left(ke^{3W+V-3U}dr\wedge \textrm{vol}_{S^3}-le^{3U+V-3W}dr\wedge \textrm{vol}_{M_3}\right)\nonumber \\
& &-\frac{8}{g}\sin\xi (k\textrm{vol}_{M_3}+l\textrm{vol}_{S^3})\wedge d\xi
\end{eqnarray}
with $d\Omega^2_{(3)}=d\hat{\mu}^id\hat{\mu}^i$ being the metric on a unit $S^3$ and
\begin{eqnarray}
\Delta&=&e^{8\phi}\cos^2\xi+e^{-2\phi}\sin^2\xi,\qquad \epsilon_{(3)}=\frac{1}{3!}\epsilon_{ijkl}\hat{\mu}^id\hat{\mu}^j\wedge d\hat{\mu}^k\wedge d\hat{\mu}^l,\nonumber \\
U&=&(e^{16\phi}-4e^{6\phi})\cos^2\xi-(e^{6\phi}+2e^{-4\phi})\sin^2\xi\, .
\end{eqnarray}
The $SO(4)$ residual symmetry of the seven-dimensional solution is the isometry of the $S^3$ inside the $S^4$. The $3$-manifold $M_3$ can be $Mkw_3$ or $AdS_3$. Due to the dyonic profile of the four-form field strength, this solution should describe a bound state of M$2$- and M$5$-branes similar to the solutions considered in \cite{7D_sol_Dibitetto}. It is also interesting to find a relation between the solution with $M_3=AdS_3$ and the $SO(2,2)\times SO(4)\times SO(4)$ symmetric solution studied in \cite{Lunin_AdS3_S3_S3}. 
\\
\indent We can repeat a similar procedure for the $SO(3)$ symmetric solutions. With the index $M=(a,4,5)$, $a=1,2,3$, the $SL(5)/SO(5)$ scalar matrix is given by
\begin{equation}
\mc{M}=\begin{pmatrix}e^{4\phi_1}\mathbf{I}_3 & 0 \\
0& e^{-6\phi_1}M_2
\end{pmatrix}
\end{equation}
with the $2\times 2$ matrix $M_2$ given by
\begin{equation}
M_2=\begin{pmatrix}e^{2\phi_3}\cosh^2\phi_2+\sinh^2\phi_2 & \sinh\phi_2\cosh\phi_2(1+e^{-2\phi_3}) \\
\sinh\phi_2\cosh\phi_2(1+e^{2\phi_3})& e^{-2\phi_3}\cosh^2\phi_2+\sinh^2\phi_2
\end{pmatrix}.
\end{equation}
We now separately discuss the uplifted solutions for the two cases with $\phi_2=0$ and $\phi_3=0$. We will also denote $k_5$ and $l_5$ simply by $k$ and $l$ with $k_4=\tanh\phi_2k$ and $l_4=\tanh \phi_2l$. Recall also that for $SO(3)$ symmetric solutions, we only have $M_3=AdS_3$. 
\\
\indent For $\phi_2=0$ and the $S^4$ coordinates
\begin{equation}
\mu^M=(\cos\xi \hat{\mu}^a,\sin\xi\cos\psi,\sin\xi\sin\psi)
\end{equation}
with $\hat{\mu}^a\hat{\mu}^a=1$, we find the eleven-dimensional metric
\begin{eqnarray}
d\hat{s}^2_{11}&=&\Delta^{\frac{1}{3}}\left(e^{2U}ds_{AdS_3}^2+e^{2V}dr^2+e^{2W}ds_{S^3}^2 \right)+\frac{16}{g^2}\Delta^{-\frac{2}{3}}\left[e^{4\phi_1}(\sin^2\xi d\xi^2\right.\nonumber \\
& &+\cos^2\xi d\hat{\mu}^ad\hat{\mu}^a)
+e^{-6\phi_1}\left\{\sin^2\xi (e^{2\phi_3}\sin^2\psi+e^{-2\phi_3}\cos^2\psi)d\psi^2 \right. \nonumber \\
& &\left.\left.-\sin2\psi \sin2\xi\sinh 2\phi_3d\xi d\psi+\cos^2\xi (e^{2\phi_3}\cos^2\psi+e^{-2\phi_3}\sin^2\psi)d\xi^2\right\}\right]\nonumber \\
& &
\end{eqnarray}
where
\begin{equation}
\Delta=e^{-4\phi_1}\cos^2\xi+e^{6\phi_1}\sin^2\xi (e^{-2\phi_3}\cos^2\psi+e^{2\phi_3}\sin^2\psi).
\end{equation}
The four-form field strength is given by
\begin{eqnarray}
\hat{F}_{(4)}&=&-2e^{6\phi_1+2\phi_3}\sin\xi \sin \psi dr\wedge (ke^{3W+V-3U}\textrm{vol}_{S^3}-le^{3U+V-3W}\textrm{vol}_{AdS_3})\nonumber \\
& &+\frac{8}{g}(k\textrm{vol}_{AdS_3}+l\textrm{vol}_{S^3})\wedge (\cos\xi \sin \psi d\xi +\sin\xi \cos \psi d\psi)\nonumber \\
& &-\frac{64}{g^3}\Delta^{-2}\epsilon_{(2)}\wedge\left[\cos^2\xi \sin\xi U d\xi \wedge d\psi +\phi_3'e^{12\phi_1}\sin^3\xi \cos^2\xi\sin2\psi dr\wedge d\xi \right.\nonumber \\
& &-e^{2\phi_1-2\phi_3}\sin\xi \cos^3\xi dr\wedge \{(6\phi_1'\sin\xi +2\phi_3'\sin\xi \cos\psi)d\psi -2\phi_3'\cos\xi\times \nonumber \\
& &\sin\psi d\xi\}-2\phi_1'e^{2\phi_1}\sin2\xi \cos^2\xi dr\wedge \left\{(e^{-2\phi_3}-e^{2\phi_3})\sin\psi \cos\psi \cos\xi d\xi 
\right.\nonumber \\
& &\left.\left.+\sin\xi (e^{2\phi_3}\sin^2\psi +e^{-2\phi_3}\cos^2\psi)d\psi\right\} \right]
\end{eqnarray}
with
\begin{eqnarray}
\epsilon_{(2)}&=&\frac{1}{2}\epsilon_{abc}\hat{\mu}^ad\hat{\mu}^b\wedge \hat{\mu}^c,\\
U&=&\frac{1}{2}e^{2\phi_1}\left[\sin^2\xi (1-e^{-4\phi_3})\{3e^{2\phi_3}\cos 2\psi-e^{10\phi_1}(1+\cos2\psi-2e^{4\phi_3}\sin^2\psi)\}\right. \nonumber \\
& &\left. +(\cos 2\xi-5)\cosh 2\phi_3 \right]-e^{-8\phi_1}\cos^2\xi\, .
\end{eqnarray}
\indent For $\phi_3=0$, we find
\begin{eqnarray}
d\hat{s}^2_{11}&=&\Delta^{\frac{1}{3}}\left(e^{2U}ds_{AdS_3}^2+e^{2V}dr^2+e^{2W}ds_{S^3}^2 \right)+\frac{16}{g^2}\Delta^{-\frac{2}{3}}\left[e^{4\phi_1}(\sin^2\xi d\xi^2\right.
\nonumber \\
& &
+\cos^2\xi d\hat{\mu}^ad\hat{\mu}^a)+e^{-6\phi_1}\sinh2\phi_2\{\sin2\psi(\cos^2\xi d\xi^2-\sin^2\xi d\psi^2)
 \nonumber \\
& &\left.+\sin2\xi\cos2\psi d\psi d\xi\}+e^{-6\phi_1}\cosh2\phi_2(\cos^2\xi d\xi^2+\sin^2\xi d\psi^2)\right]
\end{eqnarray}
and
\begin{eqnarray}
\hat{F}_{(4)}&=&2\sin\xi e^{6\phi_1+V}(\cos\psi\tanh\phi_2-\sin\psi)dr\wedge (ke^{3W-3U}\textrm{vol}_{S^3}-le^{3U-3W}\textrm{vol}_{AdS_3})\nonumber \\
& &+\frac{8}{g}(k\textrm{vol}_{AdS_3}+l\textrm{vol}_{S^3})\wedge \left[(\tanh\phi_2\cos\psi+\sin\psi)\cos\xi d\xi\right. \nonumber \\
& &\left.+\sin\xi (\cos\psi-\tanh\phi_2\sin\psi)\right]-\frac{64}{g^3}U\Delta^{-2}\sin\xi \cos^2\xi \epsilon_{(2)}\wedge d\xi \wedge d\psi \nonumber \\
& &+\frac{64}{g^3}\Delta^{-2}dr\wedge \epsilon_{(2)}\wedge \left[\frac{1}{2}e^{12\phi_1}\phi_2'\sin\xi \sin^22\xi \cos 2\psi d\xi\right. \nonumber \\
& &
+\frac{1}{2}e^{-4\phi_1}\cos^2\xi \sin2\xi \left\{\sin^2\xi \left(e^{6\phi_1}\cosh2\phi_2\right)'d\psi \right. \nonumber \\
& &\left.\left. +(e^{6\phi_1}\sinh2\phi_2\right)'(\cos\xi \cos2\psi d\xi -\sin\xi \sin 2\psi d\psi)\right\}
\nonumber \\
& &
+2\phi_1'e^{2\phi_1}\cos^2\xi \sin 2\xi \left\{\sin\xi \cosh2\phi_2 d\psi\right.\nonumber \\
& &\left.\phantom{\frac{1}{2}}\left.-\sinh2\phi_2 (\sin \xi\sin2\psi d\psi -\cos2\psi d\xi)\right\}\right]
\end{eqnarray}
where
\begin{eqnarray}
\Delta&=&e^{-4\phi_1}\cos^2\xi+e^{6\phi_1}\sin^2\xi (\cosh 2\phi_2-\sin 2\psi\sinh2\phi_2),\\
U&=&\sin^2\xi \left[3e^{2\phi_1}\sin2\psi \sinh 2\phi_2+e^{12\phi_1}(6\cosh^22\phi_2-\sin2\psi \sinh4\phi_2)\right]\nonumber \\
& &+(2e^{-4\phi_1}-3e^{-8\phi_1})\cos^2\xi+\frac{1}{2}e^{2\phi_1}\cosh2\phi_2 (\cos2\xi-5).
\end{eqnarray}
\indent All of these solutions should describe bound states of M$2$- and M$5$-branes with different transverse spaces and are expected to be holographically dual to conformal surface defects in $N=(2,0)$ SCFT in six dimensions. Solutions with $SO(2)\times SO(2)$ symmetry can similarly be uplifted, but we will not give them here due to their complexity.

\subsubsection{Uplift to type IIA theory}
We now carry out a similar analysis for solutions in $CSO(4,0,1)$ gauge group to find uplifted solutions in ten-dimensional type IIA theory. Relevant formulae are reviewed in appendix \ref{truncation_ansatz}. In the solutions we will consider, gauge fields, massive three-forms and axions $b_i=\chi_i$ vanish. The ten-dimensional fields are then given only by the metric, the dilaton and the NS-NS two-form field. Therefore, in this case, we expect the solutions to describe bound states of NS$5$-branes and the fundamental strings.
\\
\indent We begin with a simpler $SO(4)$ symmetric solution in which the $SL(4)/SO(4)$ scalar matrix is given by $\widetilde{\mc{M}}_{ij}=\delta_{ij}$. The ten-dimensional metric, NS-NS three-form flux and the dilaton are given by
\begin{eqnarray}
d\hat{s}^2_{10}&=&e^{\frac{3}{2}\phi_0}\left(e^{2U}ds_{M_3}^2+e^{2V}dr^2+e^{2W}ds_{S^3}^2 \right)+\frac{16}{g^2}e^{-\frac{5}{2}\phi_0}d\Omega^2_{(3)},\nonumber \\
\hat{H}_{(3)}&=&\frac{128}{g^3}\epsilon_{(3)}+\frac{8}{g}(k\textrm{vol}_{M_3}+l\textrm{vol}_{S^3}),\nonumber \\
\hat{\varphi}&=&5\phi_0\, .
\end{eqnarray}
It should be noted that, in this case, we have a constant NS-NS flux.
\\
\indent For $SO(3)$ symmetric solutions, we parametrize the $SL(4)/SO(4)$ scalar matrix as
\begin{equation}
\widetilde{\mc{M}}_{ij}=\textrm{diag}(e^{2\phi},e^{2\phi},e^{2\phi},e^{-6\phi})
\end{equation}
and choose the $S^3$ coordinates to be
\begin{equation}
\mu^i=(\sin\xi\hat{\mu}^a,\cos\xi),\qquad a=1,2,3
\end{equation}
with $\hat{\mu}^a$ being the coordinates on $S^2$ subject to the condition $\hat{\mu}^a\hat{\mu}^a=1$. We again recall that only solutions with $\phi_2=0$ are possible in this case.
\\
\indent With all these ingredients and writing $k=k_5$ and $l=l_5$, we find that the ten-dimensional fields are given by
\begin{eqnarray}
d\hat{s}^2_{10}&=&e^{\frac{3}{2}\phi_0}\Delta^{\frac{1}{4}}\left(e^{2U}ds_{AdS_3}^2+e^{2V}dr^2+e^{2W}ds_{S^3}^2 \right)\nonumber \\
& &+\frac{16}{g^2}e^{-\frac{5}{2}\phi_0}\Delta^{-\frac{3}{4}}\left[\left(e^{-6\phi}\sin^2\xi+e^{2\phi}\cos^2\xi\right)d\xi^2+\sin^2\xi e^{2\phi}d\hat{\mu}^ad\hat{\mu}^a\right], \\
e^{2\hat{\varphi}}&=&\Delta^{-1}e^{10\phi_0},\\
\hat{H}_{(3)}&=&\frac{64}{g^3}\Delta^{-2}\sin^3\xi\left(U\sin\xi d\xi+8e^{4\phi}\cos\xi \phi'dr\right)\wedge \epsilon_{(2)}+\frac{8}{g}(k\textrm{vol}_{AdS_3}+l\textrm{vol}_{S^3})\nonumber \\
& &
\end{eqnarray}
in which
\begin{eqnarray}
\Delta&=&e^{6\phi}\cos^2\xi+e^{-2\phi}\sin^2\xi,\qquad \epsilon_{(2)}=\frac{1}{2}\epsilon_{abc}\hat{\mu}^ad\hat{\mu}^b\wedge d\hat{\mu}^c,\nonumber \\
U&=&e^{12\phi}\cos^2\xi-e^{-4\phi}\sin^2\xi-e^{4\phi}(\sin^2\xi+3\cos^2\xi).
\end{eqnarray}
The solutions for $\phi_0$ and $\phi$ are obtained from $\phi_1$ and $\phi_3$ in section \ref{SO3_Y_DW} by the following relations
\begin{equation}
\phi=\frac{1}{4}(5\phi_1-\phi_3)\qquad \textrm{and}\qquad \phi_0=-\frac{1}{4}(\phi_3+3\phi_1).
\end{equation}
These are obtained by comparing the scalar matrices obtained from \eqref{SO3_Y_coset} and \eqref{Ti_S3}.

\section{Supersymmetric solutions from gaugings in $\overline{\mathbf{40}}$ representation}\label{Z_gauging}
In this section, we repeat the same analysis for gaugings from $\overline{\mathbf{40}}$ representation. Setting $Y_{MN}=0$, we are left with the quadratic constraint
\begin{equation}\label{RedQuadCon}
\epsilon_{MRSTU}Z^{RS,N}Z^{TU,P}=0\, .
\end{equation}
Following \cite{N4_7D_Henning}, we can solve this constraint by taking
\begin{equation}\label{pureZ}
Z^{MN,P}=v^{[M}w^{N]P}
\end{equation}
with $w^{MN}=w^{(MN)}$ and $v^M$ being a five-dimensional vector.
\\
\indent The $SL(5)$ symmetry can be used to fix the vector $v^M=\delta^M_5$. Therefore, it is useful to split the $SL(5)$ index as $M=(i,5)$. Setting $w^{55}=w^{i5}=0$ for simplicity, we can use the remaining $SL(4)\subset SL(5)$ symmetry to diagonalize $w^{ij}$ as
\begin{equation}\label{wij}
w^{ij}=\text{diag}(\underbrace{1,..,1}_p,\underbrace{-1,..,-1}_q,\underbrace{0,..,0}_r).
\end{equation}
The resulting gauge generators read
\begin{equation}\label{pureZgaugeGen}
{(X_{ij})_k}^l=2\epsilon_{ijkm}w^{ml}
\end{equation}
corresponding to a $CSO(p,q,r)$ gauge group with $p+q+r=4$.
\\
\indent With the split of $SL(5)$ index $M=(i,5)$ and the decomposition $SL(5)\rightarrow SL(4)\times SO(1,1)$, we can parametrize the $SL(5)/SO(5)$ coset representative in term of the $SL(4)/SO(4)$ one as
\begin{equation}\label{t=4decompose}
\mathcal{V}=e^{b_it^i}\widetilde{\mathcal{V}}e^{\phi_0t_0}\, .
\end{equation}
$\widetilde{\mathcal{V}}$ is the $SL(4)/SO(4)$ coset representative, and $t_0$, $t^i$ refer to $SO(1,1)$ and four nilpotent generators, respectively. The unimodular matrix $\mathcal{M}_{MN}$ is then given by
\begin{equation}
\mathcal{M}_{MN}=\begin{pmatrix}e^{-2\phi_0}\widetilde{\mathcal{M}}_{ij}+e^{8\phi_0}b_ib_j & e^{8\phi_0}b_i \\
e^{8\phi_0}b_j& e^{8\phi_0}
\end{pmatrix}
\end{equation}
with $\widetilde{\mathcal{M}}_{ij}=(\widetilde{\mathcal{V}}\widetilde{\mathcal{V}}^T)_{ij}$. Using \eqref{scalarPot}, we can compute the scalar potential for these gaugings
\begin{equation}\label{ZwithNilscalarPot}
\mathbf{V}=\frac{g^2}{4}e^{14\phi_0}b_iw^{ij}\widetilde{\mathcal{M}}_{jk}w^{kl}b_l+\frac{g^2}{4}e^{4\phi_0}\left(2\widetilde{\mathcal{M}}_{ij}w^{jk}\widetilde{\mathcal{M}}_{kl}w^{li}-(\widetilde{\mathcal{M}}_{ij}w^{ij})^2\right).
\end{equation}
The presence of the dilaton prefactor $e^{\phi_0}$ shows that this potential does not admit any critical points. Note also that we can always consistently set the nilpotent scalars $b_i$ to zero for simplicity since they do not appear linearly in any terms in the Lagrangian.
\\
\indent We will use the same ansatz as in the case of gaugings in the $\mathbf{15}$ representation to find charged domain wall solutions. However, we note here that, for gaugings in the $\overline{\mathbf{40}}$ representation, there are no massive three-form fields $S^M_{\mu\nu\rho}$. The three-form fluxes given in \eqref{fulldyonic3form} in this case correspond solely to the two-form fields $B_{\mu\nu M}$. We now consider a number of possible solutions with different symmetries.

\subsection{$SO(4)$ symmetric charged domain walls}
For $SO(4)$ residual symmetry under which only the scalar field $\phi_0$ is invariant, we have $\widetilde{\mc{M}}_{ij}=\delta_{ij}$. The only gauge group that can accommodate the $SO(4)$ unbroken symmetry is $SO(4)$ with the embedding tensor component $w^{ij}=\delta^{ij}$. The scalar potential as obtained from \eqref{ZwithNilscalarPot} takes a very simple form
\begin{equation}
\mathbf{V}=-2g^2e^{4\phi_0}
\end{equation}
which does not admit any critical points. We will consider solutions with non-vanishing $\mc{H}^{(3)}_{\mu\nu\rho 5}$ which is an $SO(4)$ singlet.
\\
\indent In this $SO(4)$ gauging, there are four massive two-form fields $B_{\mu\nu i}$, $i=1,\ldots, 4$, and one massless two-form field $B_{\mu\nu 5}$ with the latter being an $SO(4)$ singlet. We will take the ansatz for $B_{\mu\nu 5}$ as given in \eqref{sdSO(3)dyonic3form}. With the following projection conditions
\begin{equation}\label{ZDefDWProj}
\gamma_{\hat{3}}\epsilon^a_0=-{(\Gamma_5)^a}_b\epsilon^b_0=\epsilon^a_0,
\end{equation}
the BPS equations are given by
\begin{eqnarray}
U'&\hspace{-0.3cm}=&\hspace{-0.2cm}W'=\frac{1}{5}e^{V}\left(2e^{-2\phi_0}g\sec{2\theta}-e^{-U}\tau\tan{2\theta}\right),\\
\phi_0'&\hspace{-0.3cm}=&\hspace{-0.3cm}\frac{1}{10}e^{V}\left(2e^{-2\phi_0}g\sec{2\theta}-e^{-U}\tau\tan{2\theta}\right),\\
k&\hspace{-0.3cm}=&\hspace{-0.3cm}-\frac{1}{2}e^{2U-4\phi_0}\tau,\qquad \theta'=0,\\
l&\hspace{-0.3cm}=&\hspace{-0.3cm}-\frac{1}{2}e^{2U-4\phi_0}\tau\sec{2\theta}+3e^{3U-6\phi_0}g\tan{2\theta}
\end{eqnarray}
together with an algebraic constraint
\begin{equation}
\kappa=\tau\sec{2\theta}-2e^{U-2\phi_0}g\tan{2\theta}\, .
\end{equation}
In this case, we find that $\theta$ is constant. Choosing $V=0$, we find the following solution
\begin{eqnarray}
U&=&W=2\phi_0,\label{Usol_1}\\
e^{2\phi_0}&=&\frac{2}{5}gr\sec{2\theta}-\frac{1}{5}\tau r\tan{2\theta}+C,\\
 k&=&-\frac{1}{2}\tau,\\
l&=&-\frac{1}{2}\tau\sec{2\theta}+g\tan{2\theta}\label{lsol_1}
\end{eqnarray}
with an integration constant $C$. For a particular value of $\theta=0$, we find the solution
\begin{equation}
U=W=2\phi_0,\qquad e^{2\phi_0}=\frac{2}{5}gr+C,\qquad
 k=l=-\frac{1}{2}\tau\, .\label{charged_DW_SO4_Z}
\end{equation}

\subsubsection{Coupling to $SO(3)$ gauge fields}
\begin{figure}[h!]
  \centering
  \begin{subfigure}[b]{0.32\linewidth}
    \includegraphics[width=\linewidth]{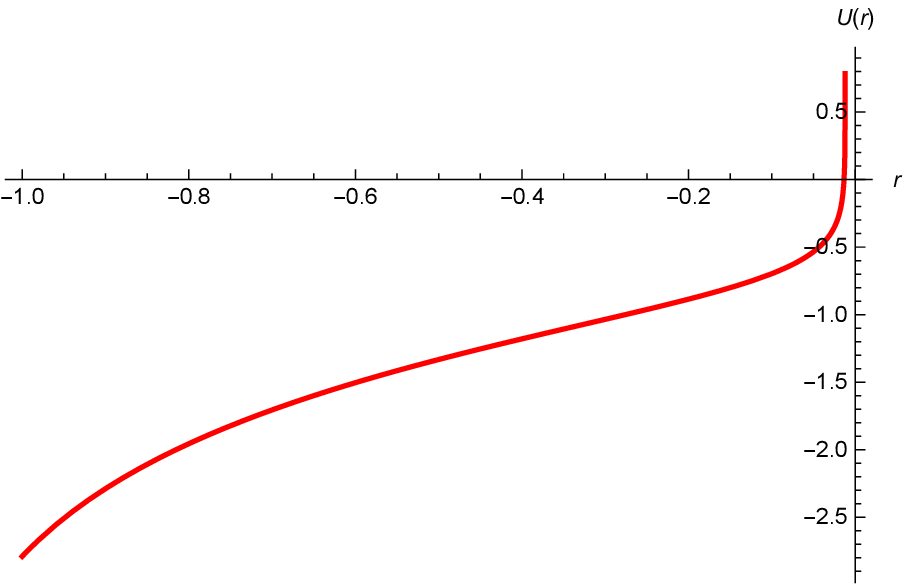}
  \caption{$U$ solution}
  \end{subfigure}
  \begin{subfigure}[b]{0.32\linewidth}
    \includegraphics[width=\linewidth]{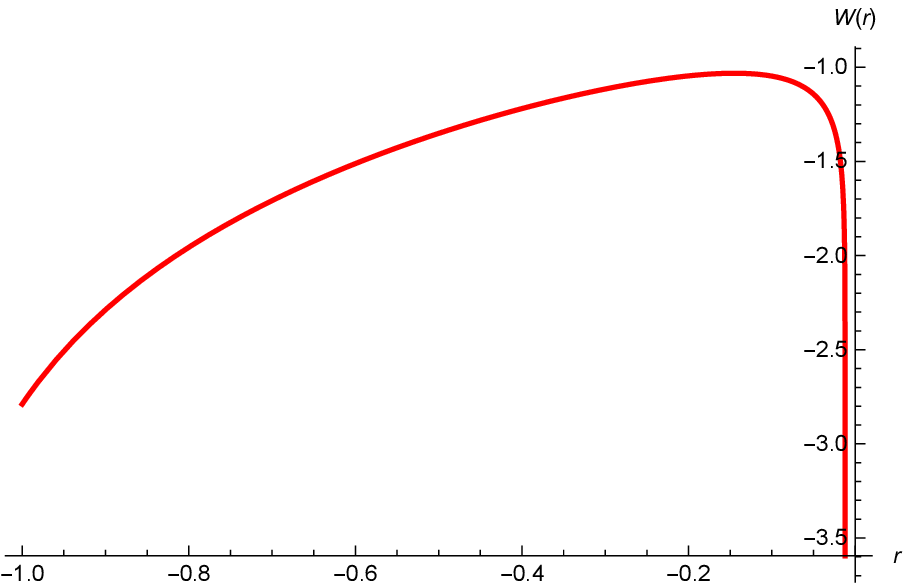}
  \caption{$W$ solution}
  \end{subfigure}
  \begin{subfigure}[b]{0.32\linewidth}
    \includegraphics[width=\linewidth]{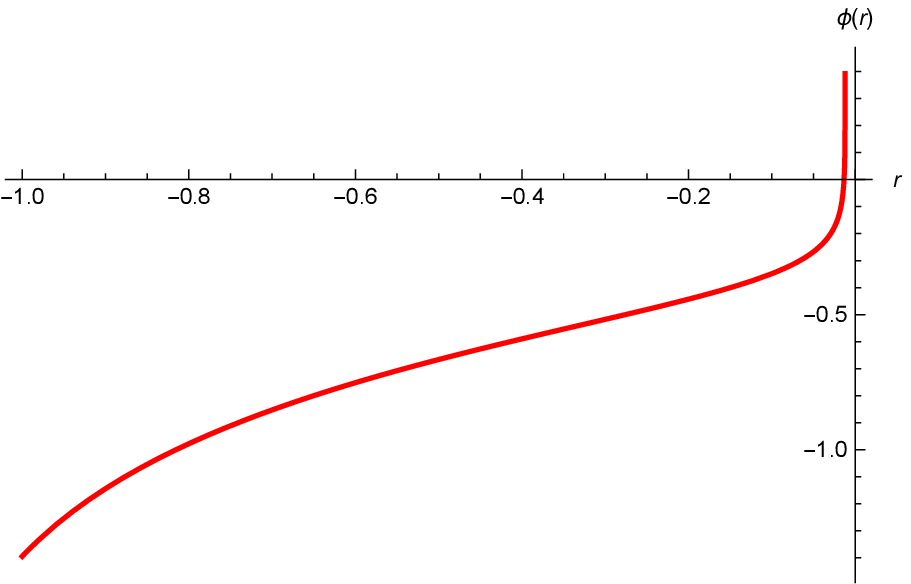}
  \caption{$\phi$ solution}
  \end{subfigure}
  \begin{subfigure}[b]{0.32\linewidth}
    \includegraphics[width=\linewidth]{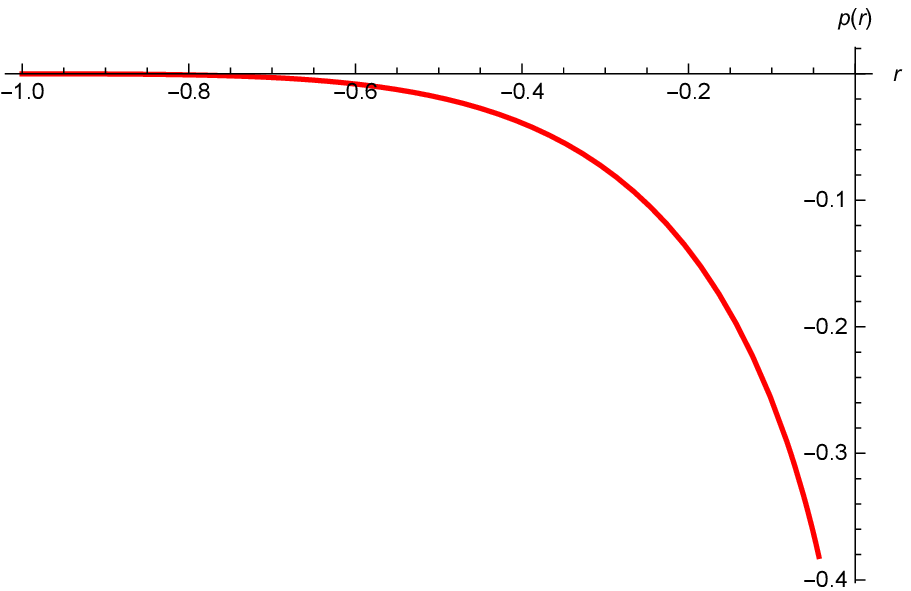}
  \caption{$p$ solution}
  \end{subfigure}
  \begin{subfigure}[b]{0.32\linewidth}
    \includegraphics[width=\linewidth]{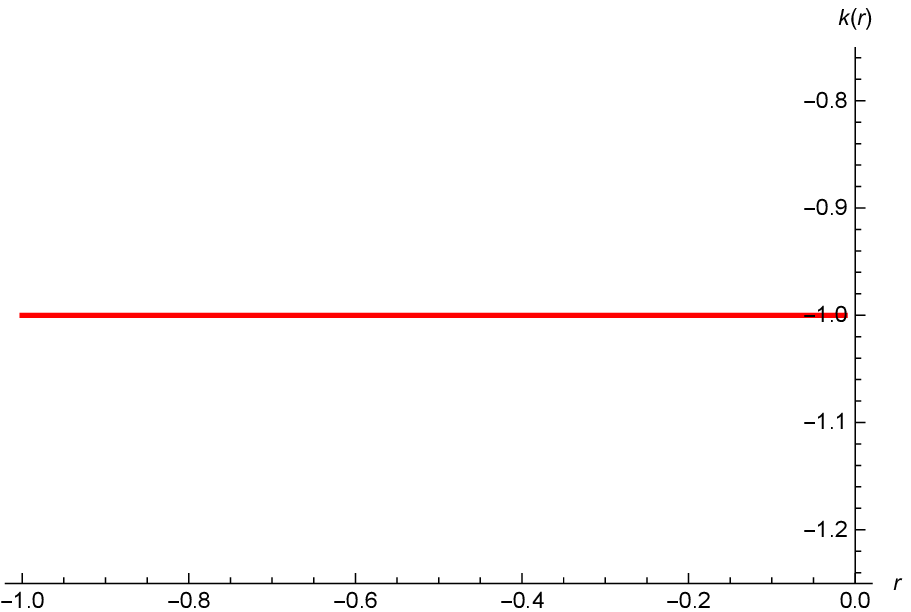}
  \caption{$k$ solution}
  \end{subfigure}
  \begin{subfigure}[b]{0.32\linewidth}
    \includegraphics[width=\linewidth]{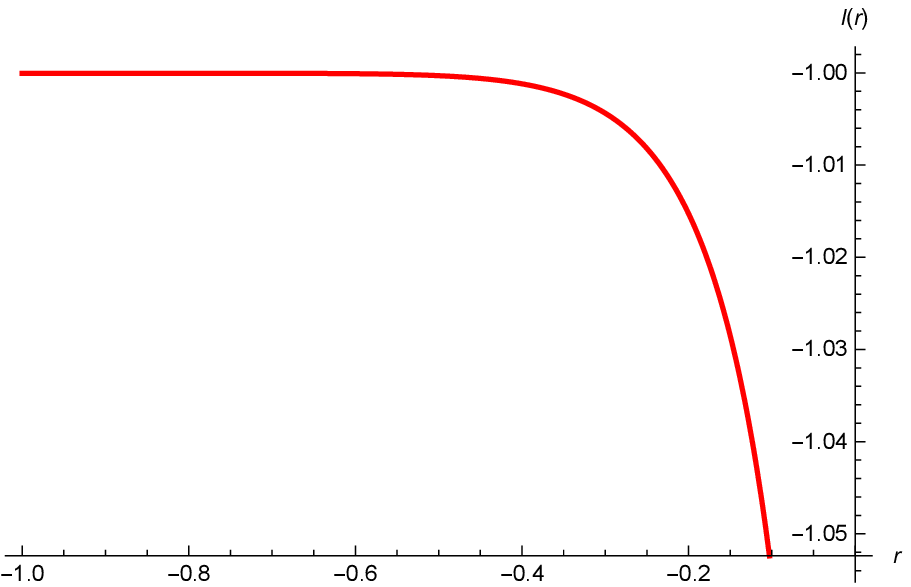}
  \caption{$l$ solution}
  \end{subfigure}
  \caption{A BPS flow from a charged domain wall at $r=-1$ to a singularity at $r=0$ in $SO(4)$ gauge group with $g=1$, $\kappa=\tau=2$ and $C=\frac{2}{5}$.}
  \label{ZSO4_new_flow}
\end{figure}

We now consider charged domain wall solutions with non-vanishing $SO(3)\subset SO(4)$ gauge fields. In this case, the projector ${(\Gamma_5)^a}_b\epsilon^b_0=-\epsilon^a_0$ implies that the non-vanishing gauge fields correspond to the self-dual $SO(3)\subset SO(4)$ given by
\begin{eqnarray}
A^{23}_{(1)}=A^{14}_{(1)}=\frac{\kappa}{16}p(r)e^{-W(r)} e^{\hat{4}},\\
A^{31}_{(1)}=A^{24}_{(1)}=\frac{\kappa}{16}p(r)e^{-W(r)} e^{\hat{5}},\\
A^{12}_{(1)}=A^{34}_{(1)}=\frac{\kappa}{16}p(r)e^{-W(r)} e^{\hat{6}}.
\end{eqnarray}
The two-form field strengths are straightforward to obtain
\begin{eqnarray}
F^{12}_{(2)}=F^{34}_{(2)}=e^{-V-W}\frac{\kappa}{16}p' e^{\hat{3}}\wedge e^{\hat{6}}+e^{-2W}\frac{\kappa^2}{32}p(2-gp)e^{\hat{4}}\wedge e^{\hat{5}},\\
F^{23}_{(2)}=F^{14}_{(2)}=e^{-V-W}\frac{\kappa}{16}p' e^{\hat{3}}\wedge e^{\hat{4}}+e^{-2W}\frac{\kappa^2}{32}p(2-gp)e^{\hat{5}}\wedge e^{\hat{6}},\\
F^{31}_{(2)}=F^{24}_{(2)}=e^{-V-W}\frac{\kappa}{16}p' e^{\hat{3}}\wedge e^{\hat{5}}+e^{-2W}\frac{\kappa^2}{32}p(2-gp)e^{\hat{6}}\wedge e^{\hat{4}}\, .
\end{eqnarray}
\indent Since the components of the embedding tensor $Z^{ij,5}$ vanish, the two-form field $B^{(2)}_5$ does not contribute to the modified two-form field strengths. Imposing the projection conditions \eqref{DefDWSO(3)Proj} and \eqref{ZDefDWProj}, we find the following BPS equations
\begin{eqnarray}
U'&\hspace{-0.2cm}=&\hspace{-0.2cm}\frac{e^{V-2(W+\phi_0)}}{80\cos{2\theta}}\left[16e^{2W}\left(g(3\cos{4\theta}-1)+2e^{2\phi_0-U}\tau\sin{2\theta}\right)\right.\nonumber\\&&\hspace{-0.2cm}\left.-3e^{4\phi_0}\left(\kappa^2p(gp-2)(\cos{4\theta}-3)-8e^{W-2\phi_0}\kappa(gp-1)\sin{4\theta}\right)\right],\\
W'&\hspace{-0.2cm}=&\hspace{-0.2cm}\frac{e^{V-2(W+\phi_0)}}{40\cos{2\theta}}\left[8e^{2W}\left(2g(2-\cos{4\theta})-3e^{2\phi_0-U}\tau\sin{2\theta}\right)\right.\nonumber\\&&\hspace{-0.2cm}\left.+e^{4\phi_0}\left(\kappa^2p(gp-2)(\cos{4\theta}-8)-8e^{W-2\phi_0}\kappa(gp-1)\sin{4\theta}\right)\right],\\
\phi_0'&\hspace{-0.2cm}=&\hspace{-0.2cm}\frac{e^{V-2(W+\phi_0)}}{160\cos{2\theta}}\left[16e^{2W}\left(g(3\cos{4\theta}-1)+2e^{2\phi_0-U}\tau\sin{2\theta}\right)\right.\nonumber\\&&\hspace{-0.2cm}\left.+3e^{4\phi_0}\left(\kappa^2p(gp-2)(3-\cos{4\theta})+8e^{W-2\phi_0}\kappa(gp-1)\sin{4\theta}\right)\right],\hspace{0.4cm}\\
\theta'&\hspace{-0.2cm}=&\hspace{-0.2cm}\frac{e^{V-2(W+\phi_0)}}{16}\left[24e^{W+2\phi_0}\left(e^{W-U}\tau+\kappa(gp-1)\cos{2\theta}\right)\right.\nonumber\\&&\hspace{-0.2cm}\left.-3\left(16ge^{2W}-e^{4\phi_0}\kappa^2p(gp-2)\right)\sin{2\theta}\right],\\
k&\hspace{-0.2cm}=&\hspace{-0.2cm}-\frac{1}{2}e^{2U-4\phi_0}\tau,\label{Zksol}\\
l&\hspace{-0.2cm}=&\hspace{-0.2cm}\frac{1}{8}e^{3W-6\phi_0}\left[-16g\tan{2\theta}+8e^{2\phi_0-U}\tau\sec{2\theta}\right.\nonumber\\&&\hspace{-0.2cm}\left.+3e^{4\phi_0-2W}\left(\kappa^2p(gp-2)\tan{2\theta}+4e^{W-2\phi_0}\kappa(gp-1)\right)\right],\label{Zlsol}\\
p'&\hspace{-0.2cm}=&\hspace{-0.2cm}\frac{e^{V-W-4\phi_0}}{2\kappa}\left[8e^{W+2\phi_0}\left(e^{W-U}\tau+\kappa(gp-1)\cos{2\theta}\right)\right.\nonumber\\&&\hspace{-0.2cm}\left.-\left(16ge^{2W}-e^{4\phi_0}\kappa^2p(gp-2)\right)\sin{2\theta}\right].\label{Zp'flow}
\end{eqnarray}
It can be verified that these BPS equations satisfy the second-order field equations without any additional constraint.
\\
\indent Since there is no an asymptotically locally $AdS_7$ configuration, we will consider flow solutions from a charged domain wall without vector fields given in \eqref{Usol_1}-\eqref{lsol_1} to a singular solution with non-vanishing gauge fields. To find numerical solutions, we will consider the charged domain wall with $\theta=0$ given in \eqref{charged_DW_SO4_Z} for simplicity. As $r\rightarrow-\frac{5C}{2g}$, we impose the following boundary conditions
\begin{eqnarray}\label{ZDefDWprofile}
U& \sim & W\sim\ln\left[\frac{2gr}{5}+C\right],\qquad \phi\sim\frac{1}{2}\ln\left[\frac{2gr}{5}+C\right],\nonumber \\
p&\sim&0, \qquad k\sim l\sim-\frac{\tau}{2}
\end{eqnarray}
with $\tau=\kappa$. An example of the BPS flows is shown in figure \ref{ZSO4_new_flow}. From this solution, it can be seen that $k$ is constant along the flow since the above BPS equations give $U'=2\phi_0'$ which implies the constancy of $U-2\phi_0$. It should also be noted that this solution is similar to that in $CSO(4,0,1)$ gauge group given in figure \ref{YCSO401_new_flow}. We also expect this solution to describe a surface defect within an $N=(2,0)$ nonconformal field theory.

\subsection{$SO(3)$ symmetric charged domain walls}
In this section, we look for more complicated solutions with $SO(3)$ residual symmetry generated by $X_{ij}$ with $i,j=1,2,3$. Gauge groups containing an $SO(3)$ subgroup are $SO(4)$, $SO(3,1)$ and $CSO(3,0,1)$. These gauge groups are described by the embedding tensor $w^{ij}$ of the form
\begin{equation}
w^{ij}=\text{diag}(+1,+1,+1,\rho)
\end{equation}
with $\rho=1,-1,0$, respectively.
\\
\indent Among the ten $SL(4)/SO(4)$ scalars, there is one $SO(3)$ singlet parametrized by the $SL(4)/SO(4)$ coset representative
\begin{equation}\label{ZSO(3)coset}
\widetilde{\mathcal{V}}=\text{diag}(e^\phi,e^\phi,e^\phi,e^{-3\phi}).
\end{equation}
We then obtain the scalar potential using \eqref{ZwithNilscalarPot}
\begin{equation}\label{ZSO(3)Pot}
\mathbf{V}=-\frac{g^2}{4}e^{-4(\phi_0+3\phi)}(3e^{16\phi}+6\rho e^{8\phi}-\rho^2).
\end{equation}
\indent To find the BPS equations, we use the same ansatz for the modified three-form field strength \eqref{sdSO(3)dyonic3form} and impose the projection conditions \eqref{ZDefDWProj}. We note here that, in this case, there are two two-form fields, $B^{(2)}_4$ and $B^{(2)}_5$, which are $SO(3)$ singlets. For $CSO(3,0,1)$ gauge group with $\rho=0$, both of them are massless while for the other two gauge groups, the former is massive while the latter is massless. However, in this case, we are not able to consistently incorporate $B^{(2)}_4$ in the BPS equations. We will accordingly restrict ourselves to the solutions with only $B^{(2)}_5$ non-vanishing.
\\
\indent Consistency with the field equations also leads to the conditions given in \eqref{SO3_SO2_con}. With all these, the resulting BPS equations are given by
\begin{eqnarray}
U'&\hspace{-0.2cm}=&\hspace{-0.2cm}W'=\frac{g}{10}e^{V-6\phi-2\phi_0}(3e^{8\phi_1}+\rho),\\
\phi'_0&\hspace{-0.2cm}=&\hspace{-0.2cm}\frac{g}{20}e^{V-6\phi-2\phi_0}(3e^{8\phi_1}+\rho),\\
\phi'&\hspace{-0.2cm}=&\hspace{-0.2cm}-\frac{g}{4}e^{V-6\phi-2\phi_0}(3e^{8\phi_1}-\rho),\\
k&\hspace{-0.2cm}=&\hspace{-0.2cm}-\frac{1}{2}e^{3U-W-4\phi_0}\kappa,\\
l&\hspace{-0.2cm}=&\hspace{-0.2cm}-\frac{1}{2}e^{2W-4\phi_0}\kappa\, .
\end{eqnarray}
Setting $W=U$ and $V=0$, we find the solutions for $U$, $\phi_0$, $l$ and $k$ as functions of $\phi$
\begin{eqnarray}
U&\hspace{-0.2cm}=&\hspace{-0.2cm}\frac{2}{5}\phi-\frac{1}{5}\ln\left(e^{8\phi}-\rho\right),\\
\phi_0&\hspace{-0.2cm}=&\hspace{-0.2cm}\frac{1}{5}\phi-\frac{1}{10}\ln\left(e^{8\phi}-\rho\right)+C_0,\\
k&\hspace{-0.2cm}=&\hspace{-0.2cm}l=-\frac{1}{2}e^{-4C_0}\kappa
\end{eqnarray}
in which $C_0$ is an integration constant.
\\
\indent The solution for $\phi(r)$ is given by
\begin{equation}
\phi=-\frac{5}{16}\ln\left[\frac{4}{5}(e^{-2C_0}gr-C_1)\right]
\end{equation}
for $\rho=0$ and
\begin{equation}
4g\rho r (e^{8\phi}-\rho)^{1/5}=5e^{2C_1+\frac{32}{5}\phi}\left[4-3(1-\rho e^{8\phi})^{1/5}{_2F_1}(\frac{1}{5},\frac{4}{5},\frac{9}{5},\rho e^{8\phi})\right]
\end{equation}
for $\rho=\pm 1$. In the last equation, $_2F_1$ is the hypergeometric function. This solution is again the domain wall found in \cite{our_7D_DW} with a non-vanishing three-form flux.
\\
\indent As in the $SO(3)$ symmetric solutions from the gaugings in the $\mathbf{15}$ representation, coupling to $SO(3)$ vector fields does not lead to new solutions. Consistency with the field equations implies either vanishing two-form fields or vanishing gauge fields. We also note that repeating the same analysis for $SO(2)\times SO(2)$ and $SO(2)$ symmetric solutions leads to the domain wall solutions given in \cite{our_7D_DW} with a constant three-form flux 
\begin{equation}
k=l=-\frac{1}{2}\tau\, . \label{constant3form}
\end{equation}
We will not give further detail for these cases to avoid a repetition.

\section{Supersymmetric solutions from gaugings in $\mathbf{15}$ and $\overline{\mathbf{40}}$ representations}\label{YZ_gauging}
In this section, we consider gaugings with both components of the embedding tensor in \textbf{15} and $\overline{\mathbf{40}}$ representations non-vanishing. We first give a brief review of these gaugings as constructed in \cite{N4_7D_Henning}. A particular basis can be chosen such that non-vanishing components of the embedding tensor are given by
\begin{equation}\label{mixembedding}
Y_{xy}, \qquad Z^{x\alpha,\beta}=Z^{x(\alpha,\beta)}, \qquad Z^{\alpha\beta,\gamma},
\end{equation}
with $x=1,...,t$ and $\alpha=t+1,...,5$. The $SL(5)$ index $M,N,\ldots$ are then split into $(x,\alpha)$.
\\
\indent In terms of these components, the quadratic constraint \eqref{QuadCon} reads
\begin{equation}\label{mixQuadCon}
Y_{xy}Z^{y\alpha,\beta}+2\epsilon_{xMNPQ}Z^{MN,\alpha}Z^{PQ,\beta}=0\, .
\end{equation}
$Y_{xy}$ is chosen to be
\begin{equation}\label{Yxy}
Y_{xy}=\text{diag}(\underbrace{1,..,1}_p,\underbrace{-1,..,-1}_q).
\end{equation}
We will consider two gauge groups namely $SO(2,1)\ltimes\mathbf{R}^4$ and $SO(2)\ltimes\mathbf{R}^4$ given in \cite{N4_7D_Henning}. The latter can be obtained from Scherk-Schwarz reduction of the maximal gauged supergravity in eight dimensions.
\\
\indent We begin with the $t=3$ case in which $Y_{xy}=\text{diag}(1,1,-1)$ corresponding to $SO(2,1)\ltimes\mathbf{R}^4$ gauge group. The corresponding gauge generators are given by
\begin{equation}\label{t=3gaugeGen}
{X_M}^N=\begin{pmatrix} \lambda^z{(t^z)_x}^y & Q^{(4)\beta}_{x} \\ 0_{2\times 3} & \frac{1}{2}\lambda^z {(\zeta^z)_\alpha}^\beta  \end{pmatrix}
\end{equation}
with $\lambda^z\in \mathbb{R}$ and ${(t^z)_x}^y=\epsilon^{zyu}Y_{ux}$ being generators of $SO(2,1)$ in the adjoint representation. The nilpotent generators $Q^{(4)\alpha}_{x}$ transform as $\mathbf{4}$ under $SO(2,1)$. In terms of $\zeta^x$, the component $Z^{x\alpha,\beta}$ of the embedding tensor takes the form
\begin{equation}\label{mixt=3Z}
Z^{x\alpha,\beta}=-\frac{1}{16}\epsilon^{\alpha\gamma}{(\zeta^x)_\gamma}^\beta.
\end{equation}
The explicit form of $\zeta^x$ can be given in terms of Pauli matrices as
\begin{equation}\label{realrepso(2,1)}
\zeta^1=\sigma_1, \qquad \zeta^2=\sigma_3, \qquad \zeta^3=i\sigma_2\, .
\end{equation}
\indent We now consider charged domain wall solutions with $SO(2)\subset SO(2,1)$ symmetry. As shown in \cite{our_7D_DW}, there are four $SO(2)$ singlet scalars corresponding to the following non-compact generators
\begin{eqnarray}
\bar{Y}_1&=& 2e_{1,1}+2e_{2,2}+2e_{3,3}-3e_{4,4}-3e_{5,5},\nonumber \\
\bar{Y}_2&=& e_{1,1}+e_{2,2}-2e_{3,3},\nonumber \\
\bar{Y}_3&=& e_{1,4}+e_{2,5}+e_{4,1}+e_{5,2},\nonumber \\
\bar{Y}_4&=& e_{1,5}-e_{2,4}-e_{4,2}+e_{5,1}\, .
\end{eqnarray}
The $SL(5)/SO(5)$ coset representative can be written as
\begin{equation}\label{YZSO(2)coset}
\mathcal{V}=e^{\phi_1\bar{Y}_1+\phi_2\bar{Y}_2+\phi_3\bar{Y}_3+\phi_4\bar{Y}_4}\, .
\end{equation}
The resulting scalar potential is given by
\begin{equation}\label{YZt=3scalarPot}
\mathbf{V}=\frac{g^2}{64}e^{-2(4\phi_1-\phi_2)}\left[6\cosh{2\phi_3}\cosh{2\phi_4}+e^{6\phi_2}\right]
\end{equation}
which does not admit any critical points.
\\
\indent We now repeat the same analysis as in the previous sections. We first discuss the three-form fluxes that are singlet under the $SO(2)$ residual symmetry. In the ungauged supergravity, the five two-forms transform as $\mathbf{5}$ under $SL(5)$. From the particular form of the gauge generators given in \eqref{t=3gaugeGen}, we can see that the $SO(2)$ symmetry under consideration here is embedded diagonally along the $1,2,4,5$ directions. Under $SO(2)\times SO(2)\subset SO(5)\subset SL(5)$, the two-forms transform as $(\mathbf{1},\mathbf{1})+(\mathbf{1},\mathbf{2})+(\mathbf{2},\mathbf{1})$. Under $SO(2)=[SO(2)\times SO(2)]_{\textrm{diag}}$, these two-forms transform as $\mathbf{1}+\mathbf{2}+\mathbf{2}$. Therefore, there is only one singlet two-form field under the $SO(2)$ unbroken symmetry. In gauged supergravity, this two-form field will be gauged away by a three-form gauge transformation due to the non-vanishing component $Y_{33}$ of the embedding tensor. The $SO(2)$ singlet is then described by a massive three-form field $S^{(3)}_3$.
\\
\indent We will take the ansatz for the three-form field strength to be
\begin{equation}\label{YZdyonic3form}
\mathcal{H}^{(3)}_{\hat{m}\hat{n}\hat{p} 3}=k(r)e^{-3U(r)}\varepsilon_{\hat{m}\hat{n}\hat{p}} \qquad \textrm{ and } \qquad
\mathcal{H}^{(3)}_{\hat{i}\hat{j}\hat{k} 3}=l(r)e^{-3W(r)}\varepsilon_{\hat{i}\hat{j}\hat{k}}\, .
\end{equation}
After imposing the following projection conditions
\begin{equation}
\gamma_{\hat{3}}\epsilon^a_0=-{(\Gamma_3)^a}_b\epsilon^b_0=\epsilon^a_0,
\end{equation}
we find the following BPS equations
\begin{eqnarray}
U'&\hspace{-0.2cm}=&\hspace{-0.2cm}W'=\frac{g}{40}e^{-2(2\phi_1+\phi_2)+V}\left(3\cosh{2\phi_3}\cosh{2\phi_4}-e^{6\phi_2}\right),\\
\phi'_1&\hspace{-0.2cm}=&\hspace{-0.2cm} \frac{g}{240}e^{-2(\phi_1+\phi_2)+V}\left(15\textrm{sech}{2\phi_3}\textrm{sech}{2\phi_4}-3\cosh{2\phi_3}\cosh{2\phi_4}-4e^{6\phi_2}\right),\hspace{0.8cm}\\
\phi'_2&\hspace{-0.2cm}=&\hspace{-0.2cm}\frac{g}{48}e^{-2(\phi_1+\phi_2)+V}\left(3\textrm{sech}{2\phi_3}\textrm{sech}{2\phi_4}+3\cosh{2\phi_3}\cosh{2\phi_4}+4e^{6\phi_2}\right),\\
\phi'_3&\hspace{-0.2cm}=&\hspace{-0.2cm}-\frac{3g}{16}e^{-2(2\phi_1+\phi_2)+V}\sinh{2\phi_3}\textrm{sech}{2\phi_4},\\
\phi'_4&\hspace{-0.2cm}=&\hspace{-0.2cm} -\frac{3g}{16}e^{-2(2\phi_1+\phi_2)+V}\cosh{2\phi_3}\sinh{2\phi_4},\\
k&\hspace{-0.2cm}=&\hspace{-0.2cm}-\frac{1}{2}e^{2U+2\phi_1-2\phi_2}\tau,\\
l&\hspace{-0.2cm}=&\hspace{-0.2cm}-\frac{1}{2}e^{3W-U+2\phi_1-2\phi_2}\tau\, .
\end{eqnarray}
In these equations, we have imposed the conditions \eqref{SO3_SO2_con} for consistency.
\\
\indent By choosing $V=4\phi_1+2\phi_2$ and taking $W=U$ for convenience, we obtain a charged domain wall solution
\begin{eqnarray}
\phi_1&\hspace{-0.2cm}=&\hspace{-0.2cm}\frac{2}{15}\phi_3+\frac{1}{5}C_2-\frac{1}{60}\ln\left[\frac{9}{16}(e^{2C_4}-e^{4\phi_3}-2e^{2C_4+4\phi_3}+e^{2C_4+8\phi_3})\right]\nonumber\\&\hspace{-0.2cm}&\hspace{-0.2cm}+\frac{1}{10}\ln\left[e^{4\phi_3}+1\right]-\frac{1}{5}\ln\left[e^{4\phi_3}-1\right],\\
\phi_2&\hspace{-0.2cm}=&\hspace{-0.2cm}-5\phi_1+C_2+\ln\left[e^{3\phi_3}+1\right]-\ln\left[e^{3\phi_3}-1\right],\\
\phi_3&\hspace{-0.2cm}=&\hspace{-0.2cm}\frac{1}{4}\ln\left[\frac{1+4e^{2C_4}-2e^{\frac{3gr}{8}}+e^{\frac{3gr}{4}}}{1+4e^{2C_4}+2e^{\frac{3gr}{8}}+e^{\frac{3gr}{4}}}\right],\\
\phi_4&\hspace{-0.2cm}=&\hspace{-0.2cm}\frac{1}{4}\ln\left[\frac{e^{2\phi_3}-e^{C_4}+e^{C_4+4\phi_3}}{e^{2\phi_3}+e^{C_4}-e^{C_4+4\phi_3}}\right],\\
U&\hspace{-0.2cm}=&\hspace{-0.2cm}-\frac{1}{5}\phi_3-\frac{1}{20}C_2+\frac{3}{20}\ln\left[e^{2C_4}-e^{4\phi_3}-2e^{2C_4+4\phi_3}+e^{2C_4+8\phi_3}\right]\nonumber\\&\hspace{-0.2cm}&\hspace{-0.2cm}-\ln\left[\frac{16}{9}\right]-\frac{1}{5}\ln\left[e^{4\phi_3}-1\right],\\
k&\hspace{-0.2cm}=&\hspace{-0.2cm}l=-\frac{e^{\frac{3}{10}(C_2+4\phi_3)}\tau}{2^{2/5}\times3^{3/10}}\frac{(e^{2C_4}-e^{4\phi_3}-2e^{2C_4+4\phi_3}+e^{2C_4+8\phi_3})^{1/10}}{(e^{4\phi_3}-1)^{4/5}}\, .\label{k_sol_15_40}
\end{eqnarray}
This is just the $\frac{1}{4}$-BPS domain wall obtained in \cite{our_7D_DW} together with the running dyonic profile of the three-form flux. It is useful to emphasize here that this solution is $\frac{1}{4}$-supersymmetric. In general, domain wall solutions from gaugings in both \textbf{15} and $\overline{\mathbf{40}}$ representations preserve only $\frac{1}{4}$ of the original supersymmetry, see a general discussion in \cite{Eric_SUSY_DW} and explicit solutions in \cite{our_7D_DW}. From the above solution, we see that the solutions with a non-vanishing three-form flux do not break supersymmetry any further.
\\
\indent We end this section by giving a comment on the $t=2$ case with $SO(2)\ltimes\mathbf{R}^4$ gauge group. Repeating the same procedure leads to a charged domain wall given by the solution found in \cite{our_7D_DW} with a constant three-form flux given in \eqref{constant3form}. In contrast to the $t=3$ case, the three-form flux $H^{(3)}_3$ is due to the massless two-form field $B^{(2)}_{3}$ since, in this case, we have $Y_{33}=0$. We will not give the full detail of this analysis here as it closely follows that of the previous cases.

\section{Conclusions and discussions}\label{conclusion}
In this paper, we have studied supersymmetric solutions of the maximal gauged supergravity in seven dimensions with various gauge groups. These solutions are charged domain walls with $M_3\times S^3$ slices, for $M_3=Mkw_3,AdS_3$, and non-vanishing three-form fluxes. All of these solutions can be obtained analytically. For $SO(4)$ residual symmetry, the charged domain wall solutions can couple to $SO(3)\subset SO(4)$ gauge fields, but the corresponding solutions can only be obtained numerically. For $SO(3)$ symmetric solutions, coupling to $SO(3)$ gauge fields does not lead to a consistent set of BPS equations that is compatible with the field equations. In this case, only solutions with either non-vanishing three-form fluxes or non-vanishing gauge fields are possible. Apart from these solutions, we have also given a number of $SO(2)\times SO(2)$ and $SO(2)$ symmetric solutions.
\\
\indent For $SO(5)$ gauge group, the gauged supergravity admits a supersymmetric $AdS_7$ vacuum dual to an $N=(2,0)$ SCFT in six dimensions. In this case, the solutions with an $AdS_3\times S^3$ slice can be interpreted as surface defects within the $N=(2,0)$ SCFT. For other gauge groups, the supersymmetric vacua, with only the metric and scalars non-vanishing, take the form of half-supersymmetric domain walls dual to $N=(2,0)$ non-conformal field theories in six dimensions. We then expect these $AdS_3\times S^3$-sliced domain wall solutions to describe $\frac{1}{4}$-BPS surface defects in the dual $N=(2,0)$ quantum field theories. For a number of solutions, we have found that the charged domain walls are simply given by the domain wall solutions given in \cite{our_7D_DW} with constant three-form fluxes. However, the charged domain walls preserve only $\frac{1}{4}$ of the original supersymmetry as opposed to the usual domain walls which are $\frac{1}{2}$-supersymmetric except for the domain walls from gaugings in both $\mathbf{15}$ and $\overline{\mathbf{40}}$ representations in which both charged and standard domain walls are $\frac{1}{4}$-supersymmetric.
\\
\indent Both gaugings in $\mathbf{15}$ and $\overline{\mathbf{40}}$ representations we have studied can respectively be uplifted to eleven-dimensional supergravity and type IIB theory as shown in \cite{Henning_Hohm1} and \cite{Henning_Emanuel}. We have performed only the uplift for solutions in $SO(5)$ and $CSO(4,0,1)$ gauge groups with $SO(4)$ and $SO(3)$ symmetries. In these cases, the complete truncation ansatze of eleven-dimensional supergravity on $S^4$ and type IIA theory on $S^3$ are known. Similar to the solutions in \cite{7D_sol_Dibitetto}, the uplifted solutions in these two gauge groups should describe bound states of M$2$- and M$5$-branes and of F$1$-strings and NS$5$-branes, respectively. It is natural to extend this study by constructing the full truncation ansatze of eleven-dimensional supergravity on $H^{p,q}\circ T^{5-p-q}$ and type IIB theory on $H^{p,q}\circ T^{4-p-q}$. These can be used to uplift the solutions in $CSO(p,q,5-p-q)$ and $CSO(p,q,4-p-q)$ gauge groups for any values of $p$ and $q$ leading to the full holographic interpretation of the seven-dimensional solutions found here.
\\
\indent Finding the description of conformal defects, dual to the supergravity solutions given in this paper, in the dual $N=(2,0)$ SCFT and $N=(2,0)$ QFT would be interesting and could provide another verification for the validity of the AdS/CFT correspondence. Finally, finding solutions of the form $AdS_d\times \Sigma^{7-d}$ in seven-dimensional gauged supergravity with various gauge groups is also of particular interest. These solutions would be dual to twisted compactifications of $N=(2,0)$ SCFT and $N=(2,0)$ QFT in six dimensions on a $(7-d)$-manifold $\Sigma^{7-d}$ to $(d-1)$-dimensional SCFT.

\begin{acknowledgments}
This work is supported by The Thailand Research Fund (TRF) under grant RSA6280022.
\end{acknowledgments}

\appendix
\section{Bosonic field equations}
In this appendix, we give the explicit form of the bosonic field equations derived from the Lagrangian \eqref{BosLag}. These equations read
\begin{eqnarray}
0&=&R_{\mu\nu}-\frac{1}{4}\mathcal{M}_{MP}\mathcal{M}_{NQ}(D_\mu\mathcal{M}^{MN})(D_\nu\mathcal{M}^{PQ})-\frac{2}{5}g_{\mu\nu}\mathbf{V}\nonumber\\&&-4\mathcal{M}_{MP}\mathcal{M}_{NQ}\left(\mathcal{H}_{\mu\rho}^{(2)MN}{{\mathcal{H}^{(2)PQ}}_{\nu}}^\rho-\frac{1}{10}g_{\mu\nu}\mathcal{H}_{\rho\sigma}^{(2)MN}\mathcal{H}^{(2)PQ\rho\sigma}\right)\\&&-\mathcal{M}^{MN}\left(\mathcal{H}_{\mu\rho\sigma M}^{(3)}{\mathcal{H}_\nu^{(3)\rho\sigma}}_N-\frac{2}{15}g_{\mu\nu}\mathcal{H}_{\rho\sigma\lambda M}^{(3)}{\mathcal{H}^{(3)\rho\sigma\lambda}}_N\right),\nonumber\\
0&=&D^\mu(\mathcal{M}_{MP}D_\mu\mathcal{M}^{PN})-\frac{g^2}{8}\mathcal{M}^{PQ}\mathcal{M}^{RN}\left(2Y_{RQ}Y_{PM}-Y_{PQ}Y_{RM}\right)\nonumber\\&&-\frac{4}{6}\mathcal{M}^{PN}\mathcal{H}_{\mu\nu\rho M}^{(3)}{\mathcal{H}^{(3)\mu\nu\rho}}_P-8\mathcal{M}_{MP}\mathcal{M}_{QR}\mathcal{H}_{\mu\nu}^{(2)PQ}\mathcal{H}^{(2)RN\mu\nu}\nonumber\\&&+4g^2Z^{QT,P}Z^{NR,S}\mathcal{M}_{QM}(2\mathcal{M}_{TR}\mathcal{M}_{PS}-\mathcal{M}_{TP}\mathcal{M}_{RS})\\&&+4g^2Z^{QT,P}Z^{RS,N}\mathcal{M}_{QS}(2\mathcal{M}_{TP}\mathcal{M}_{RM}-\mathcal{M}_{TR}\mathcal{M}_{PM})\nonumber\\&&-4g^2\delta^N_MZ^{TU,P}Z^{QR,S}\mathcal{M}_{TQ}\left(\mathcal{M}_{UR}\mathcal{M}_{PS}-\mathcal{M}_{UP}\mathcal{M}_{RS}\right) \nonumber\\&&+\frac{8}{5}\delta^N_M\left(\mathbf{V}+\mathcal{M}_{SP}\mathcal{M}_{QR}\mathcal{H}_{\mu\nu}^{(2)PQ}\mathcal{H}^{(2)RS\mu\nu }+\frac{1}{16}\mathcal{M}^{PQ}\mathcal{H}_{\mu\nu\rho P}^{(3)}{\mathcal{H}^{(3)\mu\nu\rho}}_Q\right),\nonumber\\
0&=&4D_\nu(\mathcal{M}_{MP}\mathcal{M}_{NQ}\mathcal{H}^{(2)PQ\nu\mu })-\frac{g}{2}{X_{MNP}}^Q\mathcal{M}_{QR}D^\mu\mathcal{M}^{PR}\nonumber\\&&-2\epsilon_{MNPQR}\mathcal{M}^{PS}{\mathcal{H}^{(3)\mu\nu\rho}}_S\mathcal{H}^{(2)QR}_{\nu\rho}+\frac{1}{9}e^{-1}\epsilon^{\mu\nu\rho\lambda\sigma\tau\kappa}\mathcal{H}^{(3)}_{\nu\rho\lambda M}\mathcal{H}^{(3)}_{\sigma\tau\kappa N},\\
0&=&D_\rho\left(\mathcal{M}^{MN}{\mathcal{H}^{(3)\rho\mu\nu}}_N\right)-2gZ^{NP,M}\mathcal{M}_{NQ}\mathcal{M}_{PR}\mathcal{H}^{(2)QR\mu\nu}\nonumber\\&&-\frac{1}{3}e^{-1}\epsilon^{\mu\nu\rho\lambda\sigma\tau\kappa}\mathcal{H}^{(2)MN}_{\rho\lambda}\mathcal{H}^{(3)}_{\sigma\tau\kappa N},\label{3formfieldEQ}\\
0&=&e^{-1}\epsilon^{\mu\nu\rho\lambda\sigma\tau\kappa}Y_{MN}\mathcal{H}^{(4)N}_{\lambda\sigma\tau\kappa}-6Y_{MN}\mathcal{M}^{NP}{\mathcal{H}^{(3)\mu\nu\rho}}_P.\label{fullSD}
\end{eqnarray}

\section{Truncation ansatze}\label{truncation_ansatz}
In this appendix, we collect relevant formulae for truncations of eleven-dimensional supergravity on $S^4$ and type IIA theory on $S^3$. These give rise to $SO(5)$ and $CSO(4,0,1)$ gauged supergravities in seven dimensions, respectively. The complete $S^4$ truncation of eleven-dimensional supergravity has been constructed in \cite{11D_to_7D_Nastase1,11D_to_7D_Nastase2} while the $S^3$ truncation of type IIA theory has been given in \cite{S3_S4_typeIIA}. For both truncations, we will use the convention of \cite{S3_S4_typeIIA}.

\subsection{Eleven-dimensional supergravity on $S^4$}
The ansatz for the eleven-dimensional metric is given by
\begin{equation}
d\hat{s}^2_{11}=\Delta^{\frac{1}{3}}ds^2_7+\frac{1}{\hat{g}^2}\Delta^{-\frac{2}{3}}T^{-1}_{MN}D\mu^MD\mu^N
\end{equation}
with the coordinates $\mu^M$, $M=1,2,3,4,5$, on $S^4$ satisfying $\mu^M\mu^M=1$. $T_{MN}$ is a unimodular $5\times 5$ symmetric matrix describing scalar fields in the $SL(5)/SO(5)$ coset. The warped factor is defined by
\begin{equation}
\Delta=T_{MN}\mu^M\mu^N\, .
\end{equation}
The ansatz for the four-form field strength reads
\begin{eqnarray}
\hat{F}_{(4)}&=&\frac{1}{\hat{g}^3}\Delta^{-2}\left[\frac{1}{3!}\epsilon_{M_1\ldots M_5}\mu^M\mu^NT^{M_1M}DT^{M_2N}\wedge D\mu^{M_3}\wedge D\mu^{M_4}\wedge D\mu^{M_5}\right]\nonumber \\
& &-\frac{1}{\hat{g}^3}\Delta^{-2}U\epsilon_{(4)}+\frac{1}{4\hat{g}^2}\Delta^{-1}\epsilon_{M_1\ldots M_5}F_{(2)}^{M_1M_2}\wedge D\mu^{M_3}\wedge D\mu^{M_4}T^{M_5N}\mu^N\nonumber \\
& &+\frac{1}{\hat{g}}\tilde{S}^M_{(3)}\wedge D\mu^M-T_{MN}*\tilde{S}^M_{(3)}\mu^N\, .
\end{eqnarray}
In these equations, we have used the following definitions
\begin{eqnarray}
U&=&2T_{MN}T_{NP}\mu^{M}\mu^P-\Delta T_{MM}, \\
\epsilon_{(4)}&=&\frac{1}{4!}\epsilon_{M_1\ldots M_5}\mu^{M_1}D\mu^{M_2}\wedge D\mu^{M_3}\wedge D\mu^{M_4} \wedge D\mu^{M_5},\\
D\mu^M&=&d\mu^M+\hat{g}\tilde A^{MN}_{(1)}\mu^N,\qquad F^{MN}_{(2)}=d\tilde A^{MN}_{(1)}+\hat{g}\tilde A^{MP}_{(1)}\wedge \tilde A^{PN}_{(1)},\\
DT_{MN}&=&dT_{MN}+\hat{g}\tilde A^{MP}_{(1)}T_{PN}+\hat{g}\tilde A^{NP}_{(1)}T_{MP}\, .
\end{eqnarray}
We have denoted the vector and massive three-form fields by $\tilde{A}^{MN}_{(1)}$ and $\tilde{S}^M_{(3)}$ to avoid confusion with those appearing in \eqref{BosLag}.
\\
\indent To find the identification between the seven-dimensional fields and parameters obtained from the $S^4$ truncation and those in seven-dimensional gauged supergravity of \cite{N4_7D_Henning}, we consider the kinetic terms of various fields and the scalar potential. After multiplied by $\frac{1}{2}$, the relevant terms in the seven-dimensional Lagrangian of \cite{S3_S4_typeIIA} can be written as
\begin{eqnarray}
e^{-1}\mc{L}_{S^4}&=&\frac{1}{2}R+\frac{1}{8}D_\mu T^{-1}_{MN}D^\mu T_{MN}-\frac{1}{4}\hat{g}^2\left[2T_{MN}T_{MN}-(T_{MM})^2\right]\nonumber \\
& &-\frac{1}{16}T_{MP}^{-1}T_{NQ}^{-1}F^{MN}_{\mu\nu}F^{PQ\mu\nu}-\frac{1}{24}T_{MN}\tilde{S}^M_{\mu\nu\rho}\tilde{S}^{N\mu\nu\rho}\, .
\end{eqnarray}
Comparing with \eqref{BosLag} with $Y_{MN}=\delta_{MN}$, $Z^{MN,P}=0$, we find the following identification
\begin{equation}
T_{MN}=\mc{M}^{MN},\qquad \tilde{S}^M_{(3)}=2\mc{H}^{(3)}_M,\qquad F^{MN}_{(2)}=4\mc{H}^{MN}_{(2)}, \qquad \hat{g}=\frac{1}{4}g\, .\label{7D_higher_D_rel}
\end{equation}

\subsection{Type IIA supergravity on $S^3$}
The consistent truncation of type IIA supergravity on $S^3$ has been obtained in \cite{S3_S4_typeIIA} by taking a degenerate limit of the $S^4$ truncation of eleven-dimensional supergravity. To write down this truncation ansatz, we first split the index $M$ as $M=(i,5)$, $i=1,2,3,4$. The scalar matrix of $SL(5)/SO(5)$ coset is then given by
\begin{equation}
T^{-1}_{MN}=\begin{pmatrix}  \Phi^{-\frac{1}{4}}M^{-1}_{ij}+\Phi \chi_i\chi_j& \Phi \chi_i \\
\Phi \chi_j &\Phi
\end{pmatrix}\label{Ti_S3}
\end{equation}
where $M_{ij}$ is a unimodular $4\times 4$ symmetric matrix describing the $SL(4)/SO(4)$ coset.
\\
\indent The ten-dimensional metric, dilaton and field strength tensors of various form fields are given by
\begin{eqnarray}
d\hat{s}^2_{10}&=&\Phi^{\frac{3}{16}}\Delta^{\frac{1}{4}}ds^2_7+\frac{1}{\hat{g}^2}\Phi^{-\frac{5}{16}}\Delta^{-\frac{3}{4}}M^{-1}_{ij}D\mu^iD\mu^j,\\
e^{2\hat{\varphi}}&=&\Delta^{-1}\Phi^{\frac{5}{4}}, \\
\hat{F}_{(2)}&=& G^i_{(1)}\wedge D\mu^i+\hat{g}\mu^i G^i_{(2)},\nonumber \\
\hat{H}_{(3)}&=&\frac{1}{\hat{g}^3}\Delta^{-2}\left[-U\epsilon_{(3)}+\frac{1}{2}\epsilon_{i_1i_2i_3i_4}M_{i_1j}\mu^j\mu^kDM_{i_2k}\wedge D\mu^{i_3}\wedge D\mu^{i_4}\right]\nonumber \\
& &+\frac{1}{2\hat{g}^2}\Delta^{-1}\epsilon_{ijkl}M_{im}\mu^mF^{jk}_{(2)}\wedge D\mu^l+\frac{1}{\hat{g}}\tilde{S}_{(3)},\\
\hat{F}_{(4)}&=&\frac{1}{\hat{g}^3}\Delta^{-1}M_{ij}\mu^jG^i_{(1)}\wedge \epsilon_{(3)}+\frac{1}{2\hat{g}^2}\Delta^{-1}\epsilon_{i_1i_2i_3i_4}M_{i_4j}\mu^jG^{i_1}_{(2)}\wedge D\mu^{i_2}\wedge D\mu^{i_3}\nonumber \\
& &+M_{ij}\Phi^{\frac{1}{4}}\mu^j*G^i_{(3)}+\frac{1}{\hat{g}}G^i_{(3)}\wedge D\mu^i
\end{eqnarray}
with
\begin{eqnarray}
\epsilon_{(3)}&=&\frac{1}{3!}\epsilon_{ijkl}\mu^iD\mu^j\wedge D\mu^k\wedge D\mu^l,\qquad D\mu^i=d\mu^i+\hat{g}\tilde{A}^{ij}_{(1)}\mu^j,\\
U&=&2M_{ij}M_{jk}\mu^i\mu^k-\Delta M_{ii},\qquad \Delta=M_{ij}\mu^i\mu^j,\\
G^i_{(1)}&=&D\chi_i+\hat{g}\tilde{A}^{i5}_{(1)},\qquad G^i_{(2)}=D\tilde{A}_{(1)}^{5i}+\chi_jF^{ji}_{(2)},\\
G^i_{(3)}&=&\tilde{S}^i_{(3)}-\chi_i\tilde{S}_{(3)},\qquad F^{ij}_{(2)}=d\tilde{A}^{ij}_{(1)}+\hat{g}\tilde{A}^{ik}_{(1)}\wedge \tilde{A}^{kj}_{(1)},\\
\tilde{S}_{(3)}&=&dB_{(2)}+\frac{1}{8}\epsilon_{ijkl}\left(F^{ij}_{(2)}\wedge \tilde{A}^{kl}_{(1)}-\frac{1}{3}\hat{g}\tilde{A}^{ij}_{(1)}\wedge \tilde{A}^{km}_{(1)}\wedge \tilde{A}^{ml}_{(1)}\right).
\end{eqnarray}
\indent By comparing the truncated Lagragian and the seven-dimensional gauged Lagrangian given in \eqref{BosLag} with $Y_{ij}=\delta_{ij}$ and $Y_{55}=0$, we find the following relations
\begin{eqnarray}
\Phi&=&e^{8\phi_0},\qquad \chi_i=b_i,\qquad M^{-1}_{ij}=\widetilde{\mc{M}}_{ij},\nonumber \\
\hat{g}&=&\frac{1}{4}g,\qquad \tilde{S}^i_{(3)}=2\mc{H}^{(3)}_i\qquad F^{ij}_{(2)}=4\mc{H}^{ij}_{(2)},\qquad \tilde{F}^{i}_{(2)}= 4\mc{H}^{i5}_{(2)}\, .
\end{eqnarray}
In this case, $\mu^i$ are coordinates on $S^3$ satisfying $\mu^i\mu^i=1$.

\end{document}